\pdfoutput=1

\documentclass[aps,prd,showpacs,groupedaddress]{revtex4}               %
\usepackage{hyperref}   
\begin{document}

\preprint{RCTP-1101}

\title{Intrinsic Dirac Behavior of Scalar Curvature 
in a Quaternionic Weyl-Cartan Geometry}


\author{J.E. Rankin}
\email[]{jrankin@panix.com}
\affiliation{Rankin Consulting, Concord, CA}

\date{June 27, 2017}

\begin{abstract}

The ``spin-up'' and ``spin-down'' projections of the second order, chiral form of Dirac Theory are shown to fit a superposition of forms predicted in an earlier classical, complex scalar gauge theory\cite{rankin.caqg}. In some sense, it appears to be possible to view the two component Dirac spinor as a single component, quaternionic, spacetime scalar. ``Spin space'' transformations can be considered transformations of the internal quaternion basis. Essentially, quaternionic Dirac Theory projects into the complex plane neatly, where spin becomes related to the self-dual antisymmetric part of the metric. The correct Dirac eigenvalues and well-behaved eigenfunctions project intact into a pair of complex solutions for the scalar curvature in the earlier theory's Weyl-Cartan type geometry. Some estimates are made for predicted, interesting atomic and subatomic scale phenomena. A form of electromagnetic quanta appears. A generalization of the complex geometric structure is then sketched in an appendix that allows quaternionic gauges and curvatures, and has some Weyl nonmetricity mixed with torsion. It appears to be a well defined structure, and leads to the full, second order, quaternionic Dirac Equation form, and a first order equation for a closely related, auxiliary wavefunction. A family of ``free particle'' solutions is examined in the Lorentzian limit of the symmetric part of the metric. More generally, when limited to two quaternion dimensions (just two components), reasonably similar solutions can be superposed linearly into new solutions, and separate into two families with different commutation characteristics. The integrability conditions for the equation for the auxiliary wavefunction impose six conditions on the original wavefunction, satisfied for the ``free particle'' solutions examined. Covariance is examined. The Darwin solution for the hydrogen atom is examined.
%

\end{abstract}

\maketitle


\section{Introduction}

This work presents results from an extended effort to identify spin 1/2, quantum mechanical wavefunctions with the scalar curvature of a Weyl-like Cartan geometry with a self-dual antisymmetric part to the metric\cite{rankin.caqg,rankin.9404023,rankin.mg7}. The earlier papers demonstrated that the natural geometric identifications made therein imply as a simplest case that in the limit that the spacetime is Lorentzian 
($ \hat g_{\mu \nu}\approx \hat \eta _{\mu \nu} $), 
there exists a geometric wavefunction 
$ \psi $ which obeys
{\samepage 
\begin{eqnarray}
\hat \eta^{\mu \nu} \left [ \psi_{,\mu ,\nu} + \imath q A_{\mu ,\nu} 
\psi + 2 \imath q A_{\mu} \psi_{,\nu} - q^{2} A_{\mu} A_{\nu} \psi 
\right ] 
\nonumber \\
+ M^{2} \psi 
\pm \imath q \sqrt {E^{2} - B^{2} + 2 \imath \vec E 
{\bf \cdot} \vec B } \; \, \psi 
= 0 
\label{geometric.wave.eq}
\end{eqnarray}}%
In this, $ A_{\mu} $ is the electromagnetic potential which generates fields 
$ \vec E $ and $ \vec B $, $ q = e / (\hbar c) $, 
$ M = (m_{0} c) / \hbar $, 
$ \; e $ is the electronic charge, 
$ m_{0} $ is the electron rest mass, and the notation ``
$ , \mu $'' indicates the partial derivative with respect to 
$ x^{\mu} $. The actual scalar curvature of the geometry, 
$ B $ (not to be confused with the magnitude of the magnetic field vector 
$ \vec B $), is given by
\begin{equation}
B = \psi^{-2}
\label{linear.change}
\end{equation}
and it is clearly complex valued.

The appendix in reference \cite{rankin.caqg} demonstrates that solutions of equation (\ref{geometric.wave.eq}) match solutions to Dirac's Equation for the case of uniform, constant, non-null electromagnetic fields. In that proof, the Dirac Equation itself is taken to be the (chiral) second order form of the equation,
{\samepage 
\begin{eqnarray}
\hat \eta^{\mu \nu} \left [ \psi_{,\mu ,\nu} + \imath q A_{\mu ,\nu} 
\psi + 2 \imath q A_{\mu} \psi_{,\nu} - q^{2} A_{\mu} A_{\nu} \psi 
\right ] 
\nonumber \\
+ M^{2} \psi 
+ \imath q \vec \sigma {\bf \cdot} \left ( \vec E + \imath 
\vec B \right ) \psi 
= 0 
\label{dirac.wave.eq}
\end{eqnarray}}%

As a quick reference summary\cite{rankin.caqg,rankin.9404023,rankin.mg7} (or see the appendix of this paper for a more detailed, quaternion generalization of these steps), equation (\ref{geometric.wave.eq}) follows from a Weyl-like Cartan geometric model with gauge invariant variables defined for cases in which the 
curvature $ B \neq 0 $. For the metric, that definition is
\begin{equation}
\hat g_{\mu \nu} = (B/C)\, g_{\mu \nu}
\label{def.ghat}
\end{equation}
where the constant 
$ C = \pm 1 $. This is just the product of the scalar curvature and the gauge varying metric, but the case 
$ C = -1 $ is included originally to handle cases of 
$ B < 0 $, and 
$ C $ is retained as a legitimate flexibility of equation (\ref{def.ghat}) even though 
$ C = 1 $ is used 
here\cite{galehouse.minus.C,galehouse.ijtp.1,galehouse.ijtp.2}. Other possible values for $C$ will be briefly examined later.

The analog to equation (\ref{def.ghat}) for the Weyl vector is
{\samepage 
\begin{eqnarray}
\hat v_{\mu} & = & v_{\mu} - [{\textstyle{1 \over 2}} \, ln \, 
( B / C ) ]_{,\mu}
\nonumber \\
& = & v_{\mu} - ({\textstyle{1 \over 2}} \, ln \, B )_{,\mu} 
\label{def.vhat}
\end{eqnarray}}%
where in common electromagnetic gauge choices,
{\samepage 
\begin{eqnarray}
v_{\mu} & = & \imath [ e / ( \hbar c ) ] A_{\mu} 
\nonumber \\
 & = & \imath q A_{\mu} 
\label{def.v}
\end{eqnarray}}%
with $ A_{\mu} $ real. Then
{\samepage 
\begin{eqnarray}
p_{\mu \nu} & = & v_{\nu ,\mu} - v_{\mu ,\nu} \nonumber \\
& = & \hat v_{\nu ,\mu} - \hat v_{\mu ,\nu} 
\nonumber \\
& = & \hat p_{\mu \nu}
\nonumber \\
& = & \imath q F_{\mu \nu} 
\label{def.p}
\end{eqnarray}}%
where $ F_{\mu \nu} $ is the standard Maxwell tensor, the curl of 
$ A_{\mu} $, and it is real.

These relations together with the geometric kinematics imply that the gauge invariant variables are not independent, but must instead obey the kinematic identity
\begin{equation}
\hat R + 6 \hat v^{\mu}_{\; \; \| \mu} + 6 \hat v^{\mu} 
\hat v_{\mu} 
+ \hat a^{\mu \nu} \hat p_{\mu \nu} 
 = C ( 1 + {\textstyle{1 \over 4}} \, \hat a ) 
\label{def.identity}
\end{equation}
Here the ``$ {}_{\|} $'' derivative is the Riemannian covariant derivative based on 
$ \hat g_{\mu \nu} $, $ \hat a_{\mu \nu} $ 
is the self dual antisymmetric part of the metric, and 
$\hat a = \hat a_{\mu \nu} \hat a^{\mu \nu}$ (indices raised and lowered by 
$ \hat g^{\mu \nu} $ and 
$ \hat g_{\mu \nu} $). This identity is the source of equation (\ref{geometric.wave.eq}) once the dynamics is also specified, and that dynamics is given by essentially the standard Einstein-Maxwell action\cite{adler.bazin} expressed directly in the gauge invariant variables, plus the cosmological constant term and constraints,
{\samepage 
\begin{eqnarray}
I & = & \int {[ (\hat R - 2 \sigma ) - {\textstyle{1 \over 2}}\, 
j^2 (\hat p_{\mu \nu} \hat p^{\mu \nu}) + {\textstyle{1 \over 2}} 
\{ \hat \beta [ \hat R + 6 \hat v^{\mu}_{\; \; \| \mu} 
+ 6 \hat v^\mu \hat v_{\mu} +\hat a^{\mu \nu} \hat p_{\mu \nu}}
\nonumber \\
& & - C ( 1 + {\textstyle{1 \over 4}}\, 
\hat a ) ] + 2 \hat \lambda^{\mu \nu}
\hat p_{\mu \nu} + \hat \gamma (\hat a - K^2) + CC \} ] 
\sqrt {-\hat g}\, d^4x
\label{complex.gi.ahat.action}
\end{eqnarray}}%
where the 
$ CC $ is the complex conjugate of all that precedes it in the brackets. Equation (\ref{def.identity}) is included in the action as a constraint, as it must be in order to vary the gauge invariant variables independently, 
$ \sigma $ is the dimensionless cosmological constant, 
$ K $ is a constant (equal to 
$ 6 \imath $), 
$ j^2 = -2[\hbar/(ec)]^2 G b_{0} $ 
($ G $ is Newton's Constant, and 
$ b_{0} $ is the scale factor defined in equation (\ref{b0.C}), which relates natural dimensionless Lorentzian coordinates 
$ x^{\mu} $ to Lorentzian lab coordinates 
$ x^{\mu}_{LAB} $ via 
$ x^{\mu} = \sqrt{ b_{0} } \, x^{\mu}_{LAB} $),
$ \hat \lambda^{\mu \nu} $ is real, so its variation constrains 
$ \hat p_{\mu \nu} $ to be imaginary to fit equation (\ref{def.v}), and the constraint 
$ \hat a = K^2 $ is simply a zero order approximation (based on geometric considerations) to some better theory of the antisymmetric part of the metric yet to be determined. Such a better theory should become a theory of rest mass, among other things. Note that when varying the self dual quantity 
$ \hat a_{\mu \nu} $ in the action, the identity 
$ \hat a^{\mu \nu} \hat p_{\mu \nu} = (1/2) \hat a^{\mu \nu} \hat P_{\mu \nu} $ is used before variation, where
\begin{equation}
\hat P_{\mu \nu} = \hat p_{\mu \nu} + i \, {}^{*} \hat p_{\mu \nu} 
\label{big.p.hat}
\end{equation}
and 
$ {}^{*} \hat p_{\mu \nu} $ is the dual of 
$ \hat p_{\mu \nu} $.

In the sections to follow, the scalar nature of 
$ B $ will lead to the conclusion that at least in some sense, the full Dirac wavefunction 
$ \psi $ of equation (\ref{dirac.wave.eq}) can be treated as a quaternionic spacetime scalar rather than necessarily always having spacetime spinor properties. 
It will be seen that this scalar property follows by projecting (``spin-up'' and ``spin-down'') very general solutions to the equivalent quaternionic form of Dirac Equation (\ref{dirac.wave.eq}) into the complex scalar solutions of equation (\ref{geometric.wave.eq}) in the complex plane. This projecting preserves the full content of the Dirac 
$ \psi $, with no information lost, and so it can be considered an embedding of Dirac Theory. Those sections will then be followed by a discussion of predicted quantum and other effects, and then finally the appendix will present an introduction to a generalized structure in which the scalar curvature 
$ B $ and gauge transformations may become general, quaternionic quantities. In that generalized structure, an equal mixture of nonmetricity and torsion will cause the structure to produce the full, quaternionic, second order Dirac Equation form (see equation (\ref{qdirac.wave.eq})) in place of equation (\ref{geometric.wave.eq}).

Finally, before proceeding further, a warning about some unconventional notation is in order. Throughout this paper, certain operations on quantities, such as complex conjugation, the formation of the hermitian adjoint or quaternion conjugate, or the formation of a tensor dual, will all be denoted by special symbols placed on the upper {\it left} side of the affected quantity, rather than the more conventional upper right side. Thus, instead of 
$ X^{\dagger} $, this paper will use 
$ \, {}^{\dagger} \! X $. This is done to keep those operational symbols out of the physical space occupied by tensor indices and such, and equation (\ref{big.p.hat}) above is already an example of this in practice.

\section{Quaternionic Dirac Theory}
\label{quat.dirac.theory}

This section will present an immediate and simple translation of equation (\ref{dirac.wave.eq}) into the equivalent, real quaternion form. To keep this as simple as possible, no effort will be made here to go into general formulations of quaternionic quantum mechanics\cite{finkelstein,adler.qm}. Rather, the two component chiral spinor will be translated into a single real quaternion variable obeying an equation immediately analogous and equivalent to the spinor form, and still clearly containing the spinor predecessor.

\subsection{Quaternions}

In this presentation, the basic quaternions will be taken as abstract mathematical objects similar to the number 
$ 1 $ and the imaginary unit $ \imath $ 
in complex numbers. Specifically, they are taken to be the four quantities 
$ Q_{\mu} $, where the subscript does {\it not} imply the quantities are a four vector, and where
\begin{equation}
Q_{0} = \sigma_{0}
\label{def.q0}
\end{equation}
and
\begin{equation}
Q_{k} = - \imath \sigma_{k}
\label{def.qk}
\end{equation}
for $ k = 1,2,3 $, where the $ \sigma_{k} $ 
are the standard Pauli spin matrices\cite{schiff}, and 
$ \sigma_{0} $ is the unit matrix. The basic properties of quaternions are reviewed in many references, such as Adler\cite{adler.qm}, and Morse and Feshbach\cite{morse.feshbach}, and there are many representations of them which may differ from those of equations (\ref{def.q0}) and (\ref{def.qk}), yet which are algebraically isomorphic to those quantities. Such possibly isomorphic representations will be denoted here by 
$ Q^{\prime}_{\mu} $, and they may happen to be assigned particular coordinate transformation properties for convenience, unlike the 
$ Q_{\mu} $, which are mathematical invariants.

The $ Q_{k} $ have an obvious vector form
\begin{equation}
\vec Q = \sum_{ k = 1 }^{3} \hat e_{k} Q_{k} 
\label{def.qvec}
\end{equation}
where the $ \hat e_{k} $ 
are the Cartesian unit vectors. A completely analogous form exists for the 
$ Q^{\prime}_{k} $,
\begin{equation}
\vec Q^{\prime} = \sum_{ k = 1 }^{3} \hat e^{\prime}_{k} Q^{\prime}_{k} 
\label{def.qvecprime}
\end{equation}
although now the 
$ \hat e^{\prime}_{k} $ may be unit vectors in one of the curvilinear coordinate systems in common use\cite{morse.feshbach}.

For definiteness, choose a set of 
$ Q^{\prime}_{\mu} = Q_{\mu} $ (technically, these are invariants). A general real quaternion variable will be any quantity of the form
\begin{equation}
\psi = \psi_{0 R}Q_{0} - \psi_{1 I}Q_{1} + \psi_{1 R}Q_{2} -
\psi_{0 I}Q_{3}
\label{def.qpsi}
\end{equation}
where the coefficients $ \psi_{n R} $ and 
$ \psi_{n I} $ are all real numbers. The notation has been chosen with an eye to the wavefunction values, and in matrix form it is
\begin{equation}
\psi =
\left (
\begin{array}{lr}
\psi_{0 R} + \imath \psi_{0 I} & -\psi_{1 R} + \imath \psi_{1 I} \\
\psi_{1 R} + \imath \psi_{1 I} & \psi_{0 R} - \imath \psi_{0 I} 
\end{array}
\right )
\label{def.m.qpsi}
\end{equation}
This is in fact the quaternion representation of the spinor wavefunction
\begin{equation}
\psi =
\left (
\begin{array}{c}
\psi_{0 R} + \imath \psi_{0 I} \\
\psi_{1 R} + \imath \psi_{1 I} 
\end{array}
\right )
\label{def.spsi}
\end{equation}
where the notation $ \psi_{n R} $ and 
$ \psi_{n I} $ now clearly gives the real and imaginary parts of the two rows in the spinor. Notice that the quaternion has no additional information over the spinor, because the second column only contains conjugate forms of quantities in the first column. The relationship of the columns, given in equation (\ref{def.m.qpsi}), is common to all real quaternions, including real quaternion equations.

Additionally, the inner product of the wavefunction with itself is still 
$ {}^{\dagger} \psi \psi $ (with the 
``$ {}^{\dagger} $'' to the left), which is the same as the spinor result. However, the equivalent inner product of two different wavefunctions 
$ \zeta $ and $ \psi $ will {\it not} be 
$ {}^{\dagger} \zeta \psi $, but rather will transition from spinors to quaternions via
\begin{equation}
{}^{\dagger} \zeta \psi \rightarrow 
\frac{1}{2} \left [ \, {}^{\dagger} \zeta \psi - Q_{3} 
{}^{\dagger} \zeta \psi Q_{3} \right ]
\label{def.general.inner.prod}
\end{equation}
as may be verified by direct calculation. This is the complex projection of a quaternionic value into the 
$ Q_{0} $, $ Q_{3} $ complex plane)\cite{rotelli}. Thus, the inner product of two different wavefunctions is still a complex number, just as it is in spinors. Furthermore, when 
$ \zeta = \psi $, this reduces to 
$ {}^{\dagger} \psi \psi $, as it must.

\subsection{Translation from Spinor Equation to Quaternion Equation}

Equations (\ref{def.spsi}) and (\ref{def.m.qpsi}) already give the translation of 
$ \psi $ from spinor to quaternion. A spinor equation such as equation (\ref{dirac.wave.eq}) will contain the spinor 
$ \psi $ with various operations on it, including products of it with coefficients which may be complex, and are assumed to commute with it, and also the product of 
$ \vec \sigma {\bf \cdot} \left ( \vec E + \imath \vec B \right ) $ 
with it. The factor 
$ \vec \sigma $ in the spin term is {\it not} assumed to commute with 
$ \psi $.

Now in order to handle the factor 
$ \imath $ in a coefficient, note that for the quaternion 
$ \psi $,
\begin{equation}
-\psi Q_{3} =
\left (
\begin{array}{lr}
\imath (\psi_{0 R} + \imath \psi_{0 I}) & - \imath 
(-\psi_{1 R} + \imath \psi_{1 I}) \\
\imath (\psi_{1 R} + \imath \psi_{1 I}) & - \imath 
(\psi_{0 R} - \imath \psi_{0 I}) 
\end{array}
\right )
\label{def.i.m.qpsi}
\end{equation}
But the first column of this is clearly 
$ \imath $ times the original column, while the second column maintains the correct conjugate relationships for a quaternion. In point of fact, the choice of 
$ -Q_{3} $ can be generalized to other fixed directions in 
$ \hat n {\bf \cdot} \vec Q $, where 
$ \hat n $ is an arbitrary constant unit vector. However this will not be done here in order not to lose immediate transparency between quaternion and spinor forms.

Equation (\ref{def.i.m.qpsi}) leads to a general rule for translating a factor 
$ \imath $ multiplying a spinor into the equivalent quaternion value, and that is
\begin{equation}
\imath \rightarrow |(-Q_{3})
\label{i.leap}
\end{equation}
In this, the 
``$ | $'' is the ``leap-over operator'' or ``barred operator'', which signifies that the quantity immediately following it is to be applied to the right side of any product of further quantities to the right of the term in a product. The rightmost 
$ | $ is applied first, then any additional such operations are applied in order proceeding to the left when more than one such 
$ | $ operation appears in a product, so that 
$ (| Q_{2} ) ( | Q_{3} ) \psi = \psi Q_{3} Q_{2} $. This operation is required to handle the possible two-sided nature of any quaternion product (a ``split product''), and has been in use in the literature for some time by researchers of quaternionic Dirac Theory, such as De Leo and Rotelli\cite{rotelli,De.Leo.1995}, and Schwartz\cite{schwartz.leap}. For example, the spinor eigenvalue equation related to the 
$ z $ component of angular momentum translates via
\begin{equation}
\psi_{, 3} = \imath m \psi 
\rightarrow |(-Q_{3}) m \psi = 
- m \psi Q_{3} 
\label{trans.mphi}
\end{equation}

For completeness, the leap-over operation 
``$ (X) \| $'' is also defined, and will denote that 
$ X $ is applied to the left side of the product of any quantities to the left of it in a product, with the leftmost 
$ \| $ applied first, then any additional such operations are applied in order proceeding to the right when more than one such 
$ \| $ operation appears in a product. In practice, the final position in either type of leap-over operation is fixed when the equation is evaluated explicitly or implicitly (for 
$ \psi $ in this case). If the position is then shifted somehow without reevaluating the equation, contradictory, invalid results can appear. In this regard however, multiplication ``through'' the newly fixed leap-over operator 
$ | (X) $ or $ (X) \| $  by any quantity that commutes with both its fixed and leap-over forms, 
will {\it not} force the relocation of that leap-over operator 
after its position has been fixed in order to evaluate 
the equation. Such exceptions will be seen to include phase transformations of the wavefunction (see equations (\ref{psi.phase.trans}) and (\ref{qpsi.phase.trans})), and so are not trivial.

It will be noted that 
$ |(-Q_{3}) $ does in fact effectively commute with a standard quaternion value 
$ \psi $ by definition, and so it has the same property that 
$ \imath $ had in the spinor equation vis-\`{a}-vis the spinor 
$ \psi $. This property is actually quite useful in handling spinor/quaternion expression translations back and forth. Of course, the translation back to spinor form involves merely using the first column of the quaternion expression. The spinors remain explicitly visible in the first column throughout the quaternion expression.

This still leaves the product of 
$ \psi $ with 
$ \vec \sigma {\bf \cdot} \left ( \vec E + \imath \vec B \right ) $ 
in the spin term, in which the $ \vec \sigma $ will not commute with the spinor 
$ \psi $. But equations (\ref{def.qk}) and (\ref{def.qvec}) imply that
\begin{equation}
\vec \sigma = \imath \vec Q 
\label{def.qvec.sigmavec}
\end{equation}
This and equation (\ref{i.leap}) then give immediately the full real quaternion translation of spinor equation (\ref{dirac.wave.eq}) as
{\samepage 
\begin{eqnarray}
\hat \eta^{\mu \nu} \left [ \psi_{,\mu ,\nu} - q A_{\mu ,\nu} 
\psi Q_{3} - 2 q A_{\mu} \psi_{,\nu} Q_{3} - q^{2} A_{\mu} A_{\nu} 
\psi \right ] 
\nonumber \\
+ M^{2} \psi 
- q \vec Q {\bf \cdot} \left ( \vec E \psi - \vec B \psi Q_{3} 
\right ) 
= 0
\label{qdirac.wave.eq}
\end{eqnarray}}%
This could also be written as
{\samepage 
\begin{eqnarray}
\hat \eta^{\mu \nu} \left [ \psi_{,\mu ,\nu} - q A_{\mu ,\nu} 
\psi Q_{3} -2 q A_{\mu} \psi_{,\nu} Q_{3} - q^2 A_{\mu} A_{\nu} 
\psi \right ] 
\nonumber \\
+ M^{2} \psi 
- q \vec Q {\bf \cdot} \left ( \vec E - \vec B | Q_{3} 
\right ) \psi 
= 0
\label{leap.qdirac.wave.eq}
\end{eqnarray}}%
and clearly other alternative forms using 
$ |(-Q_{3}) $ are possible. It is also possible to use 
$ \vec Q^{\prime} $ and curvilinear coordinate forms in the spin term, and later it will be clear that various ``spin space'' transformations will allow passage between different representations of the 
$ \vec Q^{\prime} $. When transforming the 
$ \vec Q^{\prime} $, the 
$ |(-Q_{3}) $ quantities all remain unprimed and unchanged, because that term is treated as a coordinate invariant, like 
$ \imath $ in the coefficients in the spinor forms.

It is also possible to translate the complex equation (\ref{geometric.wave.eq}) into quaternion form for comparison. It becomes
{\samepage 
\begin{eqnarray}
\hat \eta^{\mu \nu} \left [ \psi _{,\mu ,\nu} - q A_{\mu ,\nu}\psi
Q_{3} -2 q A_{\mu} \psi _{,\nu} Q_{3} -q^2 A_{\mu} A_{\nu} 
\psi \right ] 
\nonumber \\
+ M^{2} \psi 
\pm q \sqrt {- \left [E^2 - B^2 - 2 \vec E
{\bf \cdot} \vec B |Q_{3} \right ]} \; \, \psi
= 0
\label{qgeometric.wave.eq}
\end{eqnarray}}%
where the quaternion form of $ \psi $ here is
\begin{equation}
\psi =
\left (
\begin{array}{cc}
\psi_{R} + \imath \psi_{I} & 0 \\
0 & \psi_{R} - \imath \psi_{I} 
\end{array}
\right )
\label{def.m.qpsi.complex}
\end{equation}
or equivalently,
\begin{equation}
\psi = \psi_{R} Q_{0} - \psi_{I} Q_{3} 
\label{def.qpsi.complex}
\end{equation}
This is how a complex number appears in this quaternion form.

A comparison of equations (\ref{leap.qdirac.wave.eq}) and (\ref{qgeometric.wave.eq}) now suggests immediately that in many ways they are {\it the same} equation. This is seen by noting that
{\samepage 
\begin{eqnarray}
\left [-\vec Q^{\prime} {\bf \cdot} \left ( \vec E - \vec B | Q_{3} 
\right ) \right ] \left [-\vec Q^{\prime} {\bf \cdot} \left ( 
\vec E - \vec B | Q_{3} \right ) \right ] = 
\nonumber \\
- \left [E^2 - B^2 - 2 \vec E {\bf \cdot} \vec B |Q_{3} \right ]
\label{spin.squared}
\end{eqnarray}}%
Clearly both equations have the same general form in which the spin terms are the square root of the same quantity. However, they are different roots from the quaternion viewpoint.

Equation (\ref{qgeometric.wave.eq}) extracts the roots within the complex plane, while equation (\ref{leap.qdirac.wave.eq}) uses a root that extends into more general quaternion space. Nevertheless, the solutions of equation (\ref{leap.qdirac.wave.eq}) when the electromagnetic field is non-null can be expanded in the eigenspinors/eigenquaternions of the electromagnetic field. The ``spin-up'' and ``spin-down'' projections of the result will each separately obey a form of equation (\ref{qgeometric.wave.eq}) with a superposition of terms with both plus and minus values of the square root in the equation. This is not an eigenstate of equation (\ref{qgeometric.wave.eq}) generally, which would have to contain only a single sign for the square root term. However, generally eigenstates of equation (\ref{qgeometric.wave.eq}) do not have real energy eigenvalues. Rather, those energy eigenvalues are complex valued, with corresponding poor behavior in the eigenfunctions, although exceptions include the case of uniform, constant, non-null, electromagnetic fields\cite{rankin.caqg}. However, the energy eigenvalues of equation (\ref{dirac.wave.eq}) (or equation (\ref{leap.qdirac.wave.eq})) are well known to be satisfactory with well-behaved eigenfunctions. The ``spin-up'' and ``spin-down'' projections of those solutions inherit the eigenvalues and that good behavior. Thus, the Dirac Equation provides well-behaved solutions to equation (\ref{qgeometric.wave.eq}), a subject that will be addressed more thoroughly in the next section. However, some additional points about Dirac Theory and quaternionic Dirac Theory deserve mention first.

\subsection{Quaternion Form of Dirac Theory}
The translation from spinor to quaternion can be extended to first order Dirac Theory and a Lagrangian\cite{rotelli}, and a full formalism developed. The Dirac 
$ \gamma^{\mu} $ translate just like other expressions in spinor equations. For example, the chiral representation giving equations (\ref{dirac.wave.eq}) and (\ref{leap.qdirac.wave.eq}) corresponds to
\begin{equation}
\gamma^{0} =
\left (
\begin{array}{cc}
0 & - \imath \sigma_{0} \\
- \imath \sigma_{0} & 0 
\end{array}
\right )
\rightarrow 
\left (
\begin{array}{cc}
0 & Q^{\prime}_{0} | Q_{3} \\
Q^{\prime}_{0} | Q_{3} & 0 
\end{array}
\right )
\label{def.gamma0}
\end{equation}
and
\begin{equation}
\gamma^{k} =
\left (
\begin{array}{cc}
0 & \imath \sigma_{k} \\
- \imath \sigma_{k} & 0 
\end{array}
\right )
\rightarrow 
\left (
\begin{array}{cc}
0 & -Q^{\prime}_{k} \\
Q^{\prime}_{k} & 0 
\end{array}
\right )
\label{def.gammak}
\end{equation}
for $ k = 1,2,3 $, in the first order Dirac Equation
\begin{equation}
\gamma^{\mu} \left ( \zeta_{, \mu} - q A_{\mu} \zeta 
Q_{3} \right ) = M \zeta 
\label{def.qdirac.1st}
\end{equation}
where $ \zeta $ is a column vector containing two rows of quaternions. Clearly ``spin space'' transformations in Dirac Theory become transformations now on the
$ Q^{\prime}_{\mu} $, the internal quaternion basis vectors. Such transformations will transform between various representations of the 
$ Q^{\prime}_{\mu} $.

The particular form of 
$ \gamma^{0} $ causes the formation of 
$ \, {}^{\dagger} \gamma^{0} $ (the hermitian adjoint, with the 
``$ {}^{\dagger} $'' written to the left of its target 
$ \gamma^{0} $ rather than to the right) to present more than one suggested analog translating from spinor to quaternion form, depending on whether the 
``$ {}^{\dagger} $'' is taken before, or after, translating from spinor to quaternion. The two possibilities are always related to each other in that either one can be obtained from the other by multiplying it by 
$ [ -(Q_{3})\| \; \; |(Q_{3}) ] $. However, generally most of spinor Dirac Theory translates fairly clearly, and that will not be detailed here, but there are a few points relating to the theory which produced equation (\ref{geometric.wave.eq}) originally which will be mentioned.

From the viewpoint of the theory of equation (\ref{geometric.wave.eq}), second order Dirac Theory is expected to be more fundamental than first order Dirac Theory, which can be derived from a single, second order, chiral equation\cite{sakurai}. In that same spirit, there is a spinor action principle which leads directly to the second order, spinor Dirac Equation forms, as opposed to the standard Dirac action which leads to the first order equations\cite{bade.jehle,morse.feshbach,rotelli}. The second order equations follow from
{\samepage 
\begin{eqnarray}
I & = & \int { \{ [\hbar / ( 4 M )] [ - \imath \eta^{\mu \nu} 
( \bar{ \zeta }_{, \mu} - \imath q A_{\mu} \bar{ \zeta } ) 
( \zeta_{, \nu} + \imath q A_{\nu} \zeta ) 
+ \imath M^{2} \bar{ \zeta } \zeta } 
\nonumber \\ 
 & & - \imath \bar{ \zeta }_{, \mu} \Sigma^{\mu \nu} \zeta_{, \nu}
- q A_{\mu} ( \bar{ \zeta } \Sigma^{\mu \nu} \zeta )_{, \nu}
+ CC ] - [1 / (16 \pi c)] F_{\mu \nu} F^{\mu \nu} \} d^4 x 
\label{action.2nd}
\end{eqnarray}}%
where the 
$ CC $ is the complex conjugate of all that precedes it in the brackets, 
$ \bar{ \zeta } = {}^{\dagger} \zeta \gamma^{0} $, and
\begin{equation}
\Sigma^{\mu \nu} = - \frac{1}{2} (\gamma^{\mu} \gamma^{\nu} 
- \gamma^{\nu} \gamma^{\mu})
\label{def.sigma}
\end{equation}

However, the action of equation (\ref{action.2nd}) does {\it not} give standard Dirac Theory. This is easily seen in a chiral representation, where the two chiral parts of 
$ \zeta $ will each obey a second order Dirac Equation form, but without any first order Dirac Equations relating them to each other. The two chiral parts are actually independent of each other. J. T. Wheeler pointed out to me that this independence is equivalent to the presence of a negative energy density ghost wavefunction field, in addition to the standard positive energy density wavefunction field\cite{wheeler.trans,rankin.9404023}. Moreover, the same situation arises in the theory that leads to equation (\ref{geometric.wave.eq}), meaning that it also includes such a ghost. For that reason, the action of equation (\ref{action.2nd}) remains of interest here, since it generates the analogous Dirac Theory generalization. And, it is also amenable to modifications which yield the general relativistic ``conformal term'' inherited from equation (\ref{def.identity}), instead of the 
$ {\textstyle{{1} \over {4}}} \hat R \psi $ type of term that results from the standard general relativistic formulation of the {\it spinor} form of Dirac Theory (see the appendix in reference \cite{rankin.caqg}). Clearly equation 
(\ref{def.identity}) points to a coefficient of 
$ {\textstyle{{1} \over {6}}} $ rather than 
$ {\textstyle{{1} \over {4}}} $.

Besides the antisymmetric part of the metric, 
$ \hat a_{\mu \nu} $, this ghost, negative energy density ``Wheeler'' field, is an additional classical field that appears in this framework. And, since it is a {\it classical} framework, it is worth noting that classical physics has previously shown tolerance for a negative energy density field in its midst. Prior to the appearance of General Relativity, gravitation was described by the Newtonian Gravitational Field\cite{symon}, and that field is necessarily a negative energy density field. Work is extracted from the field as the field is built up, so it is a negative energy density field by definition, yet classical physics coexists with it without catastrophe. For this reason, I take the ghost field as seriously worth examining. Indeed, a similar pair of positive-negative energy density fields has been studied in some detail by Doroshkevich et al. in recent literature\cite{pos.neg.pair}. There will be a few more comments on this ghost, negative energy density field later.

\section{Eigenquaternions and Complex/Spin Projections}
\label{projections}

\subsection{Eigenspinors/Eigenquaternions}
Consider the eigenspinor equation
\begin{equation}
\imath \vec \sigma {\bf \cdot} \left ( \vec E + \imath \vec B \right ) \xi 
= - \vec Q^{\prime} {\bf \cdot} \left ( \vec E + \imath \vec B \right ) 
\xi = \lambda \xi 
\label{def.eig.spinor}
\end{equation}
and its quaternion equivalent
\begin{equation}
- \vec Q^{\prime} {\bf \cdot} \left ( \vec E - \vec B | Q_{3} \right ) 
\xi = \lambda \xi 
\label{def.eig.quat}
\end{equation}
In these two equations, 
$ \xi $ is a spinor and its equivalent quaternion respectively, and 
$ \lambda $ is an eigenvalue which commutes with all the other terms in both cases. These 
$ \xi $ are essentially the eigenspinors/eigenquaternions of the electromagnetic field\cite{carmeli,rankin.caqg}. Hereafter, the term ``eigenvector'' will be used to denote these quantities which can be expressed as either spinors or quaternions.

Since $ \lambda $ commutes with everything in the equation, the operator 
$ -\vec Q^{\prime} {\bf \cdot} \left ( \vec E - \vec B | Q_{3} \right ) $ can be applied to 
$ \xi $ twice. Then equation (\ref{spin.squared}) gives that 
\begin{equation}
\lambda^{2} = - \left [E^2 - B^2 - 2 \vec E {\bf \cdot} \vec B |Q_{3} \right ]
\label{eig.squared}
\end{equation}
For non-null fields, this has the two distinct values
\begin{equation}
\lambda_{\pm} = \pm \sqrt { - \left [E^2 - B^2 - 2 \vec E {\bf \cdot} 
\vec B |Q_{3} \right ]}
\label{eig}
\end{equation}
where the square root is understood to be taken as a complex number of the general form 
$ \lambda_{R} + [ |(-Q_{3}) ] \: \lambda_{I} $ with 
$ \lambda_{R} $ and 
$ \lambda_{I} $ real. But these two values, 
$ \lambda_{\pm} $, are exactly the two choices for the root in equation (\ref{qgeometric.wave.eq}). Furthermore, the values of 
$ \lambda $ do commute with everything in equation (\ref{def.eig.quat}) as required, because 
$ |(-Q_{3}) $ effectively commutes with all the other terms. This commutation includes the fact that 
$ |(-Q_{3}) $ clearly commutes with 
$ |(-Q_{3}) $, and additionally, 
$ |(-Q_{3}) \, |(-Q_{3}) = -1 $.

For each of the two values of 
$ \lambda $, the matching 
$ \xi $ is the eigenvector for that eigenvalue, either 
$ \xi_{+} $ or $ \xi_{-} $. Since the fields are assumed to be non-null, the eigenvalues are nonzero and different, and the two eigenvectors are independent\cite{carmeli,rankin.caqg}. Thus, an arbitrary spinor/quaternion 
$ \psi $ can be expanded in terms of the two as
\begin{equation}
\psi = a \xi_{+} + b \xi_{-} 
\label{expand.psi}
\end{equation}
This is most easily done with the eigenspinors, with translation to eigenquaternions following the expansion. Then the coefficients 
$ a $ and $ b $ are complex functions just as 
$ \lambda $ is, and they commute with everything at least as well as 
$ \lambda $ does in both forms.

Now define
\begin{equation}
\psi_{+} = a \xi_{+}
\label{def.psi.+}
\end{equation}
and
\begin{equation}
\psi_{-} = b \xi_{-}
\label{def.psi.-}
\end{equation}
so that
\begin{equation}
\psi = \psi_{+} + \psi_{-}
\label{def.psi.sum+-}
\end{equation}
But now, equation (\ref{def.eig.quat}) implies that
\begin{equation}
- \vec Q^{\prime} {\bf \cdot} \left ( \vec E - \vec B | Q_{3} \right ) 
\psi = \lambda_{+} \psi_{+} + \lambda_{-} \psi_{-} 
\label{eig.spin.expand}
\end{equation}
and equation (\ref{leap.qdirac.wave.eq}) becomes
{\samepage 
\begin{eqnarray}
\hat \eta^{\mu \nu} [ ( \psi_{+} & + & \psi_{-} )_{,\mu ,\nu} - 
q A_{\mu ,\nu} ( \psi_{+} + \psi_{-} )
Q_{3} \nonumber \\
& - & 2 q A_{\mu} ( \psi_{+} + \psi_{-} ) _{,\nu} Q_{3} - 
q^2 A_{\mu} A_{\nu} ( \psi_{+} + \psi_{-} ) ] 
\nonumber \\
& + & M^2 ( \psi_{+} + \psi_{-} ) 
+ q \lambda_{+} \psi_{+} + q \lambda_{-} \psi_{-} = 0
\label{exp.qdirac.wave.eq}
\end{eqnarray}}%
This is now a superposition of forms similar to equation (\ref{qgeometric.wave.eq}), but the 
$ \psi_{\pm} $ are not actually complex numbers at this point. Rather they are still more general quantities like the Dirac 
$ \psi $. If the electromagnetic field is also constant and uniform, the eigenvectors can be chosen as constant vectors, and dotted into equation (\ref{exp.qdirac.wave.eq}) to obtain an equation for either coefficient 
$ a $ or $ b $ of the form of equation (\ref{qgeometric.wave.eq}) with a single choice of the sign of the square root, a pure eigenstate of that equation\cite{rankin.caqg}. But for more general electromagnetic fields, a different technique must be used to obtain the form of equation (\ref{qgeometric.wave.eq}). The more general result will give a superposition of the forms from that equation for both values of the square root, rather than a single eigenstate.

\subsection{Complex/Spin Projections}

The standard projection operations in quantum mechanics for ``spin-up'' and ``spin-down'' are\cite{drell}
\begin{equation}
\psi_{u} = \frac{1}{2} [ \psi + \sigma_{z} \psi ]
\label{def.proj.up}
\end{equation}
and
\begin{equation}
\psi^{\prime}_{d} = \frac{1}{2} [ \psi - \sigma_{z} \psi ]
\label{def.proj.down}
\end{equation}
where the next steps will reveal why the second quantity is primed.

These have immediate quaternionic translations, with ``spin-up'' becoming
\begin{equation}
\psi_{u} = \frac{1}{2} [ \psi - Q_{3} \psi Q_{3} ]
\label{def.proj.complex.u}
\end{equation}
This is immediately recognizable as the complex projection of the quaternionic 
$ \psi $ into the complex plane defined by quantities having the form of equation (\ref{def.qpsi.complex})\cite{rotelli}. The translation of equation (\ref{def.proj.down}) can be put in this same form simply by multiplying after translation from the left by 
$ -Q_{2} $, giving
\begin{equation}
\psi_{d} = \frac{1}{2} [ (-Q_{2}) (\psi + Q_{3} \psi Q_{3}) ]
\label{def.proj.complex.d}
\end{equation}
The reason for the prime in equation (\ref{def.proj.down}) is now clear, since a full equivalence to equation (\ref{def.proj.complex.d}) would require 
$ \psi^{\prime}_{d} $ to have its nonzero component shifted to the top position by prefixing the right side of equation (\ref{def.proj.down}) with a multiplier of 
$ \imath \sigma_{y} $ to match the 
$ -Q_{2} $ in equation (\ref{def.proj.complex.d}).

In matrix form, using equation (\ref{def.m.qpsi}) as 
$ \psi $, equations (\ref{def.proj.complex.u}) and (\ref{def.proj.complex.d}) are
\begin{equation}
\psi_{u} =
\left (
\begin{array}{cc}
\psi_{0 R} + \imath \psi_{0 I} & 0 \\
0 & \psi_{0 R} - \imath \psi_{0 I} 
\end{array}
\right )
\label{m.proj.complex.u}
\end{equation}
and
\begin{equation}
\psi_{d} =
\left (
\begin{array}{cc}
\psi_{1 R} + \imath \psi_{1 I} & 0 \\
0 & \psi_{1 R} - \imath \psi_{1 I} 
\end{array}
\right )
\label{m.proj.complex.d}
\end{equation}
and both are clearly complex in the sense of equations (\ref{def.m.qpsi.complex}) and (\ref{def.qpsi.complex}). Furthermore, the original quaternionic 
$ \psi $ can be fully reconstructed from this pair as
\begin{equation}
\psi = \psi_{u} + Q_{2} \psi_{d}
\label{restore.psi}
\end{equation}
so no information is lost in these projections.

Now construct 
$ \psi_{+u} $, 
$ \psi_{+d} $, 
$ \psi_{-u} $, and 
$ \psi_{-d} $ using equations (\ref{def.proj.complex.u}) and (\ref{def.proj.complex.d}) as a pattern. Since 
$ Q_{2} $ (from the left) and 
$ Q_{3} $ applied in accordance with that pattern in equations (\ref{def.proj.complex.u}) and (\ref{def.proj.complex.d}), do in fact commute with the derivatives and the coefficients of every term in equation (\ref{exp.qdirac.wave.eq}), it is immediately clear that one can add and subtract the results of the various operations with 
$ -Q_{2} $ and $ Q_{3} $ on that equation to get
{\samepage 
\begin{eqnarray}
\hat \eta^{\mu \nu} [ ( \psi_{+u} & + & \psi_{-u} ) _{,\mu ,\nu} - 
q A_{\mu ,\nu} ( \psi_{+u} + \psi_{-u} )
Q_{3} \nonumber \\
& - & 2 q A_{\mu} ( \psi_{+u} + \psi_{-u} ) _{,\nu} Q_{3} - 
q^2 A_{\mu} A_{\nu} ( \psi_{+u} + \psi_{-u} ) ] 
\nonumber \\
& + & M^2 ( \psi_{+u} + \psi_{-u} ) 
+ q \lambda_{+} \psi_{+u} + q \lambda_{-} \psi_{-u} = 0
\label{proj.u.qdirac.wave.eq}
\end{eqnarray}}%
and
{\samepage 
\begin{eqnarray}
\hat \eta^{\mu \nu} [ ( \psi_{+d} & + & \psi_{-d} ) _{,\mu ,\nu} - 
q A_{\mu ,\nu} ( \psi_{+d} + \psi_{-d} )
Q_{3} \nonumber \\
& - & 2 q A_{\mu} ( \psi_{+d} + \psi_{-d} )_{,\nu} Q_{3} - 
q^2 A_{\mu} A_{\nu} ( \psi_{+d} + \psi_{-d} ) ] 
\nonumber \\
& + & M^2 ( \psi_{+d} + \psi_{-d} ) 
+ q \lambda_{+} \psi_{+d} + q \lambda_{-} \psi_{-d} = 0
\label{proj.d.qdirac.wave.eq}
\end{eqnarray}}%
These clearly give us two separate superpositions of the form of equation (\ref{qgeometric.wave.eq}) as solutions. Furthermore, the Dirac eigenvalues propagate intact into these solutions, so that the solutions do have good eigenvalues and well behaved functions. On the other hand, these do not represent a sum of {\it eigenfunctions} of equation (\ref{qgeometric.wave.eq}), because such eigenfunctions are generally not as well behaved as the two parts of these two results.

These two solutions should correspond to actual scalar curvatures 
$ B $ as given through equation (\ref{linear.change}) as
\begin{equation}
B_{u} = \psi_{u} ^{-2} = 
\left [ \psi_{+u} + \psi_{-u} \right ] ^{-2}
\label{B.up}
\end{equation}
and
\begin{equation}
B_{d} = \psi_{d} ^{-2} = 
\left [ \psi_{+d} + \psi_{-d} \right ] ^{-2}
\label{B.down}
\end{equation}
There is a ``spin-up'' value of 
$ B $, and a separate ``spin-down'' value of 
$ B $, both complex scalars. They hold all the information of the Dirac 
$ \psi $ between them. In that sense then, the full Dirac wavefunction 
$ \psi $ has been embedded into the basic geometric structure in the complex plane, but 
$ \psi $ viewed as a quaternionic spacetime scalar, not as a spinor. This embedding suggests that the Dirac Equation has as much in common with the complex plane as it has with quaternions. To add additional perspective on this point, the appendix to this paper discusses generalizing gauge transformations and curvatures in this structure to quaternionic values.

On the point that the Dirac wavefunction is treated as a scalar, not a spinor or bispinor, Yuri Usachev demonstrated the validity of such an approach in 1961\cite{jtep.usachev}), based as he says on an earlier suggestion of Sommerfeld that evidently only appears in the Russian versions of Somerfeld's books. However, Bade and Jehle briefly do mention a similar concept in their paper\cite{bade.jehle}, before they turn away from the idea, and adopt a (standard) spinor model of the Dirac wavefunction.

One more point deserves mention here. Both equation (\ref{geometric.wave.eq}) and equation (\ref{dirac.wave.eq}) are paired with natural conjugate wavefunction equations in their respective theoretical frameworks\cite{rankin.caqg,sakurai}. Equation (\ref{geometric.wave.eq}) pairs with
{\samepage 
\begin{eqnarray}
\hat \eta^{\mu \nu} \left [ \xi_{,\mu ,\nu} - \imath q A_{\mu ,\nu} 
\xi - 2 \imath q A_{\mu} \xi_{,\nu} - q^{2} A_{\mu} A_{\nu} \xi 
\right ] 
\nonumber \\
+ M^{2} \xi 
\pm \imath q \sqrt {E^{2} - B^{2} + 2 \imath \vec E 
{\bf \cdot} \vec B } \; \, \xi 
= 0 
\label{conj.geometric.wave.eq}
\end{eqnarray}}%
where this quantity 
$ \xi $ should {\it not} be confused with the eigenvectors in equations (\ref{def.eig.spinor})  and (\ref{def.eig.quat}). The complex conjugate of this equation gives
{\samepage 
\begin{eqnarray}
\hat \eta^{\mu \nu} \left [ {}^{\#} \xi_{,\mu ,\nu} + \imath q 
A_{\mu ,\nu} {}^{\#} \xi + 2 \imath q A_{\mu} {}^{\#} \xi_{,\nu} 
- q^{2} A_{\mu} A_{\nu} {}^{\#} \xi 
\right ] 
\nonumber \\
+ M^{2} {}^{\#} \xi 
\mp \imath q \sqrt {E^{2} - B^{2} - 2 \imath \vec E 
{\bf \cdot} \vec B } \; \, {}^{\#} \xi 
= 0 
\label{cc.conj.geometric.wave.eq}
\end{eqnarray}}%
where a 
``$ \; {}^{\#} \; $'' {\it prefixing} a quantity designates the complex conjugate of that quantity. On the other hand, the conjugate Dirac Equation form paired with equation (\ref{dirac.wave.eq}) is
{\samepage 
\begin{eqnarray}
\hat \eta^{\mu \nu} \left [ \phi_{,\mu ,\nu} + \imath q A_{\mu ,\nu} 
\phi + 2 \imath q A_{\mu} \phi_{,\nu} - q^{2} A_{\mu} A_{\nu} \phi 
\right ] 
\nonumber \\
+ M^{2} \phi 
- \imath q \vec \sigma {\bf \cdot} \left ( \vec E - \imath 
\vec B \right ) \phi 
= 0 
\label{conj.dirac.wave.eq}
\end{eqnarray}}%
Clearly one can develop a program of using the ``spin up/down'' projections of 
$ \phi $ just as was done above with 
$ \psi $. The result gives the superposition of forms of equation (\ref{cc.conj.geometric.wave.eq}) for 
$ {}^{\#} \xi $, thereby giving a solution for the conjugate 
$ \xi $ used in reference \cite{rankin.caqg}. If the (chiral) Dirac 
$ \psi $ and 
$ \phi $ are also related via the first order Dirac Equations\cite{sakurai}, then the matching family of projected scalar functions 
$ \psi $ and 
$ \xi $ might be expected to have no more symptoms of negative energy density than the original Dirac solution has. Of course, the action of equation (\ref{action.2nd}) does not require that the first order equations relate the chiral Dirac wavefunctions 
$ \psi $ and 
$ \phi $, as noted earlier. From its viewpoint, the first order equations would be additional conditions.

\section{Quantum and Other Effects}
\label{effects}

The previous sections have demonstrated that Dirac Theory does project into the complex forms of the classical gauge theory in reference \cite{rankin.caqg}. What then is predicted?

The most immediate prediction has already been mentioned, that the Dirac eigenvalues, as well as the projections of the eigenfunctions, propagate into the complex theory. That brings those quantities in the complex theory into line with Dirac, and it brings a form of spin 1/2 itself into the results. But what else does the formalism imply?

\subsection{Effects and Scale of Nonnegligible Ricci Tensor}
\label{ricci.unity}

First note that there are ``conformal terms'' in the stress tensor produced by the wavefunctions which are part of 
$ \hat \beta \, $, where $ \! \hat \beta = \xi \psi $, and 
$ \xi $ is the conjugate wavefunction in the theory derived from the action of equation (\ref{complex.gi.ahat.action})\cite{rankin.caqg}. Those terms have the form 
$ ( \hat \beta )_{ \| \mu \| \nu} - 
( \hat \beta )_{\| \gamma \| \tau} \hat g^{\gamma \tau} \hat g_{\mu \nu} $, and as noted in references \cite{rankin.mg7} and \cite{rankin.9404023}, they originate from the 
$ \hat \beta \hat R $ term in the action, or the equivalent, well-known conformal term that can be added to theories of scalar wavefunction fields\cite{wald}. However, their magnitude should be comparable to the ``ordinary'' terms in the stress tensor, since there is no higher order, multiplying factor in them. But the covariant divergence of these terms is
\begin{equation}
\left [ ( \hat \beta )_{\| \mu}^{\: \: \: \: \; \| \nu} - 
( \hat \beta )_{\| \gamma}^{\: \: \: \: \; \| 
\gamma }\delta^{\nu}_{\mu} \, \right ]_{\| \nu} = 
- \hat R_{\mu}^{\: \: \nu} ( \hat \beta )_{, \nu}
\label{div.extra}
\end{equation}
This is clearly a general relativistic quantity, which means that in regions in which 
$ \hat R_{\mu}^{\: \: \nu} \approx 0 $, these terms have vanishing covariant divergence. In that case, they cannot interact with the rest of the stress tensor except gravitationally, no matter how much energy they contain. Otherwise, they are separately conserved and invisible outside regions in which 
$ \hat R_{\mu}^{\: \: \nu} \neq 0 $. Furthermore, they stem from wavefunctions now seen to be spin 1/2 wavefunctions. In some sense, they represent nearly noninteracting, spin 1/2 matter-energy. The {\it effective} rest mass of such terms may only be clear after integrating their stress tensor terms to get a momentum four vector, and contracting that with itself.

As previously stated, these terms have long been present as ``conformal terms'' in conformal extensions of theories of scalar wavefunction fields such as the Klein-Gordon field\cite{rankin.9404023}, where the analogous terms are those in which 
$ \hat \beta \rightarrow \, {}^{\#} \psi \psi $ in the above expressions. One simply modifies the rest mass squared term in the standard action for such fields by the substitution 
$ M^2 \rightarrow M^2 + {\textstyle{1 \over 6}}\, R $, where 
$ R $ is the scalar curvature. However, it is fairly easy to demonstrate that the conformal stress tensor terms must vanish whenever the ordinary stress tensor terms of a Klein-Gordon wavefunction field vanish, because the ordinary energy density is a sum of positive definite quantities unless 
$ \psi = 0 $. But $ \psi = 0 $ implies 
$ \left ( \, {}^{\#} \psi \psi \right )_{ \| \mu \| \nu} - 
\left ( \, {}^{\#} \psi \psi \right )_{\| \gamma \| \tau} \hat g^{\gamma \tau}
\hat g_{\mu \nu} $ 
is also zero, and the conformal terms then really completely vanish also. In the conformal Klein-Gordon case, they are rather tightly linked to the presence of ``ordinary matter'', and have spin 0 as well, not spin 1/2.

In that regard, the formalism of this paper has already been noted to be spin 1/2, and because of the possibility of negative energy density, its ordinary energy density terms are no longer necessarily positive definite. It's not clear at this point if that allows the conformal stress energy terms to be nonzero while the ordinary stress energy terms vanish. However, that's the basic requirement to delink the two sets of terms from each other significantly.

Equation (\ref{div.extra}) indicates that interaction between the conformal terms and the rest of the stress tensor {\it does} occur in regions in which 
$ \left | \hat R_{\mu}^{\: \: \nu} \right | \geq 1 $. Using the Reissner-Nordstr\"{o}m solution of an electronic charge\cite{adler.bazin} as a guide to calculate 
$ \hat R_{\mu \nu} $\cite{rankin.9404023},
\begin{equation}
\left | \hat R_{\mu \nu } \right | \sim \frac{3}{4} [ (G m_{0}^{2} ) 
/ (\hbar c) ][ e^{2} / (\hbar c) ] (1/r^{4} )
\label{r.n.approx}
\end{equation}
In this, the factors of 
$ m_{0} $, $ e $, and $ \hbar $ are introduced through the scale factor 
$ b_{0} $ as well as the coefficient of 
$ A_{\mu} $ in equation (\ref{def.v}), and the scale factor is set to match the wave equation to atomic scale phenomena\cite {rankin.caqg}. The magnitude of 
$ \hat R_{\mu \nu} $ approaches unity for a dimensionless radius 
$ r_{0} = 1.8 \cdot 10^{-12} $. Convert that to lab units by dividing the dimensionless value by the square root of the scale factor
\begin{equation}
b_{0} = \frac{3}{4} \left [ \left ( m_{0} ^2 c^2 \right ) / \hbar^2 
\right ] 
\label{b0}
\end{equation}
and the result gives 
$ r_{0 \, LAB} = r_{0} / \sqrt{ b_{0} } \, \approx 8 \cdot 10^{-23} \; cm. $ 
(a much larger value than the 
$ 10^{-33} \; cm. $ at which the metric deviates significantly from flat spacetime values\cite{adler.bazin}). This represents a slight shift from values calculated in reference \cite{rankin.9404023}, partly because of a more accurate treatment of the Reissner-Nordstr\"{o}m solution, and partly because a different universal scale factor 
$ b_{0} $ is indicated by the more general formalism which includes an antisymmetric part to the metric\cite{rankin.caqg}.

As noted in references \cite{rankin.9404023} and \cite{rankin.mg7}, any corresponding interaction cross section should contain the factor 
$ \pi r_{0 \, LAB}^{2} $, so this crude estimate gives a very approximate cross section of about 
$ 2 \cdot 10^{-44} \; cm.^2 $. This is roughly the lower bound of observed neutrino cross sections in matter\cite{texas92}. However, this isn't just the cross section at which the conformal terms interact with the rest of the stress tensor. It's also the cross section at which {\it any} general relativistic effects of the Reissner-Nordstr\"{o}m solution that depend upon nonzero 
$ \hat R_{\mu \nu} $ could become important. Again, these general relativistic effects are caused by the electric charge term in the Reissner-Nordstr\"{o}m metric, which produces these effects in 
$ \hat R_{\mu \nu} $ at a much larger radius than either its effects on the metric, or the metric effects of the Schwarzchild type mass term in the solution\cite{adler.bazin}. That mass term does not contribute to a nonzero 
$ \hat R_{\mu \nu} $ anyway.

\subsection{Scale of Nonnegligible Scalar Curvature}

There are also non-electromagnetic contributions to 
$ \hat R_{\mu \nu} $ as well as 
$ \hat R $ from the $ \hat \beta $ terms in the stress tensor, but they occur at magnitudes beneath the above effects. To get a feel for their magnitude, directly compare the 
forms generated for the ``rest mass term'' in the stress tensors from the action of equations (\ref{complex.gi.ahat.action}), and the Klein-Gordon action in the appendix in reference \cite{rankin.9404023}. The Klein-Gordon action can be used for this purpose since the rest mass term in it is the same magnitude as the analogous term in the action of equation (\ref{action.2nd}), and the simple scalar nature of the earlier Klein-Gordon 
$ \psi $ is easier to handle. However, the 
Klein-Gordon case should be rendered dimensionless by multiplying the stress tensor by 
$ ( 1 / b_{0} ) $ to allow direct comparison of dimensionless quantities. That 
$ ( 1 / b_{0} ) $ multiplier is absorbed into the derivatives involved in the calculation of curvature so that the derivatives are consistently taken with respect to the same dimensionless coordinates as are used in equation (\ref{def.identity}), and throughout the formalism of the gauge invariant variables.

The pertinent terms are 
$ -{\textstyle{3 \over 2}}\, \left [ {\textstyle{1 \over 6}}\, C 
( 1 + {\textstyle{1 \over 4}}\, \hat a ) \hat \beta + CC \right ] $ 
and 
$ -[ ( k \hbar c ) / ( 2 b_{0} ) ] M_{A} 
\, {}^{\#} \psi \psi $ 
respectively, where 
$ M_{A} = ( m_{A} c ) / \hbar $, 
$ m_{A} $ is the actual rest mass of the system in question, and 
$ k = (8 \pi G ) / c^{4} $. But also, the identification is made that 
$ {\textstyle{1 \over 6}}\, C ( 1 + {\textstyle{1 \over 4}}\, \hat a ) 
= - M_{A}^{2} / b_{0} $, 
and positive energy density wavefunctions 
$ \zeta_{2} $ are chosen in 
$ \hat \beta $. Reference \cite{rankin.9404023} details the introduction of 
$ \zeta_{2} $, but in brief, 
$ \hat \beta = \xi \psi $, and 
$ \zeta_{1} $ and $ \zeta_{2} $ are defined by
\begin{equation}
\psi = ( 1 / \sqrt{2}\, ) 
( \zeta_{1} + \zeta_{2} )
\label{def.zetas.1}
\end{equation}
and
\begin{equation}
\xi = ( 1 / \sqrt{2}\, ) 
( \, {}^{\#} \zeta_{1} - \, {}^{\#} \zeta_{2} )
\label{def.zetas.2}
\end{equation}
This splits results into positive and negative energy density fields\cite{wheeler.trans}, and the negative energy density in 
$ \zeta_{1} $ is discarded for this calculation, giving 
$ ( \hat \beta + CC ) = - {}^{\#} \zeta_{2} \zeta_{2} $. All these together then give 
\begin{equation}
{}^{\#} \zeta_{2} \zeta_{2} = ( 8 \pi / 3 ) 
[ ( \hbar G ) / c^{3} ] [ \hbar / ( m_{A} c ) ] 
\; {}^{\#} \psi \psi 
\label{relate.wave.products}
\end{equation}
where the factor 
$ ( \hbar G ) / c^{3} = {\cal L}_{P}^{2} $, and 
$ {\cal L}_{P} $ is the well-known Planck length. 
Since $ \psi $ is the standard laboratory quantum mechanical wavefunction, 
$ {}^{\#} \psi \psi $ integrates to unity over its containing volume, giving 
$ {}^{\#} \psi \psi $ the dimensions of inverse volume. This is consistent with 
$ {}^{\#} \zeta_{2} \zeta_{2} $ (and $ \hat \beta $) being dimensionless, as required. But this relation is all that is needed to relate the magnitudes of 
$ \psi $ and $ \zeta_{2} $ via
\begin{equation}
\zeta_{2} = \{ ( 8 \pi / 3 ) 
[ ( \hbar G ) / c^{3} ] [ \hbar / ( m_{A} c ) ] \}^{1/2} 
\psi 
\label{relate.zeta.psi}
\end{equation}

Now ignoring the cosmological constant in reference \cite{rankin.caqg}, the expression for 
$ \hat R $ is
{\samepage 
\begin{eqnarray}
\hat R & = & \frac{1}{2} \left [ C ( 1 + {\textstyle{1 \over 4}}\, 
\hat a ) \hat \beta + CC \right ] 
\nonumber \\ 
& = & 3 ( M_{A}^{2} / b_{0} ) 
\, {}^{\#} \zeta_{2} \zeta_{2} 
\nonumber \\ 
& = & 8 \pi ( M_{A}^{2} / b_{0} ) 
[ ( \hbar G ) / c^{3} ] [ \hbar / ( m_{A} c ) ] 
\; {}^{\#} \psi \psi 
\nonumber \\ 
& = & 8 \pi ( M_{A} / b_{0} ) 
[ ( \hbar G ) / c^{3} ] 
\; {}^{\#} \psi \psi 
\label{caqg.scalar.rhat}
\end{eqnarray}}%
Unlike $ \hat R_{\mu \nu } $, 
$ \hat R $ contains no electromagnetic contribution. Thus it will be terms containing 
$ \hat R $ that are sensitive to 
$ ( \hat \beta + CC ) = - {}^{\#} \zeta_{2} \zeta_{2} $ approaching unit magnitude, such as the conformal term 
$ {\textstyle{1 \over 6}}\, \hat R \psi $ in the wave equation itself. A simple calculation reveals that an electron would have to be confined tightly within a sphere with a radius of roughly 
$ 10^{-25} cm. $ to have a wavefunction magnitude great enough to produce an 
$ \hat R $ with a magnitude approaching unity. This radius is three orders of magnitude beneath the radius at which 
$ \hat R_{\mu \nu} $ approaches unit magnitude in the Reissner-Nordstr\"{o}m solution, as discussed above. It is still about eight orders of magnitude above a Planck scale radius.

\subsection{Quantized Energy Exchange Between Fields}

What energy values are given when the 
$ \hat \beta $ terms in the stress tensor generated by equation (\ref{complex.gi.ahat.action}) are actually integrated? I investigated this for the hydrogen atom field shortly after publication of reference \cite{rankin.caqg} in hopes that the results would agree better with Dirac Theory than the complex energy eigenvalues of equation (\ref{geometric.wave.eq}). Using an adaptation of Lin's integrals as explained by Infeld and Hull\cite{infeld.hull} -- and with help from some simplifying assumptions before I understood that energy density could become negative -- I found that to lowest order, integrated energy values agreed with the energy eigenvalues of nonrelativistic Schr\"{o}dinger Theory\cite{schiff}. It's to this same order that the eigenvalues of equation (\ref{geometric.wave.eq}) are real, and the eigenfunctions well behaved, so wavefunction energy eigenvalues tracked in tandem with the integrated energy values to the same degree that the results were well behaved overall. In retrospect, this should not be surprising, since had this not been true for the similar, original field theories of quantum wavefunctions\cite{morse.feshbach} -- except that eigenfunctions for negative energy eigenvalues would be expected to integrate to the positive energy, ``antiparticle'' energies -- those theories would have been internally inconsistent with respect to definite, quantized atomic energies. Nevertheless, this has an important consequence.

Ignoring general relativistic effects (gravitation), the total stress tensor energy is conserved. Since the wavefunction portion carries the discrete energy amounts predicted by wave mechanics, it is necessarily true that the electromagnetic field {\it must} exchange energy discretely with the bound state wavefunctions. But that is a major part of Planck's original quantum postulate\cite{schiff}. All that remains is to demonstrate that the electromagnetic radiation emitted or absorbed has frequency 
$ \nu = E / h $ where $ E $ is the discrete energy amount transferred, and the full Planck postulate would be respected. That would account correctly for blackbody radiation.

Lamb and Scully demonstrate that the photoelectric effect can be explained without resort to formal quantum electrodynamics\cite{lamb.photo}, while Thorn et al.\cite{thorn.ajp} discuss just what experimental issues really do need to be addressed to measure genuine quantum electrodynamic phenomena. Lamb and Scully list four phenomena to attribute to quantum electrodynamics, including spontaneous emission, blackbody radiation, the Compton Effect, and electrodynamic level shifts. At this time, it appears possible that at least one of these might be explained by the approach of this paper, and that at least some form of electromagnetic quanta exist in this theory. This needs further examination.

\subsection{Effect of More General Values for the Constant C}
\label{rescale}

So far, the constant $ C $ has been assumed to be 
$ C = 1 $. If the magnitude of 
$ C $ differs from unity, then equation (\ref{b0}) generalizes to
\begin{equation}
b_{0} = \frac{3}{4} \left [ \left ( m_{0} ^2 c^2 \right ) / 
\left ( \left | C \right | \hbar^2 \right ) \right ] 
\label{b0.C}
\end{equation}
where $ \left | C \right | $ is the magnitude or absolute value of 
$ C $. This and other appearances of 
$ C $ in the various equations will affect the considerations of the above scale calculations.

Specifically, for positive 
$ C $, the scale of nonnegligible 
$ \hat R_{\mu \nu} $ is shifted via 
$ r_{0 \, LAB} \approx C^{1 / 4} \, 8 \cdot 10^{-23} \; cm. $, 
but the radial scale of nonnegligibile 
$ \hat R $ is instead shifted by a multiplicative factor of 
$ C^{1 / 3} $ rather than the factor 
$ C^{1 / 4} $. The fact that these two scales depend on ``Galehouse's Constant'' 
$ C $ differently, indicates that it has a nontrivial scaling effect, and values other than those with magnitude of unity may need to be considered empirically. Another way of looking at these results is that in a sense, rest mass scales like 
$ C $, but charge scales like the square root of 
$ C $ in the determination of these two scale magnitudes.

Note that
\begin{equation}
e^{2} / \left ( \hbar c \right ) = 
\sqrt { { 4 \over 3 } | C | } 
\left \{ \sqrt { b_{0} } \, \left [ e^{2} / \left ( 
m_{0} c^{2} \right ) \right ] \right \}
\label{dimensionless.classical.radius}
\end{equation}
This suggests that the fine structure constant is essentially the dimensionless form of the {\it classical} radius of the electron, 
$ e^{2} / \left ( m_{0} c^{2} \right ) $ (Gaussian units\cite{jackson.self.force}). However this simple point does not explain why that particular radius appears as an interaction coefficient in the wave equation.

\section{Conclusions}
\label{conclusions}

The second order, chiral form of the Dirac Equation in flat spacetime does indeed ``spin up/down'' project neatly into the complex plane as a superposition of forms predicted by the classical gauge theory of reference \cite{rankin.caqg}. That theory thus did include Dirac solution forms and spin 1/2 generally, {\it not} just in the special case of uniform, constant, non-null fields. The Dirac wavefunction field is seen to project into scalars in the complex plane, and in this sense, it can be viewed naturally as a quaternionic spacetime scalar field itself. It correlates via the ``spin up/down'' projections with two distinct values of the scalar curvature of the Weyl-Cartan geometry underlying the classical gauge theory. Dirac Theory becomes correlated with a theory of purely geometric quantities in this model. Since ``spin space'' transformations can be used on the quaternionic Dirac wavefunction field to generate an entire family of paired curvatures that share common Dirac eigenvalues, that becomes a degeneracy of the theory.

As it stands, this theory should account for a number of atomic phenomena, to the degree that Dirac Theory can do so. It may account for some subatomic phenomena as well. It also seems to predict a form of electromagnetic field quanta. On the other hand, the Dirac Equation is still needed for practical computation of results.

There is no evidence currently of charge quantization in this model, or direct evidence of spatial localization of charge in the sources of the electromagnetic field. Nor is there currently evidence of quantization or spatial localization of rest masses of the wavefunction field. However, until a better theory of the antisymmetric part of the metric is developed, it's not clear what this formalism has to say about variations in rest masses, or the effects of possible excursions of the value of 
$ \hat a $ away from the constant value of 
$ ( 6 \imath )^{2} = -36 $, possibly into the complex plane. Furthermore, the negative energy density possible in some solutions should accompany a charge density opposite in sign to that associated with similar positive energy density solutions (the electron wavefunctions) in Dirac Theory (see the appendix in reference \cite{rankin.9404023}), and it may simultaneously provide an attractive binding force in analogy to Newtonian Gravitation.


\appendix* 
\section{Quaternionic Gauges and Curvatures}
\label{extended}

The close relationship between equations (\ref{leap.qdirac.wave.eq}) and (\ref{qgeometric.wave.eq}) suggests that equation (\ref{leap.qdirac.wave.eq}) might be a direct result of generalizing the theory of reference \cite{rankin.caqg} from the complex numbers to the quaternions. If equations (\ref{leap.qdirac.wave.eq}) and (\ref{linear.change}) remain true in that generalization, that would imply that the scalar curvature 
$ B $ is fully quaternionic instead of simply being complex valued as in section \ref{projections}.

Now in fact, it is possible to generalize reference \cite{rankin.caqg} from complex numbers to quaternions, but equations (\ref{linear.change}) and (\ref{leap.qdirac.wave.eq}) also generalize when this is done. Basically, 
non-commutativity in the generalized structure produces a more complicated generalization of equation (\ref{def.identity}) as well as other complications not seen in the earlier work. However if some amount of Weyl's nonmetricity\cite{weyl.stm,eddington.mtr} is mixed with the torsion of the model of reference \cite{rankin.caqg}, the generalization of equation (\ref{def.identity}) will give the second order, quaternionic Dirac Equation in 
$ \psi $. The generalization of equation (\ref{linear.change}) introduces an auxiliary wavefunction 
$ \chi $  defined by a first order partial differential equation in 
$ \chi $ and 
$ \psi $. The integrability conditions for that equation impose conditions upon 
$ \psi $, although these conditions still allow solutions to exist, some of which will be examined.

This appendix will briefly sketch these results.

\subsection{Basics}

The basic rules will remain the same. There must always exist a gauge (or gauges) in which the symmetric part of the metric, 
$ g_{\mu \nu} $, is real. In that gauge, 
$ g_{\mu \nu} $ will be denoted by 
$ \tilde g_{\mu \nu} $. Unless stated otherwise, tensor indices are lowered and raised using the symmetric part of the metric, 
$ g_{\mu \nu} $, and its inverse. The antisymmetric part of the metric is still required to be self-dual (dual proportional), and of the form
\begin{equation}
a_{\mu \nu} = h_{\mu \nu} - {}^{*} h_{\mu \nu} | Q_{3} 
\label{ahat.makeup}
\end{equation}
where the ``$ \; {}^{*} \; $'' indicates the dual of the quantity following it (see equation (\ref{def.hdual})). In this, 
$ h_{\mu \nu} $ is assumed to be real in the same gauge in which 
$ g_{\mu \nu} $ is real, a requirement that will be slightly relaxed later. For all these quantities, a 
``$ \; \tilde{\;} \; $'' indicates evaluation in a gauge in which the quantity is real. Note the use of 
$ |(-Q_{3}) $ in place of the 
$ \imath $ used in this same form in reference \cite{rankin.caqg}. That will maintain a simplifying element of commutativity in this form, and elsewhere in the quaternionic generalizations of equations from the earlier work, as already seen in section \ref{quat.dirac.theory}. Clearly the full asymmetric, covariant metric is
\begin{equation}
m_{\mu \nu} = g_{\mu \nu} + a_{\mu \nu} 
\label{mhat.makeup}
\end{equation}

Likewise, there must always exist a gauge (or gauges) in which equation (\ref{def.v}) generalizes to the form
\begin{equation}
v_{\mu} = - q A_{\mu} |Q_{3} 
\label{def.qv}
\end{equation}
with $ A_{\mu} $ being the standard, real, electromagnetic potential. This gauge will generally {\it not} be the same gauge in which 
$ g_{\mu \nu} $ is real.

Since $ g_{\mu \nu} $ (and 
$ h_{\mu \nu} $) is real in some gauge, it is always possible to write
\begin{equation}
g_{\mu \nu} = \tilde g_{\mu \nu} \gamma 
\label{g.to.greal}
\end{equation}
where $ \gamma $ is some quaternionic scalar. This allows the various operations with 
$ g_{\mu \nu} $ and 
$ h_{\mu \nu} $ to be set up through easy correspondence with operations on real or complex forms. For example, 
$ g^{\mu \nu} $ is easily calculated through the standard requirement that
\begin{equation}
g^{\mu \alpha} g_{\alpha \nu} = \delta^{\mu}_{\nu} 
\label{g.ginv}
\end{equation}
where $ \delta^{\mu}_{\nu} $ is the Kronecker delta. In fact,
\begin{equation}
g^{\mu \nu} = \gamma^{-1} \tilde g^{\mu \nu} 
\label{ginv.to.ginvreal}
\end{equation}
Furthermore, equations (\ref{g.to.greal}) and (\ref{ginv.to.ginvreal}) give
{\samepage 
\begin{eqnarray}
g^{\alpha \tau} \! g_{\mu \nu} & = & 
\gamma^{-1} \tilde{g}^{\alpha \tau} \tilde{g}_{\mu \nu} \gamma
\nonumber \\
 & = & 
\tilde{g}^{\alpha \tau} \tilde{g}_{\mu \nu} 
\nonumber \\
 & = & 
\tilde{g}_{\mu \nu} \tilde{g}^{\alpha \tau} 
\nonumber \\
 & = & 
\tilde{g}_{\mu \nu} \gamma \gamma^{-1} \tilde{g}^{\alpha \tau} 
\nonumber \\
 & = & 
g_{\mu \nu} g^{\alpha \tau} 
\label{gauge.inv.gpair}
\end{eqnarray}}%
That metric combination is always real, gauge invariant, and commutes {\it as a unit} with everything.

Since gauge transformations are quaternionic, they can generally be applied to 
$ g_{\mu \nu} $ from either the left or right, as denoted by
\begin{equation}
\bar g_{\mu \nu} = \lambda g_{\mu \nu} \rho 
\label{gbar.lr}
\end{equation}
where $ \lambda $ is the left gauge transformation, and 
$ \rho $ is the right. However, more restricted cases where one of these two multipliers is always taken to be 
$ 1 $ will be most useful here. Since equation (\ref{gbar.lr}) can always be written as
\begin{equation}
\bar g_{\mu \nu} = \lambda \tilde g_{\mu \nu} \gamma \rho 
\label{gbar.lr.gamma}
\end{equation}
according to equation (\ref{g.to.greal}), and since 
$ \tilde g_{\mu \nu} $ commutes with every scalar multiplier, restricting either 
$ \lambda $ or $ \rho $ to be 
$ 1 $ may not be a serious restriction in practice. However whatever scheme is adopted, 
$ h_{\mu \nu} $ will be assumed to gauge transform by the {\it same} pattern as is adopted for 
$ g_{\mu \nu} $. The same pattern can then be shown to hold for 
$ {}^{*} h_{\mu \nu} $ provided the dual is defined as
\begin{equation}
{}^{*} h_{\mu \nu} = \frac{1}{2} g_{\mu \tau} (-g)^{-1 / 4} \, 
\epsilon^{\tau \sigma \alpha \beta} h_{\alpha \beta} 
(-g)^{-1 / 4} \, g_{\sigma \nu} 
\label{def.hdual}
\end{equation}
where 
$ \epsilon^{\tau \sigma \alpha \beta} $ is the completely antisymmetric unit symbol with 
$ \epsilon^{0 1 2 3} = 1 $, and in addition to the definition of 
$ {}^{*} h^{\tau \sigma} $, one can discern in this the symmetric manner used to raise and lower indices on 
$ h_{\mu \nu} $ and 
$ {}^{*} h_{\mu \nu} $ using 
$ g^{\mu \nu} $ and 
$ g_{\mu \nu} $. And, since the leap-over operator 
``$ | $'' keeps the 
$ Q_{3} $ out of the way to the far right, equation (\ref{ahat.makeup}) then indicates that 
$ a_{\mu \nu} $ will likewise follow the same gauge transformation pattern as 
$ g_{\mu \nu} $ and $ h_{\mu \nu} $, provided the gauge functions themselves either contain no leap-over operations, or contain none other than 
$ |Q_{3} $. This assumption will essentially always be made about the gauge functions here, although since the inverse gauge function is also needed, then in practice, the worst case scenario actually means that one of the function and inverse function pair contains nothing worse than 
$ |(Q_{3}) $, while the other may contain nothing worse than 
$ (Q_{3})\| $. The qualifier ``may'' is used in that sentence because it is not difficult to find cases where both the gauge function and its inverse may appear to contain just 
$ |(Q_{3}) $. In fact, the standard range of electromagnetic gauge transformations with real potentials 
$ A_{\mu} $, which produces a phase transformation of the wavefunction\cite{bade.jehle}, presents a case where the phase (gauge) transformation is
\begin{equation}
e^{ \imath \left ( \phi / 2 \right ) } \rightarrow 
e^{ - \left ( | Q_{3} \right ) \left ( \phi / 2 \right ) } = 
\cos{ \left ( \phi / 2 \right ) } - 
\left ( | Q_{3} \right ) \sin{ \left ( \phi / 2 \right ) }
\label{qphase.trans}
\end{equation}
The quantity 
$ e^{ (|Q_{3}) ( \phi / 2 ) } $ is actually one possible inverse to this, as may be verified by direct multiplication of the quantities. However 
\begin{equation}
\left [ e^{ - \left ( | Q_{3} \right ) \left ( \phi / 2 \right ) } 
\psi \right ]^{-1} = \psi^{-1} e^{ \left ( Q_{3} \| \right ) 
\left ( \phi / 2 \right ) } = 
e^{ Q_{3} \left ( \phi / 2 \right ) } \psi^{-1}
\label{inverse.psibar}
\end{equation}
produces an alternate form for the inverse containing 
$ Q_{3}\| $ rather than 
$ |Q_{3} $. Indeed
\begin{equation}
e^{ Q_{3} \left ( \phi / 2 \right ) } \psi^{-1} 
\psi e^{ - Q_{3} \left ( \phi / 2 \right ) } = 1
\label{inverse.psibar.psibar}
\end{equation}
and
\begin{equation}
e^{ \left ( | Q_{3} \right ) \left ( \phi / 2 \right ) } \psi^{-1} 
e^{ - \left ( | Q_{3} \right ) \left ( \phi / 2 \right ) } \psi = 1
\label{alt.inverse.psibar.psibar}
\end{equation}
are both true. This continues to demonstrate the ambiguities which can arise in results when choosing to convert a complex expression to quaternions {\it before} a second operation is performed, as opposed to choosing to convert it to quaternions {\it after} performing the same operation, such as taking the inverse, just as noted after equation (\ref{def.qdirac.1st}). Such ambiguities need further examination, although since 
$ \sin{ ( \phi / 2 ) } $ commutes with 
$ Q_{3} $, one might argue that the two results here are indistinguishable when the phase transformation inverses are compared solely to each other. At any rate, all these points will be understood to be included when it is stated that a gauge function or other quantity contains nothing worse than 
$ |Q_{3} $.

\subsection{Left and Right Covariant Derivatives / Christoffel Symbols}

Since $ g_{\mu \nu} $ can now have a limited quaternionic nature, the expression
\begin{equation}
g_{\mu \nu ; \gamma} = g_{\mu \nu , \gamma} 
- g_{\alpha \nu} \left \{ \left. {}^{\: \alpha}_{\mu \gamma } \right ] \right. 
- g_{\mu \alpha} \left \{ \left. {}^{\: \alpha}_{\nu \gamma } \right ] \right. 
= 0 
\label{def.right.cov.metric}
\end{equation}
is not necessarily the same as
\begin{equation}
g_{\mu \nu ; \gamma} = g_{\mu \nu , \gamma} 
- \left [ \left. {}^{\: \alpha}_{\mu \gamma } \right \} \right. g_{\alpha \nu} 
- \left [ \left. {}^{\: \alpha}_{\nu \gamma } \right \} \right. g_{\mu \alpha} 
= 0 
\label{def.left.cov.metric}
\end{equation}
Thus, equation (\ref{def.right.cov.metric}) defines the covariant derivative of 
$ g_{\mu \nu} $ with respect to the ``right handed Christoffel Symbol'' 
$ \left \{ \left. ^{\: \alpha}_{\mu \nu} \right ] \right. $, while equation (\ref{def.left.cov.metric}) defines the covariant derivative of 
$ g_{\mu \nu} $ with respect to the ``left handed Christoffel Symbol'' 
$ \left [ \left. ^{\: \alpha}_{\mu \nu} \right \} \right. $. Both equations actually define the associated Christoffel Symbols, giving
\begin{equation}
\left \{ \left. ^{\: \alpha}_{\mu \nu} \right ] \right. = 
\frac{1}{2} \; g^{\alpha \tau} ( g_{\mu \tau , \nu} + 
g_{\nu \tau , \mu} - g_{\mu \nu , \tau} ) 
\label{def.right.chrst}
\end{equation}
and
\begin{equation}
\left [ \left. ^{\: \alpha}_{\mu \nu} \right \} \right. = 
\frac{1}{2} \: ( g_{\mu \tau , \nu} + 
g_{\nu \tau , \mu} - g_{\mu \nu , \tau} ) \, 
g^{\alpha \tau} 
\label{def.left.chrst}
\end{equation}
respectively.

Furthermore, equation (\ref{g.ginv}) and 
$ \delta^{\mu}_{\nu , \tau} = 0 $ give
\begin{equation}
g^{\beta \mu}_{\; \; \; \; , \tau} = 
- g^{\beta \nu} g_{\nu \alpha , \tau} \, g^{\alpha \mu} 
\label{inv.met.der}
\end{equation}
This and equations (\ref{gauge.inv.gpair}), (\ref{def.right.cov.metric}), and (\ref{def.left.cov.metric}) then give
\begin{equation}
g^{\beta \xi}_{\; \; \; \; ; \tau} = g^{\beta \xi}_{\; \; \; \; , \tau} 
+ \left \{ \left. {}^{\: \beta }_{\alpha \tau } \right ] \right. g^{\alpha \xi} 
+ \left \{ \left. {}^{\: \xi }_{\alpha \tau } \right ] \right. g^{\beta \alpha} 
= 0 
\label{def.right.cov.imetric}
\end{equation}
and
\begin{equation}
g^{\beta \xi}_{\; \; \; \; ; \tau} = 
g^{\beta \xi}_{\; \; \; \; , \tau} 
+ g^{\alpha \xi} \left [ \left. {}^{\: \beta }_{\alpha \tau } \right \} \right. 
+ g^{\beta \alpha} \left [ \left. {}^{\: \xi }_{\alpha \tau } \right \} \right. 
= 0 
\label{def.left.cov.imetric}
\end{equation}
This indicates that the contravariant metric's indices interact with their associated Christoffel symbols on the opposite side from the covariant metric's indices in the definition of the covariant derivative of the metric. That same convention is now adopted as well here for the form of all covariant derivatives, and also more general affine derivatives, of {\it any} tensor quantity, where the affine derivative simply uses the more general affine connections 
$ {}_{R} \Gamma^{\alpha}_{\mu \nu} $ and 
$ {}_{L} \Gamma^{\alpha}_{\mu \nu} $ in place of the Christoffel Symbols 
$ \left \{ \left. ^{\: \alpha}_{\mu \nu} \right ] \right. $ and 
$ \left [ \left. ^{\: \alpha}_{\mu \nu} \right \} \right. $. The more general affine connections are prefixed with the lowered ``R'' and ``L'' to maintain different right and left handed forms on the more general level, just as there are right and left handed Christoffel Symbols.

For completeness, now define the 
``$ \; \tilde{;} \; $'' derivative as the reversal of the convention just given for the 
``$ \; ; \; $'' derivative. That means that all the tensor - Christoffel Symbol positions are reversed for each term for each tensor index in the covariant derivative expression. For example,
\begin{equation}
g_{\mu \nu \, {\bf \tilde{;}} \, \gamma} = g_{\mu \nu , \gamma} 
- \left \{ \left. {}^{\: \alpha}_{\mu \gamma } \right ] \right. g_{\alpha \nu} 
- \left \{ \left. {}^{\: \alpha}_{\nu \gamma } \right ] \right. g_{\mu \alpha} 
\; ( \; \neq 0 ) 
\label{def.rev.cov.metric}
\end{equation}
By definition, the Christoffel Symbols are always defined using a ``normal'' covariant derivative of the (symmetric part of the) metric tensor. Clearly, tensor - Christoffel Symbol positions in these expressions are determined {\it both} by the type of Christoffel Symbol 
(``$ \{ ] $'' or ``$ [ \} $''), {\it and} the use of 
``$ \; ; \; $'' or ``$ \; \tilde{;} \; $'' in the derivative.

With these facts established, the ``left handed'' Christoffel Symbols and more general affine connection 
$ {}_{L} \Gamma^{\alpha}_{\mu \nu} $ will now be arbitrarily dropped, and the right handed cases used in what follows. However, it should also be noted that the generalization of equation (\ref{complex.gi.ahat.action}) in this extended structure will eventually involve a quaternion conjugate 
(``$ QC $'') term to replace the 
``$ CC $'' term in equation (\ref{complex.gi.ahat.action}). That quaternion conjugate term will tend to involve left handed forms to balance the right handed forms that are now being chosen in the first part of the action. Because of that, the basic left hand should not be suppressed or shortchanged overall, although a somewhat different type of right or left handedness will also arise at the next level to be examined. There, the right and left handed forms will be found necessarily to enter asymmetrically, but they should still tend to be balanced overall by corresponding opposite handed forms in the 
$ QC $ terms in the action.

For general quaternionic 
$ M_{\mu} $ and $ N_{\nu} $,
\begin{equation}
( M_{\mu} N_{\nu} )_{; \tau} \neq M_{\mu ; \tau} N_{\nu} + 
M_{\mu} N_{\nu ; \tau}
\label{def.no.product.rule}
\end{equation}
Additionally, contraction on tensor indices inside an already evaluated covariant derivative will not necessarily equal the covariant derivative of the contracted quantity. Examples such as these limit the usefulness of these gauge varying, generalized covariant derivatives outside of gauges in which quantities commute easily in products. However, in other cases, these covariant derivatives will still be helpful, and will be used. On the other hand, any genuine physics in this structure will require gauge invariant constructions, including gauge invariant covariant derivatives. Those will be developed later, and they will involve real Christoffel Symbols that thus avoid these limitations just noted.

\subsection{Weyl-Like Connections, Gauge Properties, and Curvatures}

Since right handed forms are now chosen, such as equation (\ref{def.right.chrst}), specialize equation (\ref{gbar.lr}) to 
$ \lambda = 1 $, or
\begin{equation}
\bar g_{\mu \nu} = g_{\mu \nu} \rho 
\label{gbar.r}
\end{equation}
Corresponding to that,
\begin{equation}
\bar g^{\mu \nu} = \rho^{-1} g^{\mu \nu} 
\label{gbar.i.r}
\end{equation}
These together with equation (\ref{def.right.chrst}) then give that
{\samepage 
\begin{eqnarray}
\bar{ \{ ^{\: \alpha}_{\mu \nu} ] } & = & 
\rho^{-1} \{ ^{\: \alpha}_{\mu \nu} ] \rho + 
{\textstyle{1 \over 2}}\, \delta^{\alpha}_{\mu} \rho^{-1} \rho_{, \nu} + 
{\textstyle{1 \over 2}}\, \delta^{\alpha}_{\nu} \rho^{-1} \rho_{, \mu} - 
{\textstyle{1 \over 2}}\, \rho^{-1} g^{\alpha \tau} \! g_{\mu \nu} \, 
\rho_{, \tau} 
\nonumber \\
 & = & 
\rho^{-1} \{ ^{\: \alpha}_{\mu \nu} ] \rho + 
{\textstyle{1 \over 2}}\, \delta^{\alpha}_{\mu} \rho^{-1} \rho_{, \nu} + 
{\textstyle{1 \over 2}}\, \delta^{\alpha}_{\nu} \rho^{-1} \rho_{, \mu} - 
{\textstyle{1 \over 2}}\, g^{\alpha \tau} \! g_{\mu \nu} \, 
\rho^{-1} \rho_{, \tau} 
\label{def.barchrst}
\end{eqnarray}}%
where the second line follows from equation (\ref{gauge.inv.gpair}).

Analogy with reference \cite{rankin.caqg} would now suggest an affine connection 
$ {}_{R} \Gamma^{\alpha}_{\mu \nu} $ (hereafter denoted by 
$ \Gamma^{\alpha}_{\mu \nu} $) to exist which gauge transforms in analogy to Einstein's ``lambda invariance''\cite{einstein.meaning} via
\begin{equation}
\bar{\Gamma}^{\alpha}_{\mu \nu} = 
\rho^{-1} \Gamma^{\alpha}_{\mu \nu} \, \rho + 
{\textstyle{1 \over 2}}\, \delta^{\alpha}_{\mu} \, \rho^{-1} \rho_{, \nu} 
\label{gauge.gamma.cartan}
\end{equation}
This would be done by defining the connection as
\begin{equation}
\Gamma^{\alpha}_{\mu \nu} = 
\{ ^{\: \alpha}_{\mu \nu} ] + 
\delta^{\alpha}_{\nu} v_{\mu} - 
g^{\alpha \tau} \! g_{\mu \nu} v_{\tau}
\label{def.gamma.cartan}
\end{equation}
where $ v_{\mu} $ is the quantity of equation (\ref{def.qv}), and
{\samepage 
\begin{eqnarray}
\bar{v}_{\mu} & = & \rho^{-1} (v_{\mu} - {\textstyle{1 \over 2}}\, 
\rho_{, \mu} \rho^{-1}) \rho 
\nonumber \\
 & = & \rho^{-1} v_{\mu} \, \rho - {\textstyle{1 \over 2}}\, 
\rho^{-1} \rho_{, \mu} 
\label{def.vbar.xtnd}
\end{eqnarray}}%
Equation (\ref{def.vbar.xtnd}) will indeed be adopted, but in place of equations (\ref{gauge.gamma.cartan}) and (\ref{def.gamma.cartan}), adopt
\begin{equation}
\bar{\Gamma}^{\alpha}_{\mu \nu} = 
\rho^{-1} \Gamma^{\alpha}_{\mu \nu} \, \rho + 
k \delta^{\alpha}_{\mu} \, \rho^{-1} \rho_{, \nu} 
\label{gauge.gamma.mixed.k}
\end{equation}
where $ k $ is a constant, and
\begin{equation}
\Gamma^{\alpha}_{\mu \nu} = 
\{ ^{\: \alpha}_{\mu \nu} ] + 
n \delta^{\alpha}_{\mu} v_{\nu} +
\delta^{\alpha}_{\nu} v_{\mu} - 
g^{\alpha \tau} \! g_{\mu \nu} v_{\tau}
\label{def.gamma.mixed}
\end{equation}
where $ n $ is a constant, and 
\begin{equation}
k = \left ( 1 - n \right ) / 2
\label{def.k.n}
\end{equation}
For 
$ n \neq 0 $, some amount of nonmetricity appears, and indeed, 
$ n = 1 $ would give the quaternionic generalization of Weyl's original geometry\cite{weyl.stm} with no torsion. However, other values of 
$ n $ will give some torsion, and keep a form of Einstein's ``lambda invariance''\cite{einstein.meaning}.

Now both of the connections 
$ \Gamma^{\alpha}_{\mu \nu} $ and 
$ \{ ^{\: \alpha}_{\mu \nu} ] $ have associated curvature tensors, but those have both a ``right handed'' and a ``left handed'' form themselves, irrespective of the right-left nature of the underlying connection used in them. Specifically,
\begin{equation}
{}_{R} B^{\gamma}_{\mu \tau \sigma} = 
\Gamma^{\gamma}_{\mu \sigma , \tau} - 
\Gamma^{\gamma}_{\mu \tau , \sigma} + 
\Gamma^{\gamma}_{\eta \tau} \Gamma^{\eta}_{\mu \sigma} - 
\Gamma^{\gamma}_{\eta \sigma} \Gamma^{\eta}_{\mu \tau} 
\label{def.b.r}
\end{equation}
\begin{equation}
{}_{L} B^{\gamma}_{\mu \tau \sigma} = 
\Gamma^{\gamma}_{\mu \sigma , \tau} - 
\Gamma^{\gamma}_{\mu \tau , \sigma} + 
\Gamma^{\eta}_{\mu \sigma} \Gamma^{\gamma}_{\eta \tau} - 
\Gamma^{\eta}_{\mu \tau} \Gamma^{\gamma}_{\eta \sigma} 
\label{def.b.l}
\end{equation}
\begin{equation}
{}_{R} R^{\gamma}_{\mu \tau \sigma} = 
\{ ^{\: \gamma}_{\mu \sigma} ]_{, \tau} - 
\{ ^{\: \gamma}_{\mu \tau} ]_{, \sigma} + 
\{ ^{\: \gamma}_{\eta \tau} ] \{ ^{\: \eta}_{\mu \sigma} ] - 
\{ ^{\: \gamma}_{\eta \sigma} ] \{ ^{\: \eta}_{\mu \tau} ] 
\label{def.r.r}
\end{equation}
and
\begin{equation}
{}_{L} R^{\gamma}_{\mu \tau \sigma} = 
\{ ^{\: \gamma}_{\mu \sigma} ]_{, \tau} - 
\{ ^{\: \gamma}_{\mu \tau} ]_{, \sigma} + 
\{ ^{\: \eta}_{\mu \sigma} ] \{ ^{\: \gamma}_{\eta \tau} ] - 
\{ ^{\: \eta}_{\mu \tau} ] \{ ^{\: \gamma}_{\eta \sigma} ] 
\label{def.r.l}
\end{equation}

When equation (\ref{gauge.gamma.mixed.k}) is substituted into equations (\ref{def.b.r}) and (\ref{def.b.l}) to obtain the gauge properties of those curvature tensors, generally neither curvature form transforms in a particularly neat manner by itself. Surprisingly however, the combination
\begin{equation}
B^{\gamma}_{\mu \tau \sigma} = 
\left [ \left ( k + 1 \right ) / \left ( 2k \right ) \right ] 
{}_{R} \! B^{\gamma}_{\mu \tau \sigma} + 
\left [ \left ( k - 1 \right ) / \left ( 2k \right ) \right ] 
{}_{L} \! B^{\gamma}_{\mu \tau \sigma} 
\label{def.b.neat.k}
\end{equation}
{\it does} have neat gauge transformation properties (reverse roles of 
$ {}_{R} \! B^{\gamma}_{\mu \tau \sigma} $ and 
$ {}_{L} \! B^{\gamma}_{\mu \tau \sigma} $ if left handed forms with 
$ \rho = 1$ and 
$ \lambda \ne 1 $ are adopted initially rather than right). Specifically
\begin{equation}
\bar{B}^{\gamma}_{\mu \tau \sigma} = 
\rho^{-1} B^{\gamma}_{\mu \tau \sigma} \, \rho 
\label{def.b.neat.gauge}
\end{equation}
which is the same form as the gauge transformation of a Yang-Mills Field in SU(2) gauge theory\cite{carmeli}. Furthermore, for 
$ n = 0 $, 
$ B^{\gamma}_{\mu \tau \sigma} $ actually reduces to the curvature tensor of reference \cite{rankin.caqg} as it is restricted to the complex plane in which products like 
$ \Gamma^{\gamma}_{\eta \tau} \Gamma^{\eta}_{\mu \sigma} $ commute internally. In other words, it is the appropriate generalization of the curvature tensor of the earlier reference. To facilitate its use in what follows, the coefficients in the definition of this tensor in equation (\ref{def.b.neat.k}) are given their own symbols,
\begin{equation}
k_{+} = \left ( k + 1 \right ) / \left ( 2 k \right )
\label{def.kplus}
\end{equation}
and
\begin{equation}
k_{-} = \left ( k - 1 \right ) / \left ( 2 k \right )
\label{def.kminus}
\end{equation}
Note that 
$ k_{+} + k_{-} = 1 $, and 
$ k_{+} - k_{-} = 1 / k $.

Now note that if one attempts to use a full Weyl connection analog by setting 
$ n = 1 $ in equation (\ref{def.gamma.mixed}), that causes 
$ k $ to become zero in equation (\ref{gauge.gamma.mixed.k}), and no suitable generalized curvature 
$ B^{\gamma}_{\mu \tau \sigma} $ emerges at all. Rather, as 
$ k $ approaches $ 0 $ in equation (\ref{def.b.neat.k}), the coefficients of 
$ {}_{R} B^{\gamma}_{\mu \tau \sigma} $ and 
$ {}_{L} B^{\gamma}_{\mu \tau \sigma} $ approach equal but opposite infinite values. Essentially, equation (\ref{def.b.neat.k}) must then be replaced by
\begin{equation}
B^{\gamma}_{\mu \tau \sigma} = 
{}_{R} \! B^{\gamma}_{\mu \tau \sigma} - 
{}_{L} \! B^{\gamma}_{\mu \tau \sigma} 
\label{def.b.neat.k.0}
\end{equation}
or any simple multiple of the right side of this equation, but that will lead to the disappearance of the derivatives of 
$ \Gamma^{\alpha}_{\mu \nu} $ from the result. Without the derivative terms, this quantity cannot reduce to anything at all like the theory of reference \cite{rankin.caqg} in the complex plane, reducing to zero instead. In other words, this structure actually discriminates against the quaternionic analog of Weyl's original theory\cite{weyl.stm}, and favors the cases which have some torsion. This differs from reference \cite{rankin.caqg}, in which both the Weyl and Cartan type connections were in principle allowed.

Finally, the four vector 
$ v_{\mu} $ of equation (\ref{def.vbar.xtnd}) has its own directly associated Yang-Mills field tensor\cite{carmeli}. The gauge properties give that
{\samepage 
\begin{eqnarray}
y_{\mu \nu} & = & 
v_{\nu ,\mu} - v_{\mu ,\nu} + 
2(v_{\nu} v_{\mu} - v_{\mu} v_{\nu}) 
\nonumber \\
 & = & 
p_{\mu \nu} + 
2(v_{\nu} v_{\mu} - v_{\mu} v_{\nu}) 
\label{def.ym}
\end{eqnarray}}%
gauge transforms just as 
$ B^{\gamma}_{\mu \tau \sigma} $ does, or
\begin{equation}
\bar{y}_{\mu \nu} = 
\rho^{-1} y_{\mu \nu} \, \rho 
\label{def.ym.gauge}
\end{equation}
Since by assumption there exists a gauge in which 
$ v_{\mu} $ has the form of equation (\ref{def.qv}), then in that same gauge
\begin{equation}
y_{\mu \nu} = - q F_{\mu \nu} |Q_{3} 
\label{def.ym.magic.gauge}
\end{equation}
where $ F_{\mu \nu} $ is the standard Maxwell tensor, the curl of 
$ A_{\mu} $, and it is real. But then by equation (\ref{def.ym.gauge}), this tensor has this form in {\it all} gauges provided the gauge function 
$ \rho $ satisfies the assumptions made after equation (\ref{def.hdual}), that it either contains no leap-over operations, or contains none other than 
$ |Q_{3} $. Other than that, 
$ \rho $ is fully quaternionic, but the 
$ |Q_{3} $ stays out of the way of 
$ \rho^{-1} $ and $ \rho $, allowing them to cancel.

Thus $ y_{\mu \nu} $ is actually gauge invariant, and mathematically it remains in the complex plane as a purely imaginary tensor. The leap-over operator has simplified the model, allowing this sort of gauge transformation of one side of a two sided quaternionic product, while the still unaffected, constant unit imaginary quantity operates on the other side of the same product.

\subsection{The Makeup of the Curvature Tensor}

Equation (\ref{def.gamma.mixed}) can be written
\begin{equation}
\Gamma^{\alpha}_{\mu \nu} = 
\{ ^{\: \alpha}_{\mu \nu} ] + 
U^{\alpha}_{\mu \nu} 
\label{def.gamma.u}
\end{equation}
where
\begin{equation}
U^{\alpha}_{\mu \nu} = 
n \delta^{\alpha}_{\mu} v_{\nu} + 
\delta^{\alpha}_{\nu} v_{\mu} - 
g^{\alpha \tau} \! g_{\mu \nu} v_{\tau}
\label{def.u}
\end{equation}
Substituting these into equations (\ref{def.b.r}) and (\ref{def.b.l}) then gives the surprisingly neat results
\begin{equation}
{}_{R} B^{\gamma}_{\mu \tau \sigma} = 
{}_{R} R^{\gamma}_{\mu \tau \sigma} + 
U^{\gamma}_{\mu \sigma ; \tau} - 
U^{\gamma}_{\mu \tau ; \sigma} + 
U^{\gamma}_{\eta \tau} U^{\eta}_{\mu \sigma} - 
U^{\gamma}_{\eta \sigma} U^{\eta}_{\mu \tau} 
\label{def.b.r.u.r}
\end{equation}
and
\begin{equation}
{}_{L} B^{\gamma}_{\mu \tau \sigma} = 
{}_{L} R^{\gamma}_{\mu \tau \sigma} + 
U^{\gamma}_{\mu \sigma \, {\bf \tilde{;}} \, \tau} - 
U^{\gamma}_{\mu \tau \, {\bf \tilde{;}} \, \sigma} + 
U^{\eta}_{\mu \sigma} U^{\gamma}_{\eta \tau} - 
U^{\eta}_{\mu \tau} U^{\gamma}_{\eta \sigma} 
\label{def.b.r.u.l}
\end{equation}
These equations are one example (perhaps the best) in which the 
``$ \; ; \; $'' and 
``$ \; \tilde{;} \; $'' covariant derivatives give results that are both compact, and express useful information.

However, in order to proceed further with the evaluation of 
$ B^{\gamma}_{\mu \tau \sigma} $ via equations (\ref{def.b.neat.k}), (\ref{def.b.r.u.r}), and (\ref{def.b.r.u.l}), the covariant derivatives should be written out as partial derivatives and Christoffel Symbol terms, and equation (\ref{def.u}) should be substituted into the result. Furthermore, any resulting partial derivatives of 
$ g_{\mu \nu} $ or $ g^{\mu \nu} $ should be evaluated using equations (\ref{def.right.cov.metric}) and (\ref{def.right.cov.imetric}) to substitute terms with Christoffel Symbols in place of the partial derivative terms. In practice, the combination 
$ ( g^{\gamma \eta} g_{\mu \sigma} )_{, \tau} $ always appears as a unit, and can be eliminated using
{\samepage 
\begin{eqnarray}
( g^{\gamma \eta} g_{\mu \sigma} )_{, \tau} & = & 
g^{\gamma \eta} g_{\alpha \sigma} \{ ^{\: \alpha}_{\mu \tau} ] + 
g^{\gamma \eta} g_{\mu \alpha} \{ ^{\: \alpha}_{\sigma \tau} ] 
\nonumber \\
 & & - \{ ^{\: \gamma}_{\alpha \tau} ] g^{\alpha \eta} g_{\mu \sigma} - 
\{ ^{\: \eta}_{\alpha \tau} ] g^{\gamma \alpha} g_{\mu \sigma} 
\label{def.cov.gunit}
\end{eqnarray}}%
keeping equation (\ref{gauge.inv.gpair}) in mind for the result. The fact that the 
$ g^{\gamma \eta} g_{\mu \sigma} $ terms are real then allows equation (\ref{def.cov.gunit}) to have more than one valid form simply by varying the position of such terms in its products. However, the same form should consistently be chosen internally throughout evaluation of either one of the {\it separate} tensors in the pair 
$ {}_{R} B^{\gamma}_{\mu \tau \sigma} $ or 
$ {}_{L} B^{\gamma}_{\mu \tau \sigma} $ to avoid possibly encountering extraneous terms that should evaluate to zero with some effort, but are more easily avoided from the outset. Additionally, the full expression of equation (\ref{def.cov.gunit}) itself should be real, and could be moved around as a unit in products in its containing equation if necessary. However, all this flexibility leads to more than one expansion of 
$ B^{\gamma}_{\mu \tau \sigma} $ in gauge varying quantities like 
$ v_{\mu} $, although all the expansions are equivalent, and all will lead to the same, unique, gauge invariant result in what follows. Since the gauge invariant result contains any real physics, its uniqueness is what is important.

The result of the above substitutions and expansions gives
{\samepage 
\begin{eqnarray}
B^{\gamma}_{\mu \tau \sigma} & = & 
k_{+} \, {}_{R} \! R^{\gamma}_{\mu \tau \sigma} + 
k_{-} \, {}_{L} \! R^{\gamma}_{\mu \tau \sigma} 
\nonumber \\
 & & + \left ( k_{+} \, v_{\mu ; \tau} + 
k_{-} \, v_{\mu \, {\bf \tilde{;}} \, \tau} 
\right ) \delta^{\gamma}_{\sigma} - 
\left ( k_{+} \, v_{\mu ; \sigma} + 
k_{-} \, v_{\mu \, {\bf \tilde{;}} \, \sigma} 
\right ) \delta^{\gamma}_{\tau} 
\nonumber \\
 & & - \left ( k_{+} \, v_{\eta ; \tau} + 
k_{-} \, v_{\eta \, {\bf \tilde{;}} \, \tau} 
\right ) g^{\eta \gamma} g_{\mu \sigma} + 
\left ( k_{+} \, v_{\eta ; \sigma} + 
k_{-} \, v_{\eta \, {\bf \tilde{;}} \, \sigma} 
\right ) g^{\eta \gamma} g_{\mu \tau} 
\nonumber \\
 & & + \left ( 1 / k \right ) \left [ v_{\eta} , 
\{ ^{\: \gamma}_{\alpha \tau} ] \, \right ] 
g^{\eta \alpha} g_{\mu \sigma} - 
\left ( 1 / k \right ) \left [ v_{\eta} , 
\{ ^{\: \gamma}_{\alpha \sigma} ] \, \right ] 
g^{\eta \alpha} g_{\mu \tau} 
\nonumber \\
 & & + \left ( k_{+} \, v_{\sigma} v_{\mu} + 
k_{-} \, v_{\mu} v_{\sigma} \right ) 
\delta^{\gamma}_{\tau} - 
\left ( k_{+} \, v_{\tau} v_{\mu} + 
k_{-} \, v_{\mu} v_{\tau} \right ) 
\delta^{\gamma}_{\sigma} 
\nonumber \\
 & & - v_{\eta} v_{\beta} g^{\eta \beta} g_{\mu \sigma} 
\delta^{\gamma}_{\tau} + 
v_{\eta} v_{\beta} g^{\eta \beta} g_{\mu \tau} 
\delta^{\gamma}_{\sigma} 
\nonumber \\
 & & + \left ( k_{+} \, v_{\alpha} v_{\tau} + 
k_{-} \, v_{\tau} v_{\alpha} \right ) 
g^{\alpha \gamma} g_{\mu \sigma} - 
\left ( k_{+} \, v_{\alpha} v_{\sigma} + 
k_{-} \, v_{\sigma} v_{\alpha} \right ) 
g^{\alpha \gamma} g_{\mu \tau} 
\nonumber \\
 & & + n \left [ k_{+} \left ( v_{\sigma ; \tau} - 
v_{\tau ; \sigma} \right ) + k_{-} \left ( 
v_{\sigma \, {\bf \tilde{;}} \, \tau} - 
v_{\tau \, {\bf \tilde{;}} \, \sigma} \right ) 
\right ] \delta^{\gamma}_{\mu} 
\nonumber \\
& & - \left ( n / k \right ) \left [ v_{\sigma} , 
\{ ^{\: \gamma}_{\mu \tau} ] \, \right ] + 
\left ( n / k \right ) \left [ v_{\tau} , 
\{ ^{\: \gamma}_{\mu \sigma} ] \, \right ] 
\nonumber \\
& & - \left ( n^{2} / k \right ) \left ( v_{\sigma} 
v_{\tau} - v_{\tau} v_{\sigma} \right ) 
\delta^{\gamma}_{\mu} 
\nonumber \\
& & - \left ( n / k \right ) \left ( v_{\mu} 
v_{\tau} - v_{\tau} v_{\mu} \right ) 
\delta^{\gamma}_{\sigma} + 
\left ( n / k \right ) \left ( v_{\mu} 
v_{\sigma} - v_{\sigma} v_{\mu} \right ) 
\delta^{\gamma}_{\tau} 
\nonumber \\
& & + \left ( n / k \right ) \left ( v_{\alpha} 
v_{\tau} - v_{\tau} v_{\alpha} \right ) 
g^{\alpha \gamma} g_{\mu \sigma} - 
\left ( n / k \right ) \left ( v_{\alpha} 
v_{\sigma} - v_{\sigma} v_{\alpha} \right ) 
g^{\alpha \gamma} g_{\mu \tau} 
\label{def.expanded.b}
\end{eqnarray}}%
where the ``$ [ \; , \; ] $'' terms are conventional commutators. Those commutators will clearly vanish in gauges in which 
$ \{ ^{\: \gamma}_{\alpha \sigma} ] $ is real.

This can be contracted to give
\begin{equation}
B_{\mu \tau} = B^{\omega}_{\mu \tau \omega} 
\label{def.contract.b}
\end{equation}
and clearly equation (\ref{def.b.neat.gauge}) gives that
\begin{equation}
\bar{B}_{\mu \tau} = 
\rho^{-1} B_{\mu \tau} \, \rho 
\label{def.contract.b.gauge}
\end{equation}
The similarity between this equation and equation (\ref{def.ym.gauge}) might then raise expectations that the antisymmetric part of 
$ B_{\mu \tau} $ will be proportional to 
$ y_{\mu \tau} $. However, this is not quite the case. A check reveals that the antisymmetric part of 
$ k_{+} \, {}_{R} \! R^{\omega}_{\mu \tau \omega} + 
k_{-} \, {}_{L} \! R^{\omega}_{\mu \tau \omega} $
equals 
$ \{ 1 - [ 1 / ( 4 k ) ] \} [ \gamma^{-1} \gamma_{, \mu} , 
\gamma^{-1} \gamma_{, \tau} ] $ 
where $ \gamma $ is the gauge function in 
$ g_{\mu \nu} = \tilde{g}_{\mu \nu} \gamma $. Since this commutator does not generally vanish, then the antisymmetric part of 
$ k_{+} \, {}_{R} \! R^{\omega}_{\mu \tau \omega} + 
k_{-} \, {}_{L} \! R^{\omega}_{\mu \tau \omega} $ 
is generally not zero, and that antisymmetric component must be gauge balanced elsewhere by antisymmetric terms, even though it vanishes in gauges in which 
$ g_{\mu \nu} $ is real, and also when 
$ \gamma $ remains in the complex plane. However, the obvious exception is the case 
$ k = 1 / 4 $, which causes the term to vanish even when the commutator is nonzero. That special value of 
$ k $ corresponds to 
$ n = 1 / 2 $. The consequences of all this will become clearer in the next section.

Finally define the scalar curvature
\begin{equation}
B = B_{\mu \tau} M^{\mu \tau} 
\label{def.scalar.b}
\end{equation}
where $ M^{\mu \tau} $ is the full asymmetric inverse metric. It is given by
\begin{equation}
M^{\mu \tau} = (g^{\mu \tau} - a^{\mu \tau}) 
( 1 +{\textstyle{1 \over 4}}\, a )^{-1} 
\label{def.asym.imetric}
\end{equation}
where 
$ a = a_{\mu \nu} a^{\mu \nu} $, a gauge invariant quantity which actually lies in the complex plane still, since it is real except for including a 
$ |Q_{3} $.
Since
\begin{equation}
\bar{M}^{\mu \tau} = 
\rho^{-1} M^{\mu \tau} 
\label{def.asym.imetric.gauge}
\end{equation}
then equations (\ref{def.contract.b.gauge}) and (\ref{def.scalar.b}) give
\begin{equation}
\bar{B} = 
\rho^{-1} B 
\label{def.scalar.b.gauge}
\end{equation}
Thus $B$ is the key quantity needed to define gauge invariant variables. As conceived by Weyl and Eddington\cite{weyl.stm,eddington.mtr}, it is basically an intrinsic yardstick provided by the spacetime structure itself to reduce equations to dimensionless, gauge invariant quantities that can correspond to actual physics. For this purpose, it is assumed to be nonzero.

This key yardstick 
$ B $ can still be defined under slightly more general conditions. Up to this point, 
$ g_{\mu \nu} = \tilde g_{\mu \nu} \gamma $ has also implied
\begin{equation}
h_{\mu \nu} = \tilde h_{\mu \nu} \gamma 
\label{h.to.hreal}
\end{equation}
where 
$ \tilde h_{\mu \nu} $ is real like 
$ \tilde g_{\mu \nu} $ is. But this is a tighter restriction than is actually necessary. If equation (\ref{h.to.hreal}) is replaced by
\begin{equation}
h_{\mu \nu} = \tilde h_{\mu \nu} \beta 
\label{h.to.hreal.beta}
\end{equation}
where 
$ \gamma \neq \beta $, then 
$ g_{\mu \nu} $ and 
$ h_{\mu \nu} $ are no longer required to be real in the same gauge. To handle that generalization, define
\begin{equation}
{}^{*} \tilde h^{\mu \nu} = \frac{1}{2} (- \tilde g)^{-1 / 4} \, 
\epsilon^{\mu \nu \alpha \beta} \, \tilde h_{\alpha \beta} 
(- \tilde g)^{-1 / 4} 
\label{def.hdual.upper.real}
\end{equation}
as well as
\begin{equation}
{}^{*} \tilde h_{\mu \nu} = \tilde g_{\mu \tau} 
{}^{*} \tilde h^{\tau \sigma} \tilde g_{\sigma \nu}
\label{def.hdual.lower.real}
\end{equation}
and
\begin{equation}
\tilde h^{\mu \nu} = \tilde g^{\mu \tau} 
\tilde h_{\tau \sigma} \tilde g^{\sigma \nu}
\label{def.h.upper.real}
\end{equation}
Of course, these are all real quantities, so positions in these products are no longer mandatory, but are retained to compare easily to cases when position is important. Cases where position is still important are equation (\ref{def.hdual}), and
\begin{equation}
h^{\mu \nu} = g^{\mu \tau} 
h_{\tau \sigma} g^{\sigma \nu}
\label{def.h.upper}
\end{equation}
It is now easy to show that when 
$ \bar{g}_{\mu \nu} = g_{\mu \nu} \rho = \tilde g_{\mu \nu} \gamma \rho $, then
\begin{equation}
\bar{h}^{\mu \nu} = \rho^{-1} \, h^{\mu \nu} = \rho^{-1} 
\gamma^{-1} \beta \gamma^{-1} \, \tilde h^{\mu \nu} 
\label{h.to.hreal.beta.upper.bar}
\end{equation}
\begin{equation}
{}^{*} \bar{h}^{\mu \nu} = \rho^{-1} \, 
{}^{*} h^{\mu \nu} = \rho^{-1} \gamma^{-1} \beta 
\gamma^{-1} \, {}^{*} \tilde h^{\mu \nu} 
\label{hdual.to.hdualreal.beta.upper.bar}
\end{equation}
\begin{equation}
{}^{*} \bar{h}_{\mu \nu} = {}^{*} h_{\mu \nu} \rho = 
{}^{*} \tilde h_{\mu \nu} \beta \rho 
\label{hdual.to.hdualreal.beta.lower.bar}
\end{equation}
as well as
\begin{equation}
\bar{h}_{\mu \nu} = h_{\mu \nu} \rho = 
\tilde h_{\mu \nu} \beta \rho 
\label{h.to.hreal.beta.bar}
\end{equation}
Note that if 
$ \rho = \gamma^{-1} $, then 
$ \bar{g}_{\mu \nu} = \tilde g_{\mu \nu} $, 
$ \bar{h}_{\mu \nu} = \beta \gamma^{-1} \tilde h_{\mu \nu} $, and 
$ \bar{h}^{\mu \nu} = \beta \gamma^{-1} \tilde h^{\mu \nu} $, with the same relations also holding for the dual antisymmetric tensors. When 
$ g_{\mu \nu} $ is real, the factor 
$ \beta \gamma^{-1} $ is a common multiplier onto all the real parts of the antisymmetric part of the metric, both when indices are up, and when they are down.

Now instead of merely defining 
$ a = a_{\mu \nu} a^{\mu \nu} $, in its place define 
$ {}_{R} a = a_{\mu \nu} a^{\mu \nu} $, and 
$ {}_{L} a = a^{\mu \nu} a_{\mu \nu} $, and also define 
$ \tilde a = \tilde a_{\mu \nu} \tilde a^{\mu \nu} $. Since 
$ \tilde a $ is entirely real except for the 
$ |Q_{3} $ terms, it is the same regardless of the order of the terms in the product. On the other hand,
\begin{equation}
{}_{R} \bar{a} = \beta \gamma^{-1} \beta \gamma^{-1} 
\tilde a
\label{abar.right}
\end{equation}
and
\begin{equation}
{}_{L} \bar{a} = \rho^{-1} \gamma^{-1} \beta \gamma^{-1} \beta 
\rho \, \tilde a = \rho^{-1} \gamma^{-1} \beta \gamma^{-1} \beta 
\gamma^{-1} \gamma \rho \, \tilde a
\label{abar.left}
\end{equation}
Since 
$ \beta \gamma^{-1} $ is gauge invariant, 
$ {}_{R} a $ is at least still gauge invariant, but 
$ {}_{L} a $ does not even appear to be gauge invariant. However, if 
$ \left ( \beta \gamma^{-1} \right )^2 $ commutes with the other quaternions in 
$ {}_{L} \bar{a} $, then 
$ {}_{L} \bar{a} = {}_{R} \bar{a} $, and both are gauge invariant. But if 
$ \left ( \beta \gamma^{-1} \right )^2 $ involves forms such as those in equation (\ref{spin.squared}), the result is completely complex, with 
$ |Q_{3} $ as the unit imaginary. That does commute with the other terms, and also keeps the final result in equation (\ref{abar.right}) completely in the complex plane still. This is sufficient to keep 
$ M^{\mu \tau} $ still well behaved enough to define 
$ B $ as an intrinsic yardstick, or measuring gauge.

\subsection{Gauge Invariant Variables and Their Fundamental Identity}

Now by equations (\ref{gbar.r}) and (\ref{def.scalar.b.gauge}), equation (\ref{def.ghat}) generalizes immediately to
\begin{equation}
\hat g_{\mu \nu} = g_{\mu \nu} (B/C)
\label{def.qghat}
\end{equation}
and its inverse
\begin{equation}
\hat g^{\mu \nu} = C B^{-1} g^{\mu \nu} 
\label{def.qighat}
\end{equation}
for the symmetric part of the gauge invariant metric. Indeed, the full asymmetric metric has an analogous form as
{\samepage 
\begin{eqnarray}
\hat m_{\mu \nu} & = & m_{\mu \nu} (B/C) 
\nonumber \\
 & = & \hat g_{\mu \nu} + \hat a_{\mu \nu} 
\label{def.qmhat}
\end{eqnarray}}%
These are all gauge invariant, and all are required to be real except for the 
$ |Q_{3} $ multiplying 
$ {}^{*} \hat{h}_{\mu \nu} $ in the gauge invariant version of equation (\ref{ahat.makeup}), and except for the fact that 
$ \hat h_{\mu \nu} $ and its dual may contain a quaternionic multiplier whose square is still in the complex plane with 
$ |Q_{3} $ as the unit imaginary, such as is described in the discussion around equation (\ref{abar.right}). The full inverse asymmetric metric of equations (\ref{def.asym.imetric}) and (\ref{def.asym.imetric.gauge}) clearly generalizes to
{\samepage 
\begin{eqnarray}
\hat M^{\mu \nu} & = & C B^{-1} M^{\mu \nu} 
\nonumber \\
& = & (\hat g^{\mu \nu} - \hat a^{\mu \nu}) 
( 1 +{\textstyle{1 \over 4}}\, \hat a )^{-1} 
\label{def.hat.asym.imetric}
\end{eqnarray}}%
where as noted after equations (\ref{def.asym.imetric}) and (\ref{abar.left}), 
$ \hat a $ is a gauge invariant quantity which actually lies in the complex plane still.

There is a gauge invariant 
$ \hat{ \{ ^{\: \alpha}_{\mu \nu} \} } $ based on the real 
$ \hat g_{\mu \nu} $, and it is real,
\begin{equation}
\hat{ \{ ^{\: \alpha}_{\mu \nu} \} } = 
\frac{1}{2} \; \hat g^{\alpha \tau} ( \hat g_{\mu \tau , \nu} + 
\hat g_{\nu \tau , \mu} - \hat g_{\mu \nu , \tau} ) 
\label{def.hat.chrst}
\end{equation}
without a right-left nature any longer. The covariant derivative with respect to it is indicated by 
``$ {}_{\|} $'', and it is now quite well behaved, including obeying the product rule since 
$ \hat{ \{ ^{\: \alpha}_{\mu \nu} \} } $ commutes with everything. Both the 
``$ \; \tilde{;} \; $'' and the 
``$ \; ; \; $'' covariant derivative conventions will reduce to it. If equations (\ref{def.qghat}) and (\ref{def.qighat}) are substituted into equation (\ref{def.hat.chrst}), the result gives
{\samepage 
\begin{eqnarray}
\hat{ \{ ^{\: \alpha}_{\mu \nu} \} } & = & 
B^{-1} \{ ^{\: \alpha}_{\mu \nu} ] B + 
{\textstyle{1 \over 2}}\, \delta^{\alpha}_{\mu} B^{-1} B_{, \nu} + 
{\textstyle{1 \over 2}}\, \delta^{\alpha}_{\nu} B^{-1} B_{, \mu} - 
{\textstyle{1 \over 2}}\, B^{-1} g^{\alpha \tau} \! g_{\mu \nu} \, 
B_{, \tau} 
\nonumber \\
 & = & 
B^{-1} \{ ^{\: \alpha}_{\mu \nu} ] B + 
{\textstyle{1 \over 2}}\, \delta^{\alpha}_{\mu} B^{-1} B_{, \nu} + 
{\textstyle{1 \over 2}}\, \delta^{\alpha}_{\nu} B^{-1} B_{, \mu} - 
{\textstyle{1 \over 2}}\, g^{\alpha \tau} \! g_{\mu \nu} \, 
B^{-1} B_{, \tau} 
\label{def.expand.hat.chrst}
\end{eqnarray}}%

These real, commuting Christoffel Symbols now give us a normal, real, gauge invariant Riemannian geometry on which we can impose a standard form of General Relativity. They define 
$ \hat R^{\gamma}_{\mu \tau \sigma} $ as the Riemann Curvature Tensor, and the conventions used to define 
$ {}_{R} R^{\gamma}_{\mu \tau \sigma} $ and 
$ {}_{L} R^{\gamma}_{\mu \tau \sigma} $ now both reduce to this same tensor. There is a (now symmetric) 
$ \hat R_{\mu \tau} = \hat R^{\omega}_{\mu \tau \omega} $ , and a scalar 
$ \hat R = \hat g^{\mu \tau} \hat R_{\mu \tau} $.

By equations (\ref{def.vbar.xtnd}) and (\ref{def.scalar.b.gauge}), equation (\ref{def.vhat}) for 
$ \hat v_{\mu} $ generalizes to a gauge invariant
{\samepage 
\begin{eqnarray}
\hat v_{\mu} & = & B^{-1} (v_{\mu} - {\textstyle{1 \over 2}}\, 
B_{, \mu} B^{-1}) B 
\nonumber \\
 & = & B^{-1} v_{\mu} B - {\textstyle{1 \over 2}}\, 
B^{-1} B_{, \mu} 
\label{def.qvhat}
\end{eqnarray}}%
which is fully quaternionic generally. Then in analogy to equation (\ref{def.gamma.mixed}), define the gauge invariant
\begin{equation}
\hat \Gamma^{\alpha}_{\mu \nu} = 
\hat{\{ ^{\: \alpha}_{\mu \nu} \} } + 
n \delta^{\alpha}_{\mu} \hat v_{\nu} + 
\delta^{\alpha}_{\nu} \hat v_{\mu} - 
\hat g^{\alpha \tau} \! \hat g_{\mu \nu} \hat v_{\tau}
\label{def.gamma.hat.mixed}
\end{equation}
Note that since 
$ \hat \Gamma^{\alpha}_{\mu \nu} $ is fully quaternionic, the full affine derivative of a quantity using 
$ \hat \Gamma^{\alpha}_{\mu \nu} $ is {\it not} as well behaved as the covariant derivative using only the real 
$ \hat{ \{ ^{\: \alpha}_{\mu \nu} \} } $.

Now substituting into equation (\ref{def.gamma.hat.mixed}) from equations (\ref{def.expand.hat.chrst}) and (\ref{def.qvhat}), and using equation (\ref{def.gamma.mixed}), one sees
\begin{equation}
\hat \Gamma^{\alpha}_{\mu \nu} = 
B^{-1} \Gamma^{\alpha}_{\mu \nu} \, B + 
k \delta^{\alpha}_{\mu} \, B^{-1} B_{, \nu} 
\label{gamma.hat.gamma}
\end{equation}
But this is exactly the same form as a gauge transformation on 
$ \Gamma^{\alpha}_{\mu \nu} $ as defined in equation (\ref{gauge.gamma.mixed.k}). Thus, if one defines 
$ {}_{R} \hat B^{\gamma}_{\mu \tau \sigma} $ and 
$ {}_{L} \hat B^{\gamma}_{\mu \tau \sigma} $ using 
$ \hat \Gamma^{\alpha}_{\mu \nu} $ in full analogy to the use of 
$ \Gamma^{\alpha}_{\mu \nu} $ in 
$ {}_{R} B^{\gamma}_{\mu \tau \sigma} $ and 
$ {}_{L} B^{\gamma}_{\mu \tau \sigma} $, the result gives finally that
{\samepage 
\begin{eqnarray}
\hat B^{\gamma}_{\mu \tau \sigma} & = & 
k_{+} \, {}_{R} \! \hat B^{\gamma}_{\mu \tau \sigma} + 
k_{-} \, {}_{L} \! \hat B^{\gamma}_{\mu \tau \sigma} 
\nonumber \\
 & = & B^{-1} B^{\gamma}_{\mu \tau \sigma} B 
\label{def.bhat}
\end{eqnarray}}%

This can be expanded just like equation (\ref{def.expanded.b}), but now with so many quantities real, the much simpler result is
{\samepage 
\begin{eqnarray}
\hat B^{\gamma}_{\mu \tau \sigma} & = & 
\hat R^{\gamma}_{\mu \tau \sigma} + \hat v_{\mu \| \tau} 
\delta^{\gamma}_{\sigma} - 
\hat v_{\mu \| \sigma} 
\delta^{\gamma}_{\tau} 
\nonumber \\
 & & - \hat v_{\eta \| \tau} 
\hat g^{\eta \gamma} \hat g_{\mu \sigma} + 
\hat v_{\eta \| \sigma} 
\hat g^{\eta \gamma} \hat g_{\mu \tau} 
\nonumber \\
 & & + \left ( k_{+} \, \hat v_{\sigma} \hat v_{\mu} + 
k_{-} \, \hat v_{\mu} \hat v_{\sigma} \right ) 
\delta^{\gamma}_{\tau} - 
\left ( k_{+} \, \hat v_{\tau} \hat v_{\mu} + 
k_{-} \, \hat v_{\mu} \hat v_{\tau} \right ) 
\delta^{\gamma}_{\sigma} 
\nonumber \\
 & & - \hat v_{\eta} \hat v_{\beta} \hat g^{\eta \beta} 
\hat g_{\mu \sigma} 
\delta^{\gamma}_{\tau} + 
\hat v_{\eta} \hat v_{\beta} \hat g^{\eta \beta} \hat g_{\mu \tau} 
\delta^{\gamma}_{\sigma} 
\nonumber \\
 & & + \left ( k_{+} \, \hat v_{\alpha} \hat v_{\tau} + 
k_{-} \, \hat v_{\tau} \hat v_{\alpha} \right ) 
\hat g^{\alpha \gamma} \hat g_{\mu \sigma} - 
\left ( k_{+} \, \hat v_{\alpha} \hat v_{\sigma} + 
k_{-} \, \hat v_{\sigma} \hat v_{\alpha} \right ) 
\hat g^{\alpha \gamma} \hat g_{\mu \tau} 
\nonumber \\
 & & + n \left ( \hat v_{\sigma \| \tau} - 
\hat v_{\tau \| \sigma} \right ) \delta^{\gamma}_{\mu} 
\nonumber \\
 & & - \left ( n^{2} / k \right ) \left ( \hat v_{\sigma} 
\hat v_{\tau} - \hat v_{\tau} \hat v_{\sigma} \right ) 
\delta^{\gamma}_{\mu} 
\nonumber \\
 & & - \left ( n / k \right ) \left ( \hat v_{\mu} 
\hat v_{\tau} - \hat v_{\tau} \hat v_{\mu} \right ) 
\delta^{\gamma}_{\sigma} + 
\left ( n / k \right ) \left ( \hat v_{\mu} 
\hat v_{\sigma} - \hat v_{\sigma} \hat v_{\mu} \right ) 
\delta^{\gamma}_{\tau} 
\nonumber \\
 & & + \left ( n / k \right ) \left ( \hat v_{\alpha} 
\hat v_{\tau} - \hat v_{\tau} \hat v_{\alpha} \right ) 
\hat g^{\alpha \gamma} \hat g_{\mu \sigma} - 
\left ( n / k \right ) \left ( \hat v_{\alpha} 
\hat v_{\sigma} - \hat v_{\sigma} \hat v_{\alpha} \right ) 
\hat g^{\alpha \gamma} \hat g_{\mu \tau} 
\label{def.expanded.bhat}
\end{eqnarray}}%
Much of the right-left distinction of equation (\ref{def.expanded.b}), along with the commutators, is now gone. The main left-right distinction remaining is in the terms involving products of 
$ \hat v_{\mu} $, because that quantity is fully quaternionic still.

Now using equations (\ref{def.k.n}), (\ref{def.kplus}), and (\ref{def.kminus}), equation (\ref{def.expanded.bhat}) and equation (\ref{def.bhat}) then contract to give
{\samepage 
\begin{eqnarray}
\hat B_{\mu \tau} & = & \hat B^{\omega}_{\mu \tau \omega} 
\nonumber \\
 & = & B^{-1} B_{\mu \tau} B 
\nonumber \\
 & = & \hat R_{\mu \tau} + 
\left ( \hat v_{\mu \| \tau} + \hat v_{\tau \| \mu} \right ) + 
\hat v^{\alpha}_{\; \; \| \alpha} \hat g_{\mu \tau} 
\nonumber \\
 & & - \left ( \hat v_{\mu} \hat v_{\tau} + 
\hat v_{\tau} \hat v_{\mu} \right ) + 
2 \hat v^{\alpha} \hat v_{\alpha} 
\hat g_{\mu \tau} 
\nonumber \\
 & & + \left ( 1 + n \right ) \left ( \hat v_{\mu \| \tau} - 
\hat v_{\tau \| \mu} \right ) + 
\left [ \left ( 4 - 4 n - 2 n^{2} \right ) / 
\left ( 1 - n \right ) \right ] 
\left ( \hat v_{\mu} \hat v_{\tau} - 
\hat v_{\tau} \hat v_{\mu} \right ) 
\label{def.expanded.contract.bhat}
\end{eqnarray}}%
where the symmetric and antisymmetric parts have been clearly separated with the antisymmetric part all on the last line. Because 
$ \hat v_{\mu \| \tau} - \hat v_{\tau \| \mu} = \hat v_{\mu , \tau} - 
\hat v_{\tau , \mu} $, 
that antisymmetric part is
{\samepage 
\begin{eqnarray}
- \hat w_{\mu \tau} & = & 
\left ( 1 + n \right ) \left ( \hat v_{\mu \| \tau} - 
\hat v_{\tau \| \mu} \right ) + 
\left [ \left ( 4 - 4 n - 2 n^{2} \right ) / 
\left ( 1 - n \right ) \right ] 
\left ( \hat v_{\mu} \hat v_{\tau} - 
\hat v_{\tau} \hat v_{\mu} \right ) 
\nonumber \\
 & = & 
\left ( 1 + n \right ) \left ( \hat v_{\mu , \tau} - 
\hat v_{\tau , \mu} \right ) + 
\left [ \left ( 4 - 4 n - 2 n^{2} \right ) / 
\left ( 1 - n \right ) \right ] 
\left ( \hat v_{\mu} \hat v_{\tau} - 
\hat v_{\tau} \hat v_{\mu} \right ) 
\nonumber \\
 & = & 
- \left ( 1 + n \right ) \left \{ \hat v_{\tau , \mu} - 
\hat v_{\mu , \tau} + 2 \left ( \hat v_{\tau} \hat v_{\mu} - 
\hat v_{\mu} \hat v_{\tau} \right ) + \left [ 
\left ( 2 - 4 n \right ) / \left ( 1 - n^2 \right ) \right ] 
\left ( \hat v_{\tau} \hat v_{\mu} - 
\hat v_{\mu} \hat v_{\tau} \right ) \right \} 
\nonumber \\
 & = & 
- \left ( 1 + n \right ) \left \{ \hat y_{\mu \tau} + \left [ 
\left ( 2 - 4 n \right ) / \left ( 1 - n^2 \right ) \right ] 
\left ( \hat v_{\tau} \hat v_{\mu} - 
\hat v_{\mu} \hat v_{\tau} \right ) \right \} 
\label{def.w.hat}
\end{eqnarray}}%
where
{\samepage 
\begin{eqnarray}
\hat y_{\mu \tau} & = & 
\hat v_{\tau , \mu} - \hat v_{\mu , \tau} + 
2 (\hat v_{\tau} \hat v_{\mu} - 
\hat v_{\mu} \hat v_{\tau}) 
\nonumber \\
 & = & B^{-1} y_{\mu \tau} B 
\label{def.yhat}
\end{eqnarray}}%
by equations (\ref{def.ym}) and (\ref{def.ym.gauge}), since equation (\ref{def.qvhat}) has the same form as the gauge transformation of 
$ v_{\mu} $ in equation (\ref{def.vbar.xtnd}). Additionally, the distinct alternate contraction of 
$ \hat B^{\gamma}_{\mu \tau \sigma} $ gives
\begin{equation}
\hat B^{\gamma}_{\gamma \tau \sigma} = 
4 n \hat y_{\tau \sigma} + 4 \left [ 
\left ( 1 - 2 n \right ) / \left ( 1 - n \right ) \right ] 
\left ( \hat v_{\sigma} \hat v_{\tau} - 
\hat v_{\tau} \hat v_{\sigma} \right ) 
\label{def.expanded.alt.contract.bhat}
\end{equation}
This contraction of the curvature tensor is also important in Weyl's original theory\cite{weyl.stm,eddington.mtr}.

Provided the quaternionic value of 
$ B $ contains no leap-over operations, or contains none other than 
$ |Q_{3} $, equation (\ref{def.ym.magic.gauge}) implies that
\begin{equation}
\hat y_{\mu \nu} = - q F_{\mu \nu} |Q_{3} 
\label{def.ym.hat.magic}
\end{equation}
Likewise, explicitly choosing a gauge in which 
$ v_{\mu} = - q A_{\mu} |Q_{3} $, as in equation (\ref{def.qv}), gives
\begin{equation}
\hat v_{\mu} = - q A_{\mu} |Q_{3} - {\textstyle{1 \over 2}}\, 
B^{-1} B_{, \mu} 
\label{def.qvhat.normalgauge}
\end{equation}
and
{\samepage 
\begin{eqnarray}
\hat w_{\mu \nu} & = & 
\left ( 1 + n \right ) \left \{ \hat y_{\mu \nu} + \left [ 
\left ( 2 - 4 n \right ) / \left ( 1 - n^2 \right ) \right ] 
\left ( \hat v_{\nu} \hat v_{\mu} - 
\hat v_{\mu} \hat v_{\nu} \right ) \right \} 
\nonumber \\
 & = & - \left ( 1 + n \right ) q F_{\mu \nu} |Q_{3} + 
\left [ \left ( 1 - 2 n \right ) / \left ( 2 - 2 n \right ) \right ] 
\left [ B^{-1} B_{, \nu} , B^{-1} B_{, \mu} \right ] 
\nonumber \\
& = & - \left ( 1 + n \right ) q F_{\mu \nu} |Q_{3} - 
\left [ \left ( 1 - 2 n \right ) / \left ( 2 - 2 n \right ) \right ] 
\left [ B^{-1} B_{, \mu} , B^{-1} B_{, \nu} \right ] 
\label{def.w.hat.tail}
\end{eqnarray}}%
For $ n \neq 1 / 2 $, the quantity 
$ \hat w_{\mu \nu} $ appears to have a curvature generated, quaternionic tail on it, unlike 
$ \hat y_{\mu \nu} $, which remains in the complex plane. This was anticipated above when the antisymmetric part of 
$ \, k_{+} \, {}_{R} \! R^{\omega}_{\mu \tau \omega} + 
k_{-} \, {}_{L} \! R^{\omega}_{\mu \tau \omega} $ was noted to require additional antisymmetric terms to gauge balance it unless 
$ n = 1 / 2 $. This is the form those extra terms take in 
$ \hat w_{\mu \nu} $. Notice also that since the form of the last two lines of equation (\ref{def.w.hat.tail}) does imply a special family of gauges in which 
$ v_{\mu} = - q A_{\mu} |Q_{3} $ with 
$ A_{\mu} $ real, the particular form of the quaternionic tail in those two lines is not necessarily itself gauge invariant under gauge transformations that would violate that assumed gauge limitation.

Given that 
$ B = B_{\mu \tau} M^{\mu \tau} $, 
$ \; \hat M^{\mu \tau} = C B^{-1} M^{\mu \tau} $, and 
$ \hat B_{\mu \tau} = B^{-1} B_{\mu \tau} B $, then
{\samepage 
\begin{eqnarray}
\hat B & = & \hat B_{\mu \tau} \hat M^{\mu \tau} 
\nonumber \\
 & = & C B^{-1} B_{\mu \tau} M^{\mu \tau} 
\nonumber \\
 & = & C 
\label{def.scalar.bhat}
\end{eqnarray}}%
This is the fundamental, kinematic identity the gauge invariant variables must satisfy by virtue of their definitions, and the geometry's kinematics. Substituting from equations (\ref{def.hat.asym.imetric}), (\ref{def.expanded.contract.bhat}), and (\ref{def.w.hat}) for 
$ \hat M^{\mu \tau} $ and the expansion of 
$ \hat B_{\mu \tau} $, equation (\ref{def.scalar.bhat}) becomes
\begin{equation}
\hat R + 6 \hat v^{\mu}_{\; \; \| \mu} + 6 \hat v^{\mu} 
\hat v_{\mu} + \hat w_{\mu \nu} \hat a^{\mu \nu} 
 = C ( 1 +{\textstyle{1 \over 4}} \, \hat a ) 
\label{def.identity.q}
\end{equation}
This is almost exactly the same form as the identity in the complex plane, given in equation (\ref{def.identity}). However now, 
$ \hat w_{\mu \nu} $ replaces 
$ \hat p_{\mu \nu} $, 
$ \hat v_{\mu} $ is fully quaternionic, and unless 
$ n = 1 / 2 $, the position of 
$ \hat a^{\mu \nu} $ in the spin term appears to be important if 
$ \hat h^{\mu \nu} $ and 
$ {}^{*} \hat{h}^{\mu \nu} $ are not real, but are instead real values times a quantity 
$ \left ( \beta \gamma^{-1} \right )^{2} $ noted in conjunction with equations (\ref{abar.right}) and (\ref{spin.squared}). It is only the quaternionic nature of 
$ B $ that pushes 
$ \hat v_{\mu} $ and possibly 
$ \hat w_{\mu \nu} $ out of the complex plane and into the full quaternions, but nevertheless, equation (\ref{def.identity.q}) is substantially more complicated than the version in the complex plane.

Most significantly, the basic Ricatti Equation change of variable, 
$ B = \psi^{-2} $ used in all the earlier versions of this model\cite{rankin.caqg,rankin.ijtp}, no longer linearizes equation (\ref{def.identity.q}) into the ``wave equation'' without restricting 
$ B $ back into the complex plane. Instead, 
\begin{equation}
B = \chi^{-1} \psi^{-1}
\label{linear.change.quat}
\end{equation}
where 
$ \chi $ is defined via the first order, partial differential equation 
\begin{equation}
\chi_{, \mu} \chi^{-1} = \psi^{-1} \psi_{, \mu}
\label{chi.psi.relate}
\end{equation}
and both $ \chi $ and $ \psi $ contain no leap-over operations, or none worse than 
$ |Q_{3} $.

Given that for a general scalar quaternion 
$ A $,
\begin{equation}
\left ( A^{-1} \right )_{, \mu} = 
- A^{-1} A_{, \mu} A^{-1}
\label{def.ainverse.partial}
\end{equation}
then equations (\ref{linear.change.quat}) and (\ref{chi.psi.relate}) give
\begin{equation}
- {\textstyle{1 \over 2}}\, 
B^{-1} B_{, \mu} = 
\psi_{, \mu} \psi^{-1}
\label{linerizing.quat.to.psi}
\end{equation}
This will accomplish the same cancellation of nonlinear terms in 
$ \psi $ in equation (\ref{def.identity.q}) that 
$ B = \psi^{-2} $ accomplished in equation (\ref{def.identity}) in the complex plane. Specifically, in a gauge in which 
$ v_{\mu} = - q A_{\mu} |Q_{3} $ is true, equation (\ref{def.qvhat.normalgauge}) gives
\begin{equation}
\hat v_{\mu} = - q A_{\mu} |Q_{3} + \psi_{, \mu} \psi^{-1}
\label{def.qvhat.psi}
\end{equation}
and equation (\ref{def.identity.q}) becomes
{\samepage 
\begin{eqnarray}
& \left ( 1 / \sqrt {-\hat g} \, \right ) & \left ( 
\sqrt {-\hat g} \, \hat g^{\mu \nu} 
\psi_{, \nu} \right )_{, \mu} - 2 q \hat g^{\mu \nu} 
A_{\mu} \left ( |Q_{3} \right ) \psi_{, \nu} 
\nonumber \\
&  & - q \left ( 1 / \sqrt {-\hat g}\, \right ) \left ( 
\sqrt { - \hat g } \, \hat g^{\mu \nu} A_{\nu} \right )_{, \mu} 
\left ( |Q_{3} \right ) \psi - q^{2} \hat g^{\mu \nu} 
A_{\mu} A_{\nu} \psi 
\nonumber \\
&  & + {\textstyle{1 \over 6}}\, \hat R \psi 
+ {\textstyle{1 \over 6}}\, \hat w_{\mu \nu} 
\hat a^{\mu \nu} \psi = \left ( C / 6 \right ) 
\left ( 1 + {\textstyle{1 \over 4}} \, \hat a \right ) \psi
\label{geometric.wave.eq.q}
\end{eqnarray}}%
where equations (\ref{def.w.hat.tail}) and (\ref{linerizing.quat.to.psi}) now give
\begin{equation}
\hat w_{\mu \nu} = - \left ( 1 + n \right ) q F_{\mu \nu} |Q_{3} - 
\left [ \left ( 2 - 4 n \right ) / \left ( 1 - n \right ) \right ] 
\left [ \psi_{, \mu} \psi^{-1} , \psi_{, \nu} \psi^{-1} \right ]
\label{def.w.hat.tail.psi}
\end{equation}
This last equation in turn implies
\begin{equation}
\hat w_{\mu \nu} \hat a^{\mu \nu} = - \left ( 1 + n \right ) q \, 
\hat a^{\mu \nu} F_{\mu \nu} |Q_{3} - 
\left [ \left ( 2 - 4 n \right ) / \left ( 1 - n \right ) \right ] 
\left [ \psi_{, \mu} \psi^{-1} , \psi_{, \nu} \psi^{-1} \right ] 
\hat a^{\mu \nu} 
\label{def.w.hat.tail.psi.times.ahat}
\end{equation}
Now recall that it is identically true that 
$ \hat w_{\mu \nu} \hat a^{\mu \nu} = (1/2) \hat W_{\mu \nu} \hat a^{\mu \nu} $
where
\begin{equation}
\hat W_{\mu \nu} = \hat w_{\mu \nu} - {}^{*} \hat w_{\mu \nu} | Q_{3} 
\label{big.w.hat}
\end{equation}
Then for the purposes of ordinary atomic physics, the 
$ \hat R $ term in equation (\ref{geometric.wave.eq.q}) would be ignored, and equations (\ref{geometric.wave.eq.q}) and (\ref{def.w.hat.tail.psi.times.ahat}) do seem to have the form that might reduce in Lorentzian coordinates to the second order Dirac Equation form of equation (\ref{leap.qdirac.wave.eq}), except that the spin term is generalized unless 
$ n = 1 / 2 $.

Now a straightforward analogy to the value of 
$ \hat a^{\mu \nu} $ obtained in reference \cite{rankin.caqg} would suggest that here, the expected zero order approximation or asymptotic behavior for 
$ \hat a^{\mu \nu} $ would be
\begin{equation}
\hat a^{\mu \nu} = K \left ( \sqrt{ \hat W } \right )^{-1} 
\hat W^{\mu \nu} 
\label{tentative.ahat}
\end{equation}
where
$ \hat W_{\mu \nu} $ is defined in equation (\ref{big.w.hat}), and
\begin{equation}
\hat W = \hat W^{\mu \nu} \hat W_{\mu \nu} 
\label{big.w.hat.mag}
\end{equation}
But those results all assumed 
$ \hat w_{\mu \nu} $ itself is in the complex plane. For 
$ n = 1 / 2 $, that will still be true here. Otherwise, the more general quaternionic nature of 
$ \hat W_{\mu \nu} $ in forms like equation (\ref{tentative.ahat}) seems not even to fit an 
$ \hat h_{\mu \nu} $ that's a real antisymmetric tensor to within some scalar quaternionic multiplying factor whose square is still in the complex plane, with 
$ |Q_{3} $ as the unit imaginary. New constraints would then appear to be needed on 
$ \hat h_{\mu \nu} $ to keep it inside the assumed model of this appendix. While the integrability conditions for equation (\ref{chi.psi.relate}) will be found to eliminate the quaternionic tail in 
$ \hat w_{\mu \nu} $ for solutions for 
$ \psi $ and 
$ \chi $, even when 
$ n \ne 1 / 2 $, the case 
$ n = 1 / 2 $ still gives the simplest expression for 
$ \hat w_{\mu \nu} $ in equation (\ref{def.w.hat.tail}).

Moreover when 
$ n = 1 / 2 $ and 
$ \hat w_{\mu \nu} $ is in the complex plane, equation (\ref{tentative.ahat}) does produce an 
$ \hat h_{\mu \nu} $ that is easily shown to satisfy the conditions of this model. Furthermore, if equation (\ref{tentative.ahat}) is rewritten as
\begin{equation}
\hat a^{\mu \nu} = K \sqrt{ \hat W } \left ( \hat W 
\right )^{-1} \hat W^{\mu \nu} 
\label{tentative.ahat.rationalized}
\end{equation}
then the square root of the complex quantity 
$ \hat W $ can be relaxed to allow roots that are fully quaternionic, but that also square to a complex number with 
$ |Q_{3} $ as the unit imaginary. That square root automatically becomes the quaternionic scalar multiplying an otherwise real 
$ \hat h_{\mu \nu} $. Then everything fits into the model of this appendix without further ado. On top of that, the nonlinearities in 
$ \psi $ in equation (\ref{def.w.hat.tail.psi.times.ahat}) never threaten to arise, and if 
$ K = -4 |Q_{3} $ instead of the old value of 
$ K = -6 |Q_{3} $, then equation (\ref{geometric.wave.eq.q}) should reduce exactly to the quaternionic, second order Dirac Equation form of equation (\ref{leap.qdirac.wave.eq}) (with 
$ \psi $ treated as a quaternionic, spacetime scalar) in the Lorentzian limit of the symmetric part of the metric, 
$ \hat g_{\mu \nu} $. Clearly the case 
$ n = 1 / 2 $ produces a major simplification of the results. In fact, one more simplification worth noting is that for 
$ n = 1 / 2 $, and only for this value of 
$ n $, 
$ \hat B^{\gamma}_{\gamma \tau \sigma} $ given by equation (\ref{def.expanded.alt.contract.bhat}) is proportional to 
$ \hat w_{\tau \sigma} $ given by equation (\ref{def.w.hat}). That proportionality is a property that is true in Weyl's original theory\cite{weyl.stm}, so the case 
$ n = 1 / 2 $ is the only case that matches that property of Weyl's original theory (which had 
$ n = 1 $, a value not allowed in this model).

The cost of all these simplifications introduced by choosing 
$ n = 1 / 2 $, is that the structure now has an equal mix of Weyl's nonmetricity\cite{weyl.stm,eddington.mtr} with the torsion of reference \cite{rankin.caqg}, rather than insisting on just nonmetricity or torsion alone. That may seem unusual for a model in which the non-Riemannian behavior is primarily based on a Weyl-like four vector. Nevertheless, it does achieve a notable reduction in the complexity of the results, although as already noted, the integrability conditions for equation (\ref{chi.psi.relate}) will be found to enforce many of the same simplifications even for 
$ n \ne 1 / 2 $. Even so, it's interesting to see that quaternionic curvatures in this model seem not only to reject the quaternionic generalization of the pure Weyl model, as noted after equation (\ref{def.b.neat.k.0}), but that they also preferentially select this case with an equal balance of torsion and nonmetricity. That preference is expressed by the overall simplicity of this case, and its correspondence with the form of a well established equation of physics, the (second order) Dirac Equation. No such preferential selection between torsion and nonmetricity appears with purely complex gauges and curvatures. Nevertheless, it is admittedly counterintuitive that a structure with some nonmetricity would also produce the second order Dirac Equation, which produces sharp spectral lines.

However, note that it is {\it also} true that effective nonmetricity may not vanish even when 
$ n = 0 $, even though the full affine derivative of 
$ \hat g_{\mu \nu} $ using connection 
$ \hat \Gamma^{\alpha}_{\mu \nu} $ vanishes then, implying metric compatibility. This effective nonmetricity can be seen by looking at equation (\ref{def.expanded.alt.contract.bhat}) in the 
$ n = 0 $ case, and noting that the change in length of a vector around a closed loop involves this quantity\cite{eddington.mtr}. This may still be nonzero even in the 
$ n = 0 $ case here, because 
$ \hat v_{\mu} $ is fully quaternionic. Thus, it appears that there may be no quaternionic models in this family which are completely devoid of all aspects of nonmetricity. This result appears to follow from the fact that covariant and contravariant vectors interact with the affine connection on opposite sides of the (quaternionic) connection, and the length of a vector is a contraction of a covariant and a contravariant vector. To put this another way, the affine derivative using the full 
$ \hat \Gamma^{\alpha}_{\mu \nu} $, no longer obeys the product rule of differentiation because 
$ \hat \Gamma^{\alpha}_{\mu \nu} $ is quaternionic, not real. Thus, the calculations of Weyl and Eddington\cite{weyl.stm,eddington.mtr} which would give the change in a parallel transported vector's length around a closed loop, would no longer be completely valid.

It should also be noted that the change in the value of 
$ K $ to a new value just 
$ 2 / 3 $ of its value of 
$ 6 i $ in reference \cite{rankin.caqg}, also accompanies a change in 
$ b_{0} $ in which it will equal its former value (see equation (\ref{b0.C})) times 
$ 8 / 3 $. This will affect the scales discussed in section \ref{effects} just the same way that a value of 
$ C = 3 / 8 $ would affect the scales (quantified in subsection \ref{rescale}). The scale changes involved are very small.

In any event, 
$ \psi $ no longer gives a complete description of the curvature, 
$ B $. The auxiliary wavefunction 
$ \chi $ defined via equation (\ref{chi.psi.relate}) is now also required. Given the quaternions 
$ \chi $ and $ \psi $ in polar form as a magnitude, angle, and a unit vector with some direction in quaternion space, then
\begin{equation}
\chi = \rho \left [ \cos { \phi } + \left ( \vec Q 
{ \bf \cdot } \hat m \right ) \sin { \phi } 
\right ]
\label{polar.chi}
\end{equation}
and
\begin{equation}
\psi = r \left [ \cos { \theta } + \left ( \vec Q 
{ \bf \cdot } \hat n \right ) \sin { \theta } 
\right ]
\label{polar.psi}
\end{equation}
Then equation (\ref{chi.psi.relate}) gives as its real and imaginary parts that
\begin{equation}
\rho_{, \mu} / \rho = r_{, \mu} / r
\label{real.chi.psi.relate}
\end{equation}
and
{\samepage 
\begin{eqnarray}
\hat m \phi_{, \mu} + \hat m_{, \mu} 
\sin { \phi } \cos { \phi } - \left ( \hat m_{, \mu} 
\times \hat m \right ) \sin^2 { \phi } = 
\nonumber \\
\hat n \theta_{, \mu} + \hat n_{, \mu} 
\sin { \theta } \cos { \theta } + \left ( \hat n_{, \mu} 
\times \hat n \right ) \sin^2 { \theta } 
\label{imaginary.chi.psi.relate}
\end{eqnarray}}%
where $ \hat m { \bf \cdot } \hat m = 1 $ and 
$ \hat n { \bf \cdot } \hat n = 1 $.

Equation (\ref{real.chi.psi.relate}) gives that the magnitudes 
$ \rho $ and $ r $ are proportional, and the case 
$ \rho = r $ can be chosen without loss of generality. Thus 
$ \chi $ and $ \psi $ have the same absolute magnitude. On the other hand, equation (\ref{imaginary.chi.psi.relate}) has no obvious solution of such simplicity. Nevertheless, if 
$ \hat m_{, \mu} = 0 = \hat n_{, \mu} $, and 
$ \hat m = \hat n $, that is the case in which both 
$ \chi $ and $ \psi $ are in the same complex plane. Then equation (\ref{imaginary.chi.psi.relate}) gives 
$ \phi_{, \mu} = \theta_{, \mu} $, or 
$ \phi = \theta + \alpha $, where 
$ \alpha $ is a constant angle. But then 
$ \alpha $ simply produces another proportionality constant between 
$ \chi $ and $ \psi $ which can be chosen as unity, or equivalently, 
$ \alpha $ can be chosen as zero. Thus in the complex plane, 
$ \chi = \psi $, and equation (\ref{linear.change.quat}) correctly reduces to the result used in the complex plane, 
$ B = \psi^{-2} $, as it should.

Alternatively, the relation between 
$ \chi $ and $ \psi $ can be examined using their cartesian forms in which
\begin{equation}
\chi = \chi_{0} + \vec Q { \bf \cdot } \vec \chi
\label{cart.chi}
\end{equation}
and
\begin{equation}
\psi = \psi_{0} + \vec Q { \bf \cdot } \vec \psi
\label{cart.psi}
\end{equation}
where $ \chi_{0} $ and $ \psi_{0} $ are the real parts of these quaternions, and 
$ \vec \chi $ and $ \vec \psi $ are their imaginary components. Then equation (\ref{chi.psi.relate}) gives
\begin{equation}
\chi_{0}^2 + \vec \chi { \bf \cdot } \vec \chi = 
\psi_{0}^2 + \vec \psi { \bf \cdot } \vec \psi
\label{real.chi.psi.relate.cart.result}
\end{equation}
and
{\samepage 
\begin{eqnarray}
\chi_{0} \vec \chi_{, \mu} - \chi_{0 , \mu} \vec \chi 
- \left ( \vec \chi_{, \mu} \times \vec \chi \right ) = 
\nonumber \\
\psi_{0} \vec \psi_{, \mu} - \psi_{0 , \mu} \vec \psi 
+ \left ( \vec \psi_{, \mu} \times \vec \psi \right )
\label{imaginary.chi.psi.relate.cart.simple}
\end{eqnarray}}%
as results that are equivalent to the polar form relations above. If a solution is found to equation (\ref{imaginary.chi.psi.relate.cart.simple}), it must also be checked to verify that it satisfies equation (\ref{real.chi.psi.relate.cart.result}) in order to qualify as a true solution to equation (\ref{chi.psi.relate}).

\subsection{Flat Space `` Free Particle'' Solutions}

In the ``free particle'' (vanishing electromagnetic potentials and fields), Lorentzian limit of the symmetric part of the metric, the resulting equations  (\ref{leap.qdirac.wave.eq}) and (\ref{chi.psi.relate}) have solutions of the form
{\samepage 
\begin{eqnarray}
\psi & = & A \left [ Q_{2} \sin { \left ( k x - \omega t + 
\alpha \right ) } 
+ Q_{3} \cos { \left ( k x - \omega t + \alpha \right ) } 
\right ] 
\nonumber \\
& = & Q_{3} \left [ A e^{ Q_{1} \left ( k x - \omega t + 
\alpha \right ) } \right ] = \left [ A e^{ - Q_{1} 
\left ( k x - \omega t + 
\alpha \right ) } \right ] Q_{3}
\label{psi.free}
\end{eqnarray}}%
and
{\samepage 
\begin{eqnarray}
\chi & = & A \left [ - Q_{2} \sin { \left ( k x - \omega t + 
\delta \right ) } 
+ Q_{3} \cos { \left ( k x - \omega t + \delta \right ) } 
\right ] 
\nonumber \\
& = & Q_{3} \left [ A e^{ - Q_{1} \left ( k x - \omega t + 
\delta \right ) } \right ] = \left [ A e^{ Q_{1} 
\left ( k x - \omega t + 
\delta \right ) } \right ] Q_{3}
\label{chi.free}
\end{eqnarray}}%
where 
$ A $, $ k $, $ \omega $, $ \alpha $, and $ \delta $ are all constants, and
\begin{equation}
k^{2} = \left ( \omega^{2} / c^{2} \right ) - 
\left [ \left ( m_{0}^{2} c^{2} \right ) / \hbar^{2} \right ]
\label{k2}
\end{equation}
The function 
$ \psi $ is a right circularly polarized plane wave (viewed from the ``source'') in the 
$ + x $ direction, while 
$ \chi $ is a left circularly polarized plane wave in the same direction. Using these, equation (\ref{linear.change.quat}) then gives
\begin{equation}
B = - A^{-2} \left [ \cos { \left ( 2 k x - 2 \omega t + 
\alpha + \delta \right ) } 
+ Q_{1} \sin { \left ( 2 k x - 2 \omega t + \alpha + \delta 
\right ) } \right ]
\label{B.free}
\end{equation}
If $ \psi $ and $ \chi $ are interchanged, the result is still a solution overall, but the sign of the 
$ Q_{1} $ term in $ B $ reverses, suggesting that these two alternative configurations may correspond to spin ``parallel'', or ``antiparallel'' to the corresponding ``free particle'' direction of motion (in the 
$ Q_{1} $ direction). However, the arguments of the cosines and sines could just as well have started with 
$ k z - \omega t $ instead of 
$ k x - \omega t $, giving motion in the 
$ Q_{3} $ direction, so that interpretation can't be quite correct. What the two alternative configurations might represent, however, are spin ``up'' or ``down'' along the 
$ Q_1 $ axis, which is the direction normal to the plane containing 
$ \chi $ and 
$ \psi $. If that is the case, then the spin axis would be directly visualized as the geometric axis normal to the plane containing 
$ \chi $ and 
$ \psi $ in quaternion space, at least when 
$ \chi $ and 
$ \psi $ are totally imaginary.

If 
$ \psi $ and $ \chi $ are written in matrix form, the results are
\begin{equation}
\psi = A \left (
\begin{array}{cc}
- \imath \cos{(kx - \omega t + \alpha)} & 
- \sin{(kx - \omega t + \alpha)} \\
\sin{(kx - \omega t + \alpha)} & 
\imath \cos{(kx - \omega t + \alpha)} 
\end{array}
\right )
\label{psi.free.matrix}
\end{equation}
and
\begin{equation}
\chi = A \left (
\begin{array}{cc}
- \imath \cos{(kx - \omega t + \delta)} & 
\sin{(kx - \omega t + \delta)} \\
- \sin{(kx - \omega t + \delta)} & 
\imath \cos{(kx - \omega t + \delta)} 
\end{array}
\right )
\label{chi.free.matrix}
\end{equation}
In both cases, the first column is the equivalent spinor solution.

Note that
\begin{equation}
\psi_{, 0} = \left [ \left ( \omega / c \right ) 
Q_{1} \right ] \psi 
= \psi \left [ - \left ( \omega / c \right ) Q_{1} 
\right ]
\label{psi.comma.0}
\end{equation}
and
\begin{equation}
\chi_{, 0} = \left [ - \left ( \omega / c \right ) 
Q_{1} \right ] \chi
= \chi \left [ \left ( \omega / c \right ) Q_{1} 
\right ]
\label{chi.comma.0}
\end{equation}
which indicate equal but opposite signed energy eigenvalues for these two solutions in this pair. Similar results hold for 
$ \psi_{, 1} $ and $ \chi_{, 1} $, except then 
$ \omega / c \rightarrow - k $ in these equation forms.

Solutions of the form of equations (\ref{psi.free}) and (\ref{chi.free}) can be superposed to generate new solutions, with the nth subsolution each containing its own separate 
$ A_{n} $, $ k_{n} $, $ \omega_{n} $, 
$ \alpha_{n} $, and $ \delta_{n} $, where
\begin{equation}
k_{n}^{2} = \left ( \omega_{n}^{2} / c^{2} \right ) - 
\left [ \left ( m_{0}^{2} c^{2} \right ) / \hbar^{2} \right ]
\label{k2.n}
\end{equation}
Then equations (\ref{real.chi.psi.relate.cart.result}) and (\ref{imaginary.chi.psi.relate.cart.simple}) are still satisfied for the sum of all the separate subsolutions if
\begin{equation}
\alpha_{n} - \alpha_{m} = \delta_{n} - \delta_{m}
\label{angle.condition}
\end{equation}
or equivalently,
\begin{equation}
\alpha_{n} - \delta_{n} = \alpha_{m} - \delta_{m}
\label{angle.condition.alt}
\end{equation}
This condition will be clarified below following equations (\ref{two.dim.chi.psi.2}) and (\ref{two.dim.chi.psi.3}). Other than this restriction, this ability to superpose an arbitrary number of such subsolutions with different amplitudes, frequencies and initial phase angles, should allow fairly general wave packet solutions to be built up in this ``free particle'' case. This is true even though equation (\ref{chi.psi.relate}) appears at first sight to be nonlinear. In fact, for these cases, it behaves much like a linear equation would be expected to behave.

Indeed, any pair 
$ \chi $ and $ \psi $ satisfying equation (\ref{chi.psi.relate}), and in which both 
$ \chi $ and $ \psi $ are restricted to having only components in the same two dimensional plane in the quaternions containing also the origin, should generally support superposition of similar solutions in that same plane to form more general solutions. Such pairs will always either dwell in the family of solutions in some complex plane, where the general solution to equation (\ref{chi.psi.relate}) is simply 
$ \chi = \psi $, as noted after equation (\ref{imaginary.chi.psi.relate}), or they will be in some plane in the totally imaginary portion of quaternion space. Of course, there are an infinite number of ways to take a complex plane in the quaternions, and solutions in different complex planes are not covered by this consideration. As for those in a plane that is totally imaginary, they can always be manipulated (rotating the three 
$ Q $ axes if necessary) into the forms
\begin{equation}
\chi = Q_{2} \chi_{2} + Q_{3} \chi_{3} = 
Q_{3} \left ( \chi_{3} + Q_{1} \chi_{2} \right ) = 
\left ( \chi_{3} - Q_{1} \chi_{2} \right ) Q_{3}
\label{two.dim.chi}
\end{equation}
and
\begin{equation}
\psi = Q_{2} \psi_{2} + Q_{3} \psi_{3} = 
Q_{3} \left ( \psi_{3} + Q_{1} \psi_{2} \right ) = 
\left ( \psi_{3} - Q_{1} \psi_{2} \right ) Q_{3}
\label{two.dim.psi}
\end{equation}
Then equation (\ref{real.chi.psi.relate.cart.result}) becomes
\begin{equation}
\chi_{2}^{2} + \chi_{3}^{2} = \psi_{2}^{2} + \psi_{3}^{2}
\label{two.dim.real.part}
\end{equation}
and equation (\ref{chi.psi.relate}) becomes (with the 
``$ {}^{\dagger} $'' to the left)
{\samepage 
\begin{eqnarray}
\chi_{, \mu} \, {}^{\dagger} \chi & = & 
{}^{\dagger} \psi \psi_{, \mu} 
\nonumber \\
& = & \left ( \chi_{3} - Q_{1} \chi_{2} \right )_{, \mu} 
Q_{3} \left ( -Q_{3} \right ) \left ( \chi_{3} + 
Q_{1} \chi_{2} \right ) 
\nonumber \\
& = & \left ( \psi_{3} - Q_{1} \psi_{2} \right ) 
\left ( -Q_{3} \right ) Q_{3} \left ( \psi_{3} + 
Q_{1} \psi_{2} \right )_{, \mu} 
\nonumber \\
& = & \left ( \chi_{3} - Q_{1} \chi_{2} \right )_{, \mu} 
\left ( \chi_{3} + Q_{1} \chi_{2} \right ) 
\nonumber \\
& = & \left ( \psi_{3} - Q_{1} \psi_{2} \right ) 
\left ( \psi_{3} + Q_{1} \psi_{2} \right )_{, \mu} 
\nonumber \\
& = & \left ( \psi_{3} + Q_{1} \psi_{2} \right )_{, \mu} 
\left ( \psi_{3} - Q_{1} \psi_{2} \right ) 
\label{two.dim.chi.psi.relate}
\end{eqnarray}}%
where the last step follows from the fact that all the quantities on the last three lines commute with each other. But direct multiplication on the right side of the following equation gives
{\samepage 
\begin{eqnarray}
\left ( \psi_{3} + Q_{1} \psi_{2} \right )_{, \mu} 
\left ( \psi_{3} - Q_{1} \psi_{2} \right ) & = & 
\left \{ \left [ \psi_{3} \cos{ \eta } - 
\psi_{2} \sin{ \eta } \right ] + Q_{1} \left [ 
\psi_{2} \cos{ \eta } + \psi_{3} \sin{ \eta } 
\right ] \right \}_{, \mu} 
\left \{ \left [ \psi_{3} \cos{ \eta } \right. 
\right. 
\nonumber \\
& & \left. \left. - \psi_{2} \sin{ \eta } 
\right ] - Q_{1} \left [ 
\psi_{2} \cos{ \eta } + \psi_{3} \sin{ \eta } 
\right ] \right \} 
\label{two.dim.identity}
\end{eqnarray}}%
where $ \eta $ is a constant angle. But these last three equations have the obvious general solution
\begin{equation}
\chi_{2} = - \psi_{2} \cos{ \eta } - \psi_{3} \sin{ \eta } 
\label{two.dim.chi.psi.2}
\end{equation}
and
\begin{equation}
\chi_{3} = \psi_{3} \cos{ \eta } - \psi_{2} \sin{ \eta }
\label{two.dim.chi.psi.3}
\end{equation}
Clearly $ \eta $ is the value of 
$ \delta_{n} - \alpha_{n} $ in the ``free particle'' solutions examined above. Geometrically, 
$ \chi $ is the reflection of 
$ \psi $ through a plane which contains the 
$ x $ axis, and which makes an angle 
$ \eta / 2 $ with respect to the 
$ z $ axis, the angle being measured counterclockwise from the 
$ z $ axis as viewed from 
$ +x $ looking toward 
$ x = 0 $.

Now equations (\ref{two.dim.chi.psi.2}) and (\ref{two.dim.chi.psi.3}) are simple linear equations for any fixed value of 
$ \eta $, so solutions for a given 
$ \eta $ can be linearly superposed to give new solutions. This family of solutions is in addition to those that are the complex plane type noted above. Thus there are two families of two component solutions to equation (\ref{chi.psi.relate}), the complex plane cases with 
$ \chi = \psi $, or the purely imaginary alternative cases that lead to the family of solutions given by equations (\ref{two.dim.chi.psi.2}) and (\ref{two.dim.chi.psi.3}). These two families seem to exhaust the cases in a two dimensional plane also containing the origin in the quaternions, and all these forms support linear superposition of reasonably similar solutions to give new solutions. Furthermore, solutions in the same complex plane always commute with each other, while those in the same imaginary plane have more subtle commutation properties, as equation (\ref{B.free}) illustrates when the sign of its 
$ Q_{1} $ term reverses as 
$ \chi $ and $ \psi $ are interchanged.

The above examples should be contrasted with the standard solutions of the Dirac Equation for a free particle at rest\cite{drell}. In the standard Dirac representation in which the gamma matrices are
\begin{equation}
\gamma^{0}_{S} =
\left (
\begin{array}{cc}
\imath \sigma_{0} & 0 \\
0 & - \imath \sigma_{0} 
\end{array}
\right ) , 
\gamma^{k}_{S} =
\left (
\begin{array}{cc}
0 & \imath \sigma_{k} \\
- \imath \sigma_{k} & 0 
\end{array}
\right )
\label{def.gammas.std}
\end{equation}
for $ k = 1,2,3 $, those are
\begin{equation}
\psi_{S} = 
e^{ - \imath \omega t } \left [
\begin{array}{r}
1 \\
0 \\
0 \\
0
\end{array}
\right ] , 
\psi_{S} = 
e^{ - \imath \omega t } \left [
\begin{array}{r}
0 \\
1 \\
0 \\
0
\end{array}
\right ] , 
\psi_{S} = 
e^{ \imath \omega t } \left [
\begin{array}{r}
0 \\
0 \\
1 \\
0
\end{array}
\right ] , 
\psi_{S} = 
e^{ \imath \omega t } \left [
\begin{array}{r}
0 \\
0 \\
0 \\
1
\end{array}
\right ]
\label{bd.free.rest}
\end{equation}
Those must now be transformed to the chiral representation with the gammas given by equations (\ref{def.gamma0}) and (\ref{def.gammak}) (ignore the quaternion equivalent forms here) giving,
\begin{equation}
\psi = 
{ 1 \over \sqrt{ 2 } } \, 
e^{ - \imath \omega t } \left [
\begin{array}{r}
1 \\
0 \\
- 1 \\
0
\end{array}
\right ] , 
\psi = 
{ 1 \over \sqrt{ 2 } } \, 
e^{ - \imath \omega t } \left [
\begin{array}{r}
0 \\
1 \\
0 \\
- 1
\end{array}
\right ] , 
\psi = 
{ 1 \over \sqrt{ 2 } } \, 
e^{ \imath \omega t } \left [
\begin{array}{r}
1 \\
0 \\
1 \\
0
\end{array}
\right ] , 
\psi = 
{ 1 \over \sqrt{ 2 } } \, 
e^{ \imath \omega t } \left [
\begin{array}{r}
0 \\
1 \\
0 \\
1
\end{array}
\right ]
\label{bd.free.rest.chiral}
\end{equation}
In the chiral representation, 
$ \psi $ is subdivided into an upper and lower two spinor, labelled here as 
$ \xi $ and 
$ \zeta $. For the above,
\begin{equation}
\xi = 
{ 1 \over \sqrt{ 2 } } \, 
e^{ - \imath \omega t } \left [
\begin{array}{r}
1 \\
0
\end{array}
\right ] , 
\xi = 
{ 1 \over \sqrt{ 2 } } \, 
e^{ - \imath \omega t } \left [
\begin{array}{r}
0 \\
1
\end{array}
\right ] , 
\xi = 
{ 1 \over \sqrt{ 2 } } \, 
e^{ \imath \omega t } \left [
\begin{array}{r}
1 \\
0
\end{array}
\right ] , 
\xi = 
{ 1 \over \sqrt{ 2 } } \, 
e^{ \imath \omega t } \left [
\begin{array}{r}
0 \\
1
\end{array}
\right ]
\label{bd.free.rest.chiral.eta}
\end{equation}
The corresponding quaternionic forms of
$ \xi $ are
{\samepage 
\begin{eqnarray}
\xi & = & 
{ 1 \over \sqrt{ 2 } } \, 
\left [ 
Q_{0} \cos { \omega t } + Q_{3} \sin { \omega t } 
\right ] , 
\xi = 
{ 1 \over \sqrt{ 2 } } \, 
\left [ 
Q_{2} \cos { \omega t } + Q_{1} \sin { \omega t } 
\right ] , 
\nonumber \\
\xi & = & 
{ 1 \over \sqrt{ 2 } } \, 
\left [ 
Q_{0} \cos { \omega t } - Q_{3} \sin { \omega t } 
\right ] , 
\xi = 
{ 1 \over \sqrt{ 2 } } \, 
\left [ 
Q_{2} \cos { \omega t } - Q_{1} \sin { \omega t } 
\right ]
\label{bd.free.rest.chiral.eta.quat}
\end{eqnarray}}%
For
$ \zeta $, the only changes will be that the first two cases are the negatives of their corresponding values for 
$ \xi $.

It is clear that all these cases are two dimensional solutions of the types discussed above, and thus have immediate solutions for the auxiliary wavefunction 
$ \chi $. However, there is a remarkable asymmetry between the ``spin up'' and ``spin down'' solutions from the viewpoint of this paper. The two ``spin up'' cases are in the complex plane defined by 
$ Q_{0} $ and 
$ Q_{3} $, and thus have 
$ \chi = \xi $. However, the two ``spin down'' cases are totally imaginary, and thus have 
$ \chi $ equal to a reflection of 
$ \xi $ in some plane containing the 
$ Q_{3} $ axis. This major difference is not what one would reasonably expect for solutions differing only by their spin direction, at least from this paper's viewpoint. On the other hand, the values of 
$ \xi_{, 0} $ give
{\samepage 
\begin{eqnarray}
\xi_{, 0} & = & 
\left [ - { \omega \over c } \, 
\left ( - | Q_{3} \right ) \right ] \xi , \; 
\xi_{, 0} = 
\left [ - { \omega \over c } \, 
\left ( - | Q_{3} \right ) \right ] \xi , 
\nonumber \\
\xi_{, 0} & = & 
\left [ { \omega \over c } \, 
\left ( - | Q_{3} \right ) \right ] \xi , \; 
\xi_{, 0} = 
\left [ { \omega \over c } \, 
\left ( - | Q_{3} \right ) \right ] 
\xi
\label{bd.free.rest.chiral.eta.quat.eig}
\end{eqnarray}}%
Since we know 
$ \imath \rightarrow - | Q_{3} $, These (right) eigenvalues do match the standard eigenvalues for the four cases. However, this leads to another asymmetry between the ``spin up'' and ``spin down'' cases. The right eigenvalue (with no leap-over operator) commutes with 
$ \xi $ in the two ``spin up'' cases, but anticommutes in the two ``spin down'' cases. Thus, the ``spin up'' cases have only that single eigenvalue, while the two ``spin down'' cases also have an opposite signed left eigenvalue. This is again, oddly asymmetric from the viewpoint of this paper.

A ``free particle'' generalization of free spherical waves also is worth mentioning. Equations (\ref{psi.free}) and (\ref{chi.free}) have ``spherical'' wave matching solutions expressed as
{\samepage 
\begin{eqnarray}
\psi & = & { A \over r } \left [ Q_{2} \sin { \left ( k r - 
\omega t + \alpha \right ) } 
+ Q_{3} \cos { \left ( k r - \omega t + \alpha \right ) } 
\right ]
\nonumber \\
& = & Q_{3} \left [ { A \over r } \, e^{ Q_{1} \left ( k r - 
\omega t + \alpha \right ) } \right ] = \left [ { A \over r } 
\, e^{ - Q_{1} \left ( k r - \omega t + \alpha \right ) } 
\right ] Q_{3}
\label{psi.free.spherical}
\end{eqnarray}}%
and
{\samepage 
\begin{eqnarray}
\chi & = & { A \over r } \left [ - Q_{2} \sin { \left ( k r - 
\omega t + \delta \right ) } 
+ Q_{3} \cos { \left ( k r - \omega t + \delta \right ) } 
\right ]
\nonumber \\
& = & Q_{3} \left [ { A \over r } \, e^{ - Q_{1} \left ( k r - 
\omega t + \delta \right ) } \right ] = \left [ { A \over r } 
\, e^{ Q_{1} \left ( k r - \omega t + \delta \right ) } 
\right ] Q_{3}
\label{chi.free.spherical}
\end{eqnarray}}%
These are only spherically symmetric in a sense, because they are not completely spherically symmetric in terms of vectors in the spatial part of quaternion space, a limitation that seems natural enough for a solution with spin.

\subsection{Integrability Conditions}

Differentiating equation (\ref{chi.psi.relate}) and then using equation (\ref{chi.psi.relate}) gives
{\samepage 
\begin{eqnarray}
\chi_{, \mu , \nu} \chi^{-1} - \chi_{, \mu} \chi^{-1} 
\chi_{, \nu} \chi^{-1} 
& = & \chi_{, \mu , \nu} \chi^{-1} - \psi^{-1} \psi_{, \mu} 
\psi^{-1} \psi_{, \nu}
\nonumber \\
& = & \psi^{-1} \psi_{, \mu , \nu} - \psi^{-1} \psi_{, \nu} 
\psi^{-1} \psi_{, \mu}
\label{chi.psi.relate.second}
\end{eqnarray}}%
Then since 
$ \hat g^{\mu \nu} $ is symmetric in its tensor indices,
\begin{equation}
\hat g^{\mu \nu} \chi_{, \mu , \nu} \chi^{-1} = 
\hat g^{\mu \nu} \psi^{-1} \psi_{, \mu , \nu}
\label{chi.psi.relate.second.contr}
\end{equation}
Now if equation (\ref{geometric.wave.eq.q}) is multiplied from the left by 
$ \psi^{-1} $, then equations (\ref{chi.psi.relate}) and (\ref{chi.psi.relate.second.contr}) allow the result eventually to be written as
{\samepage 
\begin{eqnarray}
& \left ( 1 / \sqrt {-\hat g} \, \right ) & \left ( 
\sqrt {-\hat g} \, \hat g^{\mu \nu} 
\chi_{, \nu} \right )_{, \mu} \chi^{-1} - 
2 q \hat g^{\mu \nu} A_{\mu} \chi_{, \nu} \chi^{-1} 
Q_{3} 
\nonumber \\
&  & - q \left ( 1 / \sqrt {-\hat g}\, \right ) \left ( 
\sqrt { - \hat g } \, \hat g^{\mu \nu} A_{\nu} \right )_{, \mu} 
Q_{3} - q^{2} \hat g^{\mu \nu} A_{\mu} A_{\nu} 
\nonumber \\
&  & + {\textstyle{1 \over 6}}\, \hat R 
+ {\textstyle{1 \over 6}}\, \psi^{-1} 
\hat w_{\mu \nu} \hat a^{\mu \nu} \psi 
= \left ( C / 6 \right ) \left ( 1 + {\textstyle{1 \over 4}} 
\, \hat a \right )
\label{geometric.wave.eq.q.chi}
\end{eqnarray}}%
Unfortunately, this cannot simply be multiplied from the right by 
$ \chi $ to get an equation for 
$ \chi $ resembling equation (\ref{geometric.wave.eq.q}) for 
$ \psi $, because 
$ \psi $ has already been evaluated, and thus the rightmost 
$ Q_{3} $ is locked in place, and cannot be moved to the right of the trailing 
$ \chi $. However, that objection will not apply when the 
$ A_{\mu} $ vanish, assuming the spin term also vanishes then, since then right multiplication by 
$ \chi $ gives a meaningful, familiar result.

Now because the integrability of equation (\ref{chi.psi.relate}) requires that both 
$ \chi_{, \mu , \nu} = \chi_{, \nu , \mu} $ and 
$ \psi_{, \mu , \nu} = \psi_{, \nu , \mu} $, it is possible to make a stronger statement than equation (\ref{chi.psi.relate.second.contr}). Reverse the indices in equation (\ref{chi.psi.relate.second}), and subtract that from the original equation. The result gives that
\begin{equation}
\psi^{-1} \psi_{, \mu} \psi^{-1} \psi_{, \nu} = 
\psi^{-1} \psi_{, \nu} \psi^{-1} \psi_{, \mu}
\label{integrability.conds}
\end{equation}
must be true in order for equation (\ref{chi.psi.relate}) to have valid solutions. Then that in turn gives
\begin{equation}
\chi_{, \mu , \nu} \chi^{-1} = 
\psi^{-1} \psi_{, \mu , \nu}
\label{chi.psi.relate.second.free}
\end{equation}
from equation (\ref{chi.psi.relate.second}) for those valid solutions. Furthermore, equation (\ref{integrability.conds}) imposes conditions that 
$ \psi $ must satisfy in order for 
$ \chi $ to exist, and for them both to give 
$ B $ via equation (\ref{linear.change.quat}). Those conditions easily simplify to (with the 
``$ {}^{\dagger} $'' to the left)
\begin{equation}
\psi_{, \mu} {}^{\dagger} \psi \psi_{, \nu} = 
\psi_{, \nu} {}^{\dagger} \psi \psi_{, \mu}
\label{integrability.conds.short}
\end{equation}
These conditions are clearly satisfied if 
$ \psi $ is restricted to a complex plane in the quaternions, and it is easy to verify that any 
$ \psi $ of the form in equation (\ref{two.dim.psi}) also satisfies equation (\ref{integrability.conds.short}). Thus, the two dimensional solutions satisfy equation (\ref{integrability.conds.short}) generally. Furthermore, if 
$ \psi_{, \mu} = \lambda_{\mu} \psi $ and 
$ \lambda_{\mu} \lambda_{\nu} = \lambda_{\nu} \lambda_{\mu} $ (left eigenvalues commute), or if 
$ \psi_{, \mu} = \psi \beta_{\mu} $ and 
$ \beta_{\mu} \beta_{\nu} = \beta_{\nu} \beta_{\mu} $ (right eigenvalues commute), then the integrability conditions specified by equation 
(\ref{integrability.conds.short}) are satisfied.

Note that if equation (\ref{integrability.conds}) is true, then it is also true that
\begin{equation}
\psi_{, \mu} \psi^{-1} \psi_{, \nu} \psi^{-1} = 
\psi_{, \nu} \psi^{-1} \psi_{, \mu} \psi^{-1}
\label{integrability.conds.alt}
\end{equation}
But as already noted above, that condition will zero out the quaternionic tail in equation (\ref{def.w.hat.tail.psi}), even when 
$ n \ne 1 / 2 $. However, I do not at this time see that this would rule out a solution of equation (\ref{def.identity.q}) directly for a 
$ B $ that is not found using the form 
$ B = \chi^{-1} \psi^{-1} $, and their associated equations. If such a special solution for 
$ B $ exists, then equation (\ref{def.w.hat.tail}) might still contain a quaternionic tail unless 
$ n = 1 / 2 $. However, it is admittedly not clear at this time that the case 
$ n \ne 1 / 2 $ can be totally ruled out as unrealistic.

In terms of 
$ \psi_{0} $ and $ \vec \psi $, equation (\ref{integrability.conds.short}) is equivalent to
\begin{equation}
\vec \psi { \bf \cdot } \left ( \vec \psi_{, \mu} 
\times \vec \psi_{, \nu} \right ) = 0
\label{integrability.conds.scalar}
\end{equation}
and
\begin{equation}
\psi_{0 , \mu} \left ( \vec \psi \times \vec 
\psi_{, \nu} \right ) -  \psi_{0 , \nu} \left 
( \vec \psi \times \vec \psi_{, \mu} \right ) - 
\psi_{0} \left ( \vec \psi_{, \mu} 
\times \vec \psi_{, \nu} \right ) = 0
\label{integrability.conds.vector}
\end{equation}
If $ \psi_{0} = 0 $, only equation (\ref{integrability.conds.scalar} remains, and it is clearly satisfied by the type of two dimensional, totally imaginary solutions examined above. In fact, the case 
$ \psi_{0} = 0 $ can always be achieved by a simple phase/gauge transformation of the wavefunction and electromagnetic potentials.

As already noted in equation (\ref{qphase.trans}), in standard Dirac Theory it is customary to define the {\it spinor} wavefunction following a phase/gauge transformation\cite{bade.jehle} by
\begin{equation}
\bar \psi = \psi e^{\imath \left ( \phi / 2 \right ) }
\label{psi.phase.trans}
\end{equation}
The matching electromagnetic potentials are
\begin{equation}
\bar A_{\mu} = A_{\mu} - \left [ 1 / 
\left ( 2 q \right ) \right ] \phi_{, \mu}
\label{A.mu.phase.trans}
\end{equation}
As also noted in equation (\ref{qphase.trans}), this can be immediately generalized to the quaternion form of 
$ \psi $ via
\begin{equation}
\bar \psi = \psi 
e^{ - \left ( | Q_{3} \right ) \left ( \phi / 2 \right ) }
\label{qpsi.phase.trans}
\end{equation}
Now adopt the additional conventions
\begin{equation}
\bar \psi^{ - 1 } = \psi^{ - 1 } 
e^{ \left ( | Q_{3} \right ) \left ( \phi / 2 \right ) }
\label{qpsiinv.phase.trans}
\end{equation}
\begin{equation}
\bar \chi = \chi 
e^{ - \left ( | Q_{3} \right ) \left ( \phi / 2 \right ) }
\label{qchi.phase.trans}
\end{equation}
and
\begin{equation}
\bar \chi^{ \, - 1 } = \chi^{ - 1 } 
e^{ \left ( | Q_{3} \right ) \left ( \phi / 2 \right ) }
\label{qchiinv.phase.trans}
\end{equation}
These imply the additional transformations
\begin{equation}
\bar B = B 
e^{ \left ( | Q_{3} \right ) \phi }
\label{qB.phase.trans}
\end{equation}
and
\begin{equation}
\bar B^{ - 1 } = B^{ - 1 } 
e^{ - \left ( | Q_{3} \right ) \phi }
\label{qBinv.phase.trans}
\end{equation}
A quick check shows that these preserve the form of equation (\ref{chi.psi.relate}) for the transformed quantities, a necessary property.

Now using equation (\ref{def.spsi}), note
\begin{equation}
\psi_{0 R} + \imath \psi_{0 I} = 
\beta_{0} e^{ \imath \alpha_{0} }
\label{polar.psizero}
\end{equation}
where 
$ \beta_{0} = \sqrt{ \psi^{2}_{0 R} + \psi^{2}_{0 I} } \, $, and 
$ \alpha_{0} = \arctan{ \left ( \psi_{0 I} / \psi_{0 R} \right ) } $. Thus, for
\begin{equation}
\phi = - 2 \alpha_{0} + \pi
\label{cancelling.phase}
\end{equation}
$ \bar \psi_{ 0 R } = 0 $, and 
$ \bar \psi_{ 0 I } = \beta_{0} $. In this gauge, the quaternion component
$ \bar \psi_{0} = 0 $, and only the integrability condition of equation (\ref{integrability.conds.scalar}) survives. Since it is antisymmetric in the indices 
$ \mu $ and $ \nu $, it consists of six differential conditions.

Finally, if the imaginary three space part of quaternion space is mapped by orthogonal, curvilinear coordinates instead of the standard Cartesian coordinates, then 
$ Q_{\mu} \rightarrow Q^{'}_\mu $ as discussed originally in the discussion of equation (\ref{def.qvecprime}), and 
$ \psi \rightarrow \psi^{'} $. The vector portion of 
$ \psi^{'} $ is computed assuming unit basis vectors 
$ \hat e^{'} $ as noted in equation (\ref{def.qvecprime}), so the nth component of 
$ \vec \psi^{'} $, 
$ \psi^{'}_{n} = \left ( \vec \psi^{'} \right )_{n} $, is neither a covariant nor a contravariant three space vector, but is instead halfway between those quantities. More specifically, if the metric for the coordinates mapping the imaginary part of quaternion space is 
$ q^{'}_{m n} $, then the quantity 
$ \sqrt { \left | q^{'}_{n n} \right | } \psi^{'}_{n} $ (no sum on 
$ n $) is a covariant vector, and equations (\ref{integrability.conds.scalar}) and (\ref{integrability.conds.vector}) remain true for 
$ \psi^{'} $ provided
\begin{equation}
\psi_{\mu , \nu} \rightarrow \left ( 1 / \sqrt { \left 
| q^{'}_{\mu \mu} \right | } \right ) \left ( 
\sqrt { \left | q^{'}_{\mu \mu} \right | } 
\psi^{'}_{\mu} \right )_{ {\bf \bar{;}} \nu}
\label{psi.comma.mu.curvilinear}
\end{equation}
with no sum on 
$ \mu $. The 
``$ \; \bar ; \; $'' is the covariant derivative with respect to the quaternion space metric $ q^{'}_{\mu \nu} $, and for these purposes, 
$ q^{'}_{0 0} = 1 $, and 
$ q^{'}_{0 n} = q^{'}_{n 0} = 0 $, where 
$ n $ runs from 1 to 3. The absolute values are not actually needed here, but are used anyway as a reminder that they would be needed for a Lorentzian signature metric. The quaternion space metric 
$ q^{'}_{\mu \nu} $ is apparently the metric of an uncurved, flat Euclidian space. This three space corresponding to the totally imaginary part of the quaternion can be mapped by spherical coordinates, or any other standard, orthogonal, curvilinear coordinate system\cite{morse.feshbach}. All these points raise the issue of the covariance of such quantities in quaternion space when there are spacetime transformations, including Lorentz transformations.

\subsection{Covariance and Quaternion Space}

This subsection deals with the transformation properties of equation (\ref{def.identity.q}), and the equations derived from that. To keep this as simple as possible, it will be assumed that $ n = 1 / 2 $ in equation (\ref{def.w.hat.tail})), and that the zero order approximation 
$ \hat a = K^{2} $ holds for 
$ \hat a $.

At first sight, this appears to be a trivial issue, since equation (\ref{def.identity.q}) is manifestly generally covariant, and is even valid when spacetime is not flat. Furthermore, if the quaternion basis is the set of 
$ Q_{\mu} $, a set of mathematical scalar invariants just like 
$ \imath $ is in the complex numbers, then quaternionic quantities will be simple spacetime scalars just as the complex 
$ \psi $ was in reference \cite{rankin.caqg}. However, standard treatments of Dirac Theory\cite{bade.jehle,drell} consider 
$ \psi $ to be a spacetime spinor, with its own, nonscalar transformation properties. This springs from a demand that equation (\ref{leap.qdirac.wave.eq}) be invariant in form in that the relativistic three vector that makes up 
$ \vec Q $ in the expression 
$ \vec Q {\bf \cdot} \left ( \vec E - \vec B | Q_{3} \right ) $, must retain the same representation in terms of Pauli Matrices in all locally Lorentzian Frames\cite{bade.jehle}. However in the formalism of this paper, the spin term in equation (\ref{def.identity.q}) becomes 
$ {1 \over 2} \hat a^{\mu \nu} \hat Y_{\mu \nu} = {K \over 2} \sqrt{ \hat Y } $, where 
$ \hat Y = \hat Y^{\mu \nu} \hat Y_{\mu \nu} $, and 
$ \hat Y_{\mu \nu} = \hat y_{\mu \nu} - {}^{*} \hat y_{\mu \nu} | Q_{3} $. Thus the quantity 
$ \sqrt{ \hat Y } $ is manifestly a scalar (more precisely, a mixture of a scalar and a pseudoscalar). Therefore when the square root is extracted as in equation (\ref{spin.squared}) as the quaternionic quantity which is multiplied by itself on the left side of that equation, that will require the use of the 
$ \vec Q^{'} $ of equation (\ref{def.qvecprime}), where 
$ \vec Q^{'} $ must be a relativistic three vector, like the electromagnetic field quantities into which it is dotted. Otherwise, the square root will no longer have transformation properties that match the square root before the root is actually extracted. This use of a true relativistic three vector form of 
$ \vec Q^{'} $ produces a scalar wavefunction instead of the usual spinor wavefunction of standard Dirac Theory, just as the wave function of Usachev's version of Dirac Theory\cite{jtep.usachev} is a scalar rather than a spinor. A brief review of Dirac formalism will illustrate this.

Standard Dirac Theory in the chiral representation\cite{bade.jehle, sakurai} in either a spinor or real quaternion formulation uses the gamma matrices given in equations (\ref{def.gamma0}) and (\ref{def.gammak}). These can conveniently be rewritten by using 
$ p^{\mu} = \left [ - Q^{'}_{0} | Q_{3} , \vec Q^{'} \right ] $, and 
$ q^{\mu} = \left [ Q^{'}_{0} | Q_{3} , \vec Q^{'} \right ] $, where now
\begin{equation}
\gamma^{\mu} =
\left (
\begin{array}{cc}
0 & - p^{\mu} \\
q^{\mu} & 0 
\end{array} 
\right )
\label{def.gammamu}
\end{equation}
In a locally Lorentzian frame, these gammas equal the metric via
\begin{equation}
\hat \eta^{\mu \nu} = - \frac{1}{2} (\gamma^{\mu} \gamma^{\nu} 
+ \gamma^{\nu} \gamma^{\mu})
\label{def.metric.gammas}
\end{equation}
and this becomes
\begin{equation}
\hat \eta^{\mu \nu} =
{ 1 \over 2 } \, 
\left (
\begin{array}{cc}
p^{\mu} q^{\nu} + p^{\nu} q^{\mu} & 0 \\
0 & q^{\mu} p^{\nu} + q^{\nu} p^{\mu} 
\end{array} 
\right )
\label{def.metric.our}
\end{equation}
They also produce the antisymmetric spin tensor of equation (\ref{def.sigma}), which becomes
\begin{equation}
\Sigma^{\mu \nu} =
{ 1 \over 2 } \, 
\left (
\begin{array}{cc}
p^{\mu} q^{\nu} - p^{\nu} q^{\mu} & 0 \\
0 & q^{\mu} p^{\nu} - q^{\nu} p^{\mu} 
\end{array} 
\right )
\label{def.sigma.our}
\end{equation}
The quantity
\begin{equation}
{ 1 \over 2 } \, 
\left ( p^{\mu} q^{\nu} + p^{\nu} q^{\mu} 
\right ) = \hat \eta^{\mu \nu}
\label{def.pq.plus.pq}
\end{equation}
and the quantity
\begin{equation}
{ 1 \over 2 } \, 
\left ( p^{\mu} q^{\nu} - p^{\nu} q^{\mu} 
\right ) = 
\left (
\begin{array}{cccc}
0 & - Q^{'}_{1} | Q_{3} & - Q^{'}_{2} | Q_{3} & - Q^{'}_{3} | Q_{3} \\
 Q^{'}_{1} | Q_{3} & 0 & Q^{'}_{3} & - Q^{'}_{2} \\
 Q^{'}_{2} | Q_{3} & - Q^{'}_{3} & 0 & Q^{'}_{1} \\
 Q^{'}_{3} | Q_{3} & Q^{'}_{2} & -Q^{'}_{1} & 0 
\end{array} 
\right )
\label{def.pq.minus.pq}
\end{equation}
This in fact has the form of a relativistic three vector in 
$ \vec Q^{'} $, as desired earlier, although as noted, standard Dirac Theory assigns it an invariant form under Lorentz Transformations, rather than treating it as a true three vector. This particular antisymmetric form corresponds to the second order equation containing the effect of the upper row of 
$ \Sigma^{\mu \nu} $, an equation which is by itself a foundation for Dirac Theory\cite{sakurai}, and adequate to arrive at the alternate second order equation containing the effect of the lower row of 
$ \Sigma^{\mu \nu} $. At this point, standard Dirac Theory makes contact with equation (\ref{leap.qdirac.wave.eq}).

Standard Dirac Theory then formalizes the requirement that 
$ \vec Q^{'} = \vec Q $ in all locally Lorentzian Frames\cite{bade.jehle, drell} (in Cartesian coordinates) as follows. If a boost along the 
$ + x $ axis for a contravariant four vector is 
$ \bar V^{\mu} = a^{\mu}_{\; \; \nu} V^{\nu} $, then with 
$ \tanh{\phi} = ( v / c ) $,
\begin{equation}
a^{\mu}_{\; \; \nu} = 
\left (
\begin{array}{cccc}
\cosh{\phi} & - \sinh{\phi} & 0 & 0 \\
 - \sinh{\phi} & \cosh{\phi} & 0 & 0 \\
0 & 0 & 1 & 0 \\
0 & 0 & 0 & 1 
\end{array} 
\right )
\label{def.lorentz.boost.px.contra}
\end{equation}
and the corresponding transformation for a covariant spacetime four vector is
\begin{equation}
A_{\mu}^{\; \; \nu} = 
\left (
\begin{array}{cccc}
\cosh{\phi} & \sinh{\phi} & 0 & 0 \\
\sinh{\phi} & \cosh{\phi} & 0 & 0 \\
0 & 0 & 1 & 0 \\
0 & 0 & 0 & 1 
\end{array} 
\right )
\label{def.lorentz.boost.px.co}
\end{equation}
Then let
\begin{equation}
\Lambda = Q_{0} \cosh{ ( \phi / 2 ) } + Q_{1} 
\sinh{ ( \phi / 2 ) } | Q_{3}
\label{def.lambda}
\end{equation}
where the real quaternion form of this expression is adapted for the notation of this paper from the real quaternion forms investigated by De Leo\cite{De.Leo.1995} (for boosts along the other two axes, replace 
$ Q_{1} $ by the appropriate 
$ Q_{j} $). In all these cases, the inverse transformation is found by letting the hyperbolic sine terms reverse sign.

Then for
\begin{equation}
U = 
\left (
\begin{array}{cc}
\Lambda & 0 \\
0 & \Lambda^{-1} 
\end{array} 
\right )
\label{def.unitary.U}
\end{equation}
it is true that 
\begin{equation}
\gamma^{\mu} = \bar \gamma^{\mu} = a^{\mu}_{\; \; \nu} U \gamma^{\nu} U^{-1}
\label{gammamu.lorentz}
\end{equation}
or
\begin{equation}
p^{\mu} = \bar p^{\mu} = a^{\mu}_{\; \; \nu} \Lambda p^{\nu} \Lambda
\label{pmu.lorentz}
\end{equation}
and
\begin{equation}
q^{\mu} = \bar q^{\mu} = a^{\mu}_{\; \; \nu} \Lambda^{-1} q^{\nu} \Lambda^{-1}
\label{qmu.lorentz}
\end{equation}
Then overall consistency in equation (\ref{leap.qdirac.wave.eq}) gives 
$\bar \psi = \Lambda \psi $. This is the standard picture of 
$ \psi $ transforming as a spinor, with 
$ \vec Q $ invariant in form under the associated Lorentz Transformation\cite{bade.jehle}.

If instead of a boost, the transformation is a spatial rotation about the 
$ + x $ axis by angle 
$ \phi $, measured counterclockwise from the 
$ + z $ axis viewed looking down 
$ + x $ axis toward the origin, then
\begin{equation}
a^{\mu}_{\; \; \nu} = 
\left (
\begin{array}{cccc}
1 & 0 & 0 & 0 \\
0 & 1 & 0 & 0 \\
0 & 0 & \cos{\phi} & \sin{\phi} \\
0 & 0 & - \sin{\phi} & \cos{\phi} 
\end{array} 
\right )
\label{def.rotation.around.px.contra}
\end{equation}
Matching this now is
\begin{equation}
\Lambda = Q_{0} \cos{ ( \phi / 2 ) } - Q_{1} 
\sin{ ( \phi / 2 ) }
\label{def.lambda.rot}
\end{equation}
and for rotations about the other two axes, replace 
$ Q_{1} $ by the appropriate 
$ Q_{j} $. The inverse transformations are found by letting the sine terms reverse sign. Just as with the boost, 
$ \psi $ transforms as a spinor via 
$\bar \psi = \Lambda \psi $.

However, if 
$ \vec Q^{'} $ in equation (\ref{def.pq.minus.pq}) must transform as a relativistic three vector, then the picture of Dirac Theory developed in 1961 by Usachev\cite{jtep.usachev} is called for. This treats the wavefunction as a scalar (and thus both 
$ \psi $ and 
$ \chi $ as scalars), and the gammas as four vectors. That also means treating the antisymmetric tensor in equation (\ref{def.pq.minus.pq}) as an antisymmetric tensor under the transformation in equation (\ref{def.lorentz.boost.px.contra}) for a boost along the 
$ + x $ axis, as well as the transformation in equation (\ref{def.rotation.around.px.contra}) for a rotation about the 
$ + x $ axis. If before the boost, 
$ \vec Q^{'} = \vec Q $, then after the boost,
\begin{equation}
\bar Q^{'}_{1} = Q^{'}_{1} = Q_{1}
\label{qprime.1.boost}
\end{equation}
\begin{equation}
\bar Q^{'}_{2} = Q^{'}_{2} \cosh{\phi} - 
Q^{'}_{3} \sinh{\phi} | Q_{3} = 
Q_{2} \cosh{\phi} - Q_{3} \sinh{\phi} | Q_{3}
\label{qprime.2.boost}
\end{equation}
\begin{equation}
\bar Q^{'}_{3} = Q^{'}_{3} \cosh{\phi} + 
Q^{'}_{2} \sinh{\phi} | Q_{3} = 
Q_{3} \cosh{\phi} + Q_{2} \sinh{\phi} | Q_{3}
\label{qprime.3.boost}
\end{equation}
Likewise for the rotation about the 
$ + x $ axis,
\begin{equation}
\bar Q^{'}_{1} = Q^{'}_{1} = Q_{1}
\label{qprime.1.rot}
\end{equation}
\begin{equation}
\bar Q^{'}_{2} = Q^{'}_{2} \cos{\phi} + 
Q^{'}_{3} \sin{\phi} = 
Q_{2} \cos{\phi} + Q_{3} \sin{\phi}
\label{qprime.2.rot}
\end{equation}
\begin{equation}
\bar Q^{'}_{3} = Q^{'}_{3} \cos{\phi} - 
Q^{'}_{2} \sin{\phi} = 
Q_{3} \cos{\phi} - Q_{2} \sin{\phi}
\label{qprime.3.rot}
\end{equation}
These resulting 
$ \bar Q^{'}_{j} $ may be easily shown to be isomorphic to the 
$ Q_{j} $ by direct multiplication, so they are acceptable forms to serve as 
$ \bar Q^{'}_{j} $. Note that 
$ \Lambda \bar Q^{'}_{j} \Lambda^{-1} = Q_{j} $.

In practice in Cartesian coordinates, the root 
$ \sqrt{ \hat Y } $, extracted as in equation (\ref{spin.squared}), will be specified using 
$ \vec Q^{'} = \vec Q $ in every frame, which effectively treats 
$ \psi $ as a spinor. Nevertheless, Usachev's work shows that this is simply a convention, not a fundamental requirement. Rather, for the model of this paper the above transformations are the formally correct values for the 
$ \bar Q^{'}_{j} $ after the transformations specified. Usachev's paper\cite{jtep.usachev} actually treats such cases much more fully and neatly. However the above results correspond to his tetrad 
$ n^{\mu}_{a} $ having the specific values 
$ \delta^{\mu}_{a} $ in the pre-transformation frame. That will give the above forms.

\subsection{Comments on the Darwin Hydrogen Atom Solution}

Besides ``free particle'' solutions of Dirac's Equation, physics is concerned with bound state solutions of that equation. The best known such solution is for the hydrogen atom problem, and C. G. Darwin provided the original solution of this problem in terms of a standard wave function 
$ \psi $, rather than through use of non-commutative algebra\cite{darwin}. To translate his notation into the framework of this paper, his symbols 
$ p_{\mu} $ can be understood to be
\begin{equation}
p^{\mu} = \hat \eta^{\mu \nu} \left [ - { \hbar \over \imath } \, 
\left ( { { \partial \; } \over { \partial x^{\nu} } } + 
\imath q A_{\nu} \right ) \right ]
\label{darwin.p}
\end{equation}
where 
$ \hat \eta^{\mu \nu} $ is the flat spacetime Lorentz metric with signature 
$ ( + , - , - , - ) $, and 
$ A_{\mu} = ( \Phi , - \vec A ) $, with the electromagnetic potentials 
$ \Phi $ and $ \vec A $ defined as in Jackson\cite{jackson.self.force} (as assumed throughout this paper). Furthermore, Darwin uses notation 
($ P_{k}^{u} $) and conventions for the spherical harmonics that need translation into more familiar, recent notation. To do this, replace Darwin's notation of 
$ P_{k}^{u} $ for spherical harmonics with a more familiar looking 
$ Y^{'}_{l m} $ where 
$ k \rightarrow l $ and 
$ u \rightarrow m $ (this is {\it not} the half-integral index 
$ m $ Darwin uses), and let 
$ P_{l}^{m} $ be {\it redefined} instead as Associated Legendre Functions via
\begin{equation}
P_{l}^{m} \! ( x ) = { 1 \over { 2^{l} l ! } } \left ( 
1 - x^{2} \right )^{ m / 2 } 
{ { d^{ l + m } \; } \over { d x^{ l + m } } } 
\left ( x^{2} - 1 \right )^{l}
\label{m.and.m.alp}
\end{equation}
This definition is almost the same as Jackson's definition\cite{jackson.self.force}, except that he includes a factor of 
$ ( - 1 )^{m} $ in his definition that this omits. With these changes, then Darwin's spherical harmonics that he denotes by 
$ P_{k}^{u} $, become instead
\begin{equation}
Y^{'}_{l m} = \left ( l - m \right ) ! \, P_{l}^{m} \! ( 
\cos{ \theta } ) e^{ \imath m \phi }
\label{darwin.sp.h}
\end{equation}
A prime is used in this definition to denote the fact that this definition of spherical harmonics still differs from Jackson's definition of 
$ Y_{l m} $ in its ``normalizing'' factor, and a factor of 
$ ( - 1 )^{m} $ in Jackson's definition.

Besides those notational differences, Darwin also uses a different representation of Dirac Theory in which his choice of the gamma matrices, 
$ \gamma^{\mu}_{D} $, differs from the gamma matrices in equation (\ref{def.gammas.std}) in that 
$ \gamma^{0}_{D} = - \gamma^{0}_{S} $, and 
$ \gamma^{k}_{D} = \gamma^{k}_{S} $ for 
$ k = 1,2,3 $. Furthermore, the desired Dirac representation in this paper is the chiral one of equations (\ref{def.gamma0}) and (\ref{def.gammak}), which is also different from the representation of equation (\ref{def.gammas.std}). The unitary transformation
\begin{equation}
U =
{ 1 \over \sqrt{2} } \, 
\left (
\begin{array}{cc}
- \imath \sigma_{0} & - \imath \sigma_{0} \\
\imath \sigma_{0} & - \imath \sigma_{0} 
\end{array} 
\right )
\label{def.darwin.to.chiral}
\end{equation}
is a valid (although more complicated than necessary) 
$ U $ such that the original chiral gammas of equations (\ref{def.gamma0}) and (\ref{def.gammak}) are given by 
$ \gamma^{\mu} = {}^{\dagger} U \gamma^{\mu}_{D} U $ (with the 
``$ {}^{\dagger} $'' to the left). Then the chiral representation of Darwin's four spinor 
$ \psi_{D} $ is 
$ \psi_{C} = {}^{\dagger} U \psi_{D} $, again with the 
``$ {}^{\dagger} $'' to the left. If the top two rows of 
$ \psi_{C} $ are denoted by the two component spinor 
$ \zeta $, then 
$ \zeta $ is the quantity that satisfies equation (\ref{dirac.wave.eq}), and whose quaternion equivalent satisfies equation (\ref{leap.qdirac.wave.eq}), as may be seen by examining the discussion leading to equation(\ref{def.pq.minus.pq}).

Calculating 
$ \zeta $ using 
$ \psi_{C} = {}^{\dagger} U \psi_{D} $ then gives
\begin{equation}
\zeta = { 1 \over \sqrt{2} } 
\left (
\begin{array}{c}
\imath \psi_{D 1} - \imath \psi_{D 3} \\
\imath \psi_{D 2} - \imath \psi_{D 4} 
\end{array}
\right )
\label{def.darwin.zeta}
\end{equation}
where the components of Darwin's four spinor are still numbered 
$ 1 $ to $ 4 $ just as he numbers them. Darwin then considers two classes of his solutions. Only the first of those two cases is needed here, and gives the two components of 
$ \zeta $ (indexing from 
$ 0 $) as
\begin{equation}
\zeta_{0} = { { \left ( l - m \right ) ! } 
\over { \sqrt{2} } } 
\left [ \left ( l - m + 1 \right ) F_{l} 
P_{ l + 1 }^{m} \! ( \cos{ \theta } ) 
- \imath \left ( l + m + 1 \right ) G_{l} 
P_{l}^{m} \! ( \cos{ \theta } ) \right ] 
e^{ \imath \left ( m \phi - \omega t 
\right ) }
\label{def.chiral.zeta0}
\end{equation}
and
\begin{equation}
\zeta_{1} = { { \left ( l - m \right ) ! } 
\over { \sqrt{2} } } \left [ F_{l} 
P_{ l + 1 }^{ m + 1 } \! ( \cos{ \theta } ) 
+ \imath G_{l} 
P_{l}^{ m + 1 } \! ( \cos{ \theta } ) \right ] 
e^{ \imath \left [ \left ( m + 1 \right ) 
\phi - \omega t \right ] }
\label{def.chiral.zeta1}
\end{equation}
where the 
$ e^{ - \imath \omega t } $ in these expressions is implicit in Darwin's solution.

This spinor 
$ \zeta $ has the equivalent, factored quaternion form
\begin{equation}
\zeta = \beta 
e^{ - \left ( | Q_{3} \right ) \left ( m 
\phi - \omega t \right ) }
\label{def.chiral.quat.zeta}
\end{equation}
where the quaternion 
\begin{equation}
\beta = \sum_{ \mu = 0 }^{3} Q_{\mu} \beta_{\mu}
\label{def.quat.beta}
\end{equation}
with components
\begin{equation}
\beta_{0} = { { \left ( l - m + 1 \right ) ! } 
\over { \sqrt{2} } } F_{l} P_{ l + 1 }^{m} \! 
( \cos{ \theta } )
\label{def.quat.beta0}
\end{equation}
\begin{equation}
\beta_{1} = - { { \left ( l - m \right ) ! } 
\over { \sqrt{2} } } \left [ F_{l} 
P_{ l + 1 }^{ m + 1 } \! ( \cos{ \theta } ) 
\sin{ \phi } + G_{l} P_{l}^{ m + 1 } \! 
( \cos{ \theta } ) \cos{ \phi } \right ]
\label{def.quat.beta1}
\end{equation}
\begin{equation}
\beta_{2} = { { \left ( l - m \right ) ! } 
\over { \sqrt{2} } } \left [ F_{l} 
P_{ l + 1 }^{ m + 1 } \! ( \cos{ \theta } ) 
\cos{ \phi } - G_{l} P_{l}^{ m + 1 } \! 
( \cos{ \theta } ) \sin{ \phi } \right ]
\label{def.quat.beta2}
\end{equation}
and
\begin{equation}
\beta_{3} = { { \left ( l + m + 1 \right ) 
\left ( l - m \right ) ! } \over 
{ \sqrt{2} } } G_{l} P_{l}^{m} \! 
( \cos{ \theta } )
\label{def.quat.beta3}
\end{equation}
Of course, there is some ambiguity in equation (\ref{def.chiral.quat.zeta}) because the 
$ ( | Q_{3} ) $ in the exponential can be replaced by 
$ ( Q_{3} ) $ without the leap-over bar, for one or both terms. If the leap-over operator is used in any part of the exponential, such as a general form 
$ e^{ - \left ( | Q_{3} \right ) \eta} $, then the 
``$ {}^{\dagger} $'' of that exponential form is assumed to be 
$ e^{ \left ( | Q_{3} \right ) \eta} $, still located on the right. Exponential forms without the leap-over operator follow conventional rules for the 
``$ {}^{\dagger} $'' of a product of quaternions. All of this has some effect on what follows.

Note that 
$ \zeta $ and 
$ \beta $ are indeed scalars. 
$ \beta $ is obviously the scalar inner product of the Cartesian quaternion basis with the four Cartesian components of 
$ \beta $, and this quaternion basis is independent of coordinates. Thus the partial derivatives of 
$ \beta $ are found by simply taking the partial derivatives of components 
$ \beta_{\mu} $ in equation (\ref{def.quat.beta}), and 
$ \zeta $ is just as simple using the product rule for derivatives. The simplicity of this procedure provides ample reason for using these Cartesian components of 
$ \beta $, rather than spherical spatial coordinate components. While a check will show that such spherical coordinate components of 
$ \beta $ are in some ways simpler in form than the Cartesian components (for example, the residual dependence on 
$ \phi $ is completely absorbed into the basis in 
$ \beta $), the basis quaternions then become functions of the coordinates. These also contribute to partial derivatives of 
$ \beta $, so partial derivatives become more complicated. Now with this understanding, the integrability conditions of equation (\ref{integrability.conds.short}) for a wavefunction can be checked for this family of hydrogen atom solutions given by Darwin. The derivatives involved can be taken directly in the variables 
$ ct, r, \theta, \phi $.

These now give
\begin{equation}
\zeta_{ , 0 } = \zeta \left [ { { \omega } 
\over { c } } \left ( | Q_{3} \right ) 
\right ]
\label{zeta.comma.0}
\end{equation}
and
\begin{equation}
\zeta_{ , 3 } = { 1 \over 2 } Q_{3} \zeta - 
{ 1 \over 2 } \zeta Q_{3} - m \zeta 
\left ( | Q_{3} \right )
\label{zeta.comma.3}
\end{equation}
where 
$ | Q_{3} \rightarrow Q_{3} $ is allowed in either equation, as noted above concerning ambiguity in equation (\ref{def.chiral.quat.zeta}). The equivalent spinor form of equation (\ref{zeta.comma.3}) is easily seen to be the eigenvalue equation
\begin{equation}
\left [ { \hbar \over \imath } { 
{ \partial \; } \over { \partial \phi } } + { \hbar 
\over 2 } \, \sigma_{3} \right ] \zeta = 
\left ( m + { 1 \over 2} \right ) \hbar \zeta
\label{spinor.zeta.comma.3}
\end{equation}
which is an expected result for the 
$ z $ component of orbital plus spin angular momentum\cite{drell,darwin}.

Equation (\ref{zeta.comma.0}) immediately gives that the integrability conditions (\ref{integrability.conds.short}) containing the 
$ 0 $ coordinate are trivially true, since the leap-over operator keeps the 
$ ( | Q_{3} ) $ to the far right. Thus, if the time derivative is to be important in the integrability conditions, the leap-over operator must be dropped by having 
$ ( | Q_{3} ) \rightarrow ( Q_{3} ) $ in the 
$ \omega t $ term in the exponential. If this is done, then the integrability condition (with the 
``$ {}^{\dagger} $'' to the left)
\begin{equation}
\zeta_{, 0} {}^{\dagger} \! \zeta \zeta_{, 3} = 
\zeta_{, 3} {}^{\dagger} \! \zeta \zeta_{, 0}
\label{integrability.conds.short.0.3}
\end{equation}
will give
\begin{equation}
Q_{3} {}^{\dagger} \! \beta Q_{3} \beta = 
{}^{\dagger} \! \beta Q_{3} \beta Q_{3}
\label{integrability.conds.short.0.3.result}
\end{equation}
regardless of whether a leap-over operator is retained or dropped in the 
$ m \phi $ portion of the exponential. But equation (\ref{integrability.conds.short.0.3.result}) merely says that 
$ {}^{\dagger} \! \beta Q_{3} \beta Q_{3} $ must be real. But a check will show that cannot be satisfied for nontrivial (nonzero) quantities 
$ F_{l} $ and 
$ G_{l} $. Thus, the integrability conditions fail to be satisfied in this case.

This leaves the case when the leap-over operator is used with the 
$ \omega t $ term in the exponential, and thus only the three integrability conditions involving spatial coordinates remain to be checked. The condition
\begin{equation}
\zeta_{, 1} {}^{\dagger} \! \zeta \zeta_{, 3} = 
\zeta_{, 3} {}^{\dagger} \! \zeta \zeta_{, 1}
\label{integrability.conds.short.1.3}
\end{equation}
then leads to an equation among commutators,
\begin{equation}
\beta_{, 1} {}^{\dagger} \! \beta \left [ Q_{3} , 
\beta \right ] + \beta \left [ Q_{3} , 
{}^{\dagger} \! \beta \right ] \beta_{, 1} - 
2 m \beta \left [ {}^{\dagger} \! \beta 
\beta_{, 1} , Q_{3} \right ] = 0
\label{integrability.conds.short.1.3.result}
\end{equation}
where the third term on the left side is only present when the leap-over operator is {\it not} used with the 
$ m \phi $ term in the exponential in 
$ \zeta $, and even then, it vanishes for any 
$ m = 0 $ case. Regardless, the Darwin solution still appears {\it not} to satisfy this relation in a nontrivial way. Thus, if the (very important) Darwin solution is to be represented as a solution in this framework, it would seem to be necessary to return to the ``spin-up'' and ``spin-down'' projections of it into the complex plane, as outlined in the main body of this paper. Such representations of this classic solution to Dirac's Equation will not run afoul of the integrability conditions of equation (\ref{integrability.conds.short}), and will still allow the solution's necessary existence in the family of valid solutions.

I suspect this feature is common to many bound state solutions to Dirac's Equation, and that further examination of such projective, complex solutions is in order. That topic appears appropriate for a separate paper, including energy and charge integrals for the Darwin solution for the ``spin-up'' and ``spin-down'' complex projections, and also superpositions of those complex solutions, since such superpositions appear to be valid in the treatment of this paper's body.

\subsection{Conclusions}

From this sketch of quaternionic gauges and curvatures, it appears that the excursion into quaternions will produce the Dirac Equation form of equation (\ref{leap.qdirac.wave.eq}) if an equal mixture of nonmetricity and torsion is introduced 
($ n = 1 / 2 $). A second, auxiliary wavefunction 
$ \chi $ then also plays a role that is of interest in its own right. At the very least, its required coexistence with 
$ \psi $ imposes integrability conditions upon 
$ \psi $, conditions easily satisfied for ``free particle'' cases examined. Both wavefunctions are treated as quaternionic spacetime scalars rather than spinors, a step supported by the 1962 paper of Usachev\cite{jtep.usachev}. Bound state solutions seem more likely to appear in terms of projections into the complex plane, as described in the main body of this paper.

\begin{acknowledgments}

I would like to thank Howard Brandt and William Baylis, both members of the audience at my presentation of the Rankin-Taylor contributed talk at the April, 1997 APS Spring Meeting\cite{taylor.rankin.1997}, as well as Engelbert Sch\"{u}cking, all for suggesting that I look at Clifford Algebra generalizations of reference \cite{rankin.caqg}. Additionally, I would like to thank Jim Wheeler, David Finkelstein, Daniel Galehouse, Egon Marx, and James Anderson for helpful discussions and points that have contributed significantly to this paper. I would also like to thank Frank Taylor, who coauthored several conference contributions on this theory with me, including demonstrations of inflationary 
($ \hat a = K^{2} = 0 $) and standard 
($ \hat a = K^{2} = -36 $) cosmological solutions\cite{taylor.rankin}. And, I would like to thank Wayne Bowers, Thomas J. Boyd, Jr., Bob Levine, and Bob Rankin for helpful discussions and general support during various stages of this work, and also S.R. for support. I would also like to thank Robert H. Rohrer for encouraging me to investigate units of measure in physics.

\end{acknowledgments}



\begin{thebibliography}{99}
\bibitem{galehouse.minus.C}  My thanks to Daniel Galehouse for correctly insisting that cases which correspond to negative values of B (which thus require $ C = -1 $) are legitimate.
\end{thebibliography}
\end{document}